\documentclass[12pt,reqno]{amsart}
\usepackage{fullpage}
\usepackage{amsfonts}
\usepackage{mathbbol}
\usepackage{amssymb}
\usepackage{times}
\usepackage{graphicx}

\usepackage{mathtools}
\usepackage{multirow}
\usepackage{enumerate}

\usepackage{epsfig}

\usepackage{subcaption}
\usepackage{xcolor}

\graphicspath{ {MATLAB/} }

\newtheorem{theorem}{Theorem}
\numberwithin{theorem}{section}

\numberwithin{stheorem}{subsection}

\newtheorem{lemma}[theorem]{Lemma}

\newtheorem{proposition}[theorem]{Proposition}
{ \theoremstyle{remark} }
\newtheorem*{theorem*}{Theorem}
\newtheorem*{lemma*}{Lemma}

\DeclareMathOperator*\Tr{Tr}

\DeclareMathOperator*\erfc{erfc}

\DeclareMathOperator*\Mat{Mat}
\DeclareMathOperator*\diag{diag}

\DeclareMathOperator*\Vol{Vol}
\DeclareMathOperator*\Pf{Pf}
\DeclareMathOperator*\Sp{Sp}

\renewcommand*{\vec}[1]{\boldsymbol{#1}}

\begin{document}

\title[Article]{Truncations of Random symplectic unitary  Matrices}
\author{Boris A. Khoruzhenko and Serhii Lysychkin}
\address{School of Mathematical Sciences, Queen Mary University of London, London E1 4NS, United Kingdom}
\date{\today}
\email{b.khoruzhenko@qmul.ac.uk, s.lysychkin@qmul.ac.uk}

\begin{abstract}
This paper is concerned with complex eigenvalues of truncated unitary quaternion   matrices equipped with the Haar measure. The joint eigenvalue probability density function is obtained for truncations of any size. We also obtain the spectral density and the eigenvalue correlation functions in various scaling limits. In the limit of strong non-unitarity the universal complex Ginibre form of the correlation functions is recovered in the spectral bulk off the real line after unfolding the spectrum. In the limit of weak non-unitarity we obtain the spectral density and eigenvalue correlation functions for all regions of interest. Off the real line the obtained expressions coincide with those previously obtained for truncations of Haar unitary complex matrices. 
\end{abstract}

\maketitle

\section{Introduction}

This paper is a contribution to the studies of universality of eigenvalue statistics in the complex plane.
Interest to complex eigenvalues of random matrices goes back to the pioneering work of Ginibre  \cite{Gin1965}. 
Although Ginibre was driven by mathematical curiosity, the 
non-Hermitian random matrices introduced in his paper, and the like, have found many applications. For example, they proved to be a useful analytical tool for studying stability of large complex systems  in linear \cite{May1972} and non-linear settings \cite{FK2016,BFK2021}, the dissipative quantum maps \cite{HaakeBook}, statistics of resonances in open chaotic systems \cite{FSReview1997,FK99,FSReview2003}, and the chiral transition at non-zero chemical potential in quantum chromodynamics \cite{S96}. Eigenvalue statistics of non-Hermitian random matrices appear in studies  of synchronisation in complex networks \cite{TWT2004}, localisation of a quantum particle in disordered media subject to an imaginary vector potential \cite{HN1996} and two-dimensional diffusion with random initial conditions \cite{PS2018}. 

An important practical problem of interest in random matrix theory is the asymptotic evaluation of the eigenvalue correlation functions in various scaling limits when the matrix dimension tends to infinity. In this context, the three Gaussian ensembles of complex, real and quaternion-real matrices introduced by Ginibre have been the subject of numerous investigations, see  \cite{KSReview2011} for a short review. 
The picture which emerged from these investigations is that in the bulk of the 
spectrum away from the real line
there is no difference between complex, real and quaternion-real ensembles. 
This is in contrast to Hermitian random matrices where the eigenvalue correlations depend on the symmetry class.

To clarify this point, we recall that the three Ginibre ensembles 
are defined by the joint density 
\begin{align}\label{eq:1G}
p_N(G)=c_{\beta,N}e^{-\frac{ \beta}{2}  \Tr G^{\dagger}G} 
\end{align} 
of matrix elements  on the spaces of $N\times N$ matrices with complex, real and real quaternion matrix elements,
Here, $\beta$ is the real dimension of a single matrix entry and  $G^{\dagger}$ is the matrix transpose of $G$ for matrices with real entries ($\beta=1$) and the Hermitian conjugate transpose for matrices with complex ($\beta=2$)  and real quaternion entries ($\beta=4$). 

In the complex Ginibre ensemble the eigenvalue correlation functions  $R_n(z_1, \ldots, z_n)$  have a determinantal form \cite{Gin1965}. The determinantal kernel is just a truncated exponential series and its evaluation in various scaling limits (e.g., in the bulk or at the the edge of the eigenvalue distribution) is fairly routine. For example, in the limit of infinite matrix dimension $N\to\infty$, the one-point correlation function $R_1(z)$ (which is the average eigenvalue density with the normalisation $\int_{\mathbb{C}} R_1(z) d^2z =N$) 
is asymptotically $1/\pi$ 
inside the circle $|z|=\sqrt{N}$, i.e., for every $|u|<1$,
\begin{align}\label{eq00a}
\lim_{N\to\infty} R_1\big(\sqrt{N}u\big)=1/\pi\, , 
\end{align}
 falling sharply to zero at the boundary:
\begin{align}\label{eq:0}
\lim_{N\to\infty}  R_1\big(\big(\sqrt{N} + s \big)e^{i\varphi}\big) = \frac{1}{2\pi} \erfc \big(\sqrt{2} s \big), \qquad s\in \mathbb{R}. 
\end{align}
Locally in the bulk of the spectrum (at $z=\sqrt{N}u$, $|u|<1$), the two-point eigenvalue correlation function is given by 
\begin{align}\label{eq:1}
R_2 ( z_1, z_2)  \sim \big(R_1(z)\big)^2 \Big[1- e^{-\pi R_1(z) |z_1-z_2|^2}\Big], \quad z_1-z_2=O(1), 
\end{align}
and, more generally \cite{Gin1965}, 
\begin{align}\label{Gin_corr}
\lim_{N\to\infty}
\frac{1}{\big(R_1(z)\big)^n}R_n \left(z+\frac{s_1}{\sqrt{R_1(z)}}, \ldots z+\frac{s_n}{\sqrt{R_1(z)}}\right) = \det{ \left[ e^{ \left( \pi (s_k \overline{s_l} - \frac{1}{2} |s_k|^2 -\frac{1}{2} |s_l|^2) \right) } \right] }_{k, l = 1}^n.
\end{align}
%
We shall refer to the functional form of the eigenvalue correlation functions on the right-hand side in Eq.~ \eqref{Gin_corr} as the Ginibre correlations.

Turning to the quaternion-real Ginibre ensemble, 
the complex eigenvalues of quaternion matrices $G$ 
drawn from this ensemble appear as $N$ pairs of complex conjugated points $z_j$ and ${z}_j^*$ which 
are the eigenvalues of 
the $2N\times 2N$ complex matrix representation of $G$.  Each of these pairs is uniquely characterised by the eigenvalue in the upper half of the complex plane. This gives rise to two trivially related pictures of the eigenvalue distribution: one is of $N$ pairs in the complex plane 
with `eigenpair' correlation functions
 $R_n(z_1, \ldots , z_n)$, $z_j\in \mathbb{C}$,  and the other is of $N$ points in the upper half of the complex plane 
with the eigenvalue correlation functions
 \begin{align}\label{eq:6}
 R_n^{(+)}(z_1, \ldots , z_n) = 2^n R_n(z_1, \ldots , z_n), \quad z_j\in \mathbb{C}_{+}. 
 \end{align}
The former is prevailing in the literature on the quaternion-real ensembles, however,  we shall use both, as 
the latter is more suitable for  comparison of the eigenvalue correlation functions\footnote{$R_n(z_1, \ldots , z_n)$ is the probability density of finding a pair of complex conjugated eigenvalues around each of $2n$ points $z_j$, ${z}_j^*$ with the position of the remaining $N-n$ pairs being unobserved. Consequently, the normalisation of $R_n(z_1, \ldots , z_n)$ is slightly mismatched relative to the matrix dimension $2N$, e.g. $\int_{\mathbb{C}} R_1(z) d^2z =N$ and not $2N$. The correlation functions $R_n^{(+)}(z_1, \ldots , z_n)$ are free from this issue.}. 
The joint eigenvalue density in the latter picture can be interpreted as the Boltzmann factor for one-component plasma system 
with a neutralising background
\cite{F2016} and Chapter 15.9 in \cite{ForresterBook}.

In the limit of infinite matrix dimension, the eigenvalue density $R_1(z)$ in the quaternion-real Ginibre ensemble is the same as in the complex ensemble except for a small (relative to the typical eigenvalue) neighbourhood of the real line.
For every non-real $u$, $\lim_{N\to\infty} R_1\big( \sqrt{N} u \big) = 1/\pi$ inside the circle $|u|=1$  \cite{H2000,KSReview2011}, falling sharply to zero at the boundary \cite{Thesis,KL_Gin}
\begin{align}\label{eq:3}
\lim_{N\to\infty} R_1\big(\big(\sqrt{N} + s\big)e^{\pm i\varphi}\big) = \frac{1}{2\pi} \erfc (2s), \qquad 0<\varphi <\pi, \quad s\in \mathbb{R}.
\end{align}
On the local scale, away from the real line in the bulk of the spectrum (at $z=\sqrt{N}u$ with $|u|<1$, $\Im u \not= 0$),
the two-point eigenpair correlation function in the quaternion-real Ginibre ensemble is asymptotically given by 
\cite{H2000,KL_Gin}
\begin{align}\label{eq:5}
R_2(z_1,z_2)\sim (R_1(z))^2\big(1-e^{-2\pi R_1(z) |z_1-z_2|^2}\big), \quad z_1-z_2=O(1), \, \, \Im z_{1,2} \propto \sqrt{N}\, .
\end{align}
Switching to the eigenvalue point process in the upper half-plane,    
Eq. \eqref{eq:5} transforms into Eq. \eqref{eq:1} with $R_n$ replaced by $R_n^{(+)}$, except for the restriction $ \Im z_{1,2} \propto \sqrt{N}$ which does not apply in the complex Ginibre ensemble. 
Thus, away from the real line in the bulk of the spectrum (complex bulk) the unfolded two-point correlation function in the quaternion-real Ginibre is identical to the one in complex ensemble. This is also true of all the higher order correlation functions  \cite{akemann2019universal,Thesis, KL_Gin}.

In the real Ginibre ensemble, the average eigenvalue density $R_1(z)$ contains two parts, a smooth part $R^{(C)}_1(z)$ which describes the average density of complex eigenvalues and a singular part 
which describes the average eigenvalue density of real eigenvalues. 
The smooth part, $R^{(C)}_1(z)$, vanishes on the real line and off the real line its asymptotic behaviour for $N$ large is described by Eqs  \eqref{eq00a} and \eqref{eq:0} \cite{Edelman1997}. Similar to the average eigenvalue density, the two-point  correlation function in the real Ginibre ensemble contains a smooth part $R^{(C)}_2(z_1,z_2)$ which describes the correlations of complex eigenvalues and singular parts which describe the correlations between real eigenvalues and real and complex eigenvalues. Away from the real line, the smooth part of the two-point correlation function is given, after trivial unfolding, by the same asymptotic expression \eqref{eq:1} as in the complex ensemble 
 \cite{Som2007}.  The same is also true of the smooth part of all the higher order eigenvalue correlation functions \cite{BorodinPaper}.


Thus, in the complex bulk of the spectrum  the average eigenvalue density and the local eigenvalue correlations in all three Ginibre ensembles are identical whilst differences between these ensembles appear near the real line\footnote{A heuristic explanation of this phenomenon based on the structure of the joint density of eigenvalues can be found in \cite{akemann2019universal}. }. This poses the question about universality of the Ginibre eigenvalue correlations \eqref{Gin_corr} in the complex bulk. 

The Gaussian distribution \eqref{eq:1G} is characterised by two properties: the statistical independence of matrix entries and the invariance of the matrix distribution with respect to changing the basis in the underlying vector space. Correspondingly, one can consider two different classes of ensembles of random matrices which intersect at the Gaussian distribution \eqref{eq:1G}. One consists of ensembles whose matrix entires in a certain basis are independent random variables. And the other one consists of ensembles with invariant matrix distribution.

The question of universality of the local eigenvalue statistics in the first class was answered affirmatively in \cite{TaoVu2015}. 
%
 Although only complex and real matrices were considered in this work it is almost certain that the techniques used in  \cite{TaoVu2015} can be extended to quaternion-real matrices as well. 
 
 Universality of the Ginibre eigenvalue correlations \eqref{Gin_corr} in the class of non-Hermitian random matrices with invariant matrix distributions is not yet well understood. In this context, most of the advances so far have been for matrices with complex entries. After unfolding, the Ginibre eigenvalue correlations in the bulk of the spectrum were recovered for truncated Haar unitary matrices 
 \cite{ZS2000}, the spherical ensemble \cite{K2009} and the induced versions of these two ensembles \cite{JonitThesis}  along with the induced Ginibre ensemble  \cite{FBKSZ2012}. The joint density of matrix entries in these three families of random matrix ensembles is given by 
%
\begin{align}\label{IndGin} 
p_N(G)\propto  \det (GG^{\dagger})^{M-N} h(G), \quad M\ge N, 
\end{align} 
where $N$ is the matrix dimension 
and $h(G)= e^{-\Tr GG^{\dagger} }$ for the induced Ginibre ensemble; $h(G)= {\det} (I_N-GG^{\dagger})^{K-M-N}$, $K\ge M+N$, for the induced truncations; and $h(G)= \det (I_N+GG^{\dagger})^{-(K+M)}$, $K\ge N$,   for the induced spherical ensemble. 
The underlying matrix models leading to the matrix distributions (\ref{IndGin})
and the meaning of parameters are discussed in  refs \cite{FBKSZ2012,JonitThesis,F2016}. In the scaling regime when $M-N$  stays finite in the limit $N\to\infty$, the origin of the complex plane is a special point: the eigenvalue correlations in the neighbourhood of this point are given by an expression which depends on $M-N$ and coincides with the Ginibre correlations \eqref{Gin_corr} only if $M=N$. However, away from the origin, and thus in the \emph{complex} bulk,
the three families of complex induced ensembles exhibit the Ginibre correlations regardless of whether $M-N$ stays finite or grows with $N$ \cite{JonitThesis}. The Ginibre correlations were also obtained for products of finite number of independent samples from matrix distributions \eqref{IndGin} \cite{AB2012,ABKN2014,LW2016}, and, also, for complex \emph{normal} matrices with joint density of matrix elements in the form $p_N(G)\propto e^{-N\Tr V(GG^{\dagger})}$  under fairly general conditions on the potential $V$ 
\cite{CZ1998,AHM2011}. 

It is also worth mentioning another class of complex random matrices which arise in the study of resonances in open quantum chaotic systems \cite{FSReview1997}.
These are the skew-Hermitian deformations
\begin{align}\label{Res} 
H-iW^{\dagger}W
\end{align} 
 of the Gaussian Unitary Ensemble (GUE). Here,  $H$ is a Hermitian matrix drawn from the GUE and $W^{\dagger}W$ is fixed matrix. In the weakly non-Hermitian limit when the rank of $W^{\dagger}W$ stays finite whilst the matrix dimension grows to infinity, the appropriately scaled (unfolded) eigenvalue correlation functions of \eqref{Res} are available in a closed form \cite{FK99}. In the special case when all of the non-zero eigenvalues of $W^{\dagger}W$ are equal, these correlation functions converge to the Ginibre form, Eqs~\eqref{eq:1} -- \eqref{Gin_corr} when one lets the rank of perturbation grow to infinity \cite{FSReview2003}. This naturally suggests that in the strongly non-Hermitian limit when the rank of $W^{\dagger}W$ is proportional to the matrix dimension, the random matrix ensemble \eqref{Res} belongs to the complex Ginibre universality class for some broad class of the skew-Hermitian perturbations. Proving this remains a challenging problem. In the weakly non-Hermitian case the skew-Hermitian deformations of the GUE belong to the universality class of sub-unitary matrices \cite{FSReview2003}.  The latter contains aforementioned truncations of unitary matrices as a special case. Namely, in the nomenclature
 of random matrices \eqref{Res}, truncations of unitary matrices correspond to all non-zero eigenvalues of $W^{\dagger}W$ being equal to 1.

As far as real non-Hermitian matrices with invariant matrix distributions are concerned, the real matrix analogues of the three induced ensembles, including the real spherical ensemble \cite{FM2012} and the truncations of Haar orthogonal matrices \cite{KSZ2010}, were studied in \cite{JonitThesis} and the Ginibre correlations recovered in the complex bulk. However, we are not aware of asymptotic evaluations of the correlation functions in the complex bulk for products of real random matrices even though finite $N$ expressions for correlation functions in such ensembles are known \cite{IK2014,FI2016,FIK2020}.

%

Turning now to the non-Hermitian quaternion-real matrices with invariant matrix distributions, these are the least developed ensembles in terms of the asymptotic evaluation of the expected eigenvalue density and scaling limits of the eigenvalue correlation functions. In this context, most of the results available in the literature are for the Gaussian ensembles: the quaternion-real Ginibre ensemble, 
its elliptic deformation 
and its chiral partner.

In the quaternion-real case the correlation functions, including the average eigenvalue density, are given in terms of a quaternion determinant (or equivalently Pfaffian) \cite{Mehta2ed,KanzPaper} with self-dual quaternion kernel expressed in terms of skew-orthogonal polynomials \cite{KanzPaper,AEP2021}. One strategy for calculating its scaling limits is to write a differential equation for the kernel  
suitable for asymptotic analysis. Such an approach was first used in \cite{MehtaShri1965} and then in different forms in \cite{H2000} and  \cite{KanzPaper} and yielded the eigenvalue correlation functions in the quaternion-real Ginibre ensemble  locally  at the origin and, after additional analysis in complex bulk  \cite{akemann2019universal}, and also near the real line including the real edges \cite{akemann2021scaling,SBE2021}.
This approach also works for other Gaussian matrix distributions, e.g., the elliptic deformation of the Ginibre ensemble \cite{SBE2021}, and, locally at the origin, for its chiral partner  \cite{AEP2021}. At the origin it also works for some non-Gaussian distributions, see \cite{akemann2021scaling} and references therein.

A different strategy for the asymptotic evaluation of the kernel in the Gaussian case was employed in \cite{KSReview2011} where the kernel was related to a product of two spectral determinants averaged over the ensemble distribution with a subsequent asymptotic evaluation of this average via Grassmann integration. Such an approach works well for the Ginibre ensemble and also for its elliptic deformation. In the latter case it reproduces the asymptotic form of the eigenvalue correlation functions found in the regime of weak non-Hermiticity in  \cite{KanzPaper} and in the regime of strong non-Hermiticity it yields, after a trivial rescaling, the same asymptotic form of the eigenvalue correlation functions as in the quaternion-real Ginibre ensemble both in the complex bulk and near the real axis. This approach is also hard to extend to non-Gaussian distributions. 

One source of difficulty of extending the asymptotic analysis of the eigenvalue correlation functions to non-Gaussian ensembles of quaternion matrices is that the determinantal/Pfaffian kernel is not isotropic (rotationally invariant) due to the eigenvalues occurring in complex conjugated pairs and its evaluation in various scaling limits is a harder analytical problem compared to ensembles of complex matrices. If one is interested in statistics of moduli of eigenvalues, e.g. the radial eigenvalue density, the (circular) hole probabilities, or the distribution of the largest modulus of the eigenvalues then the mathematics of the asymptotic analysis of quaternion matrices becomes similar to that of the complex matrices and one is able to advance it to non-Gaussian ensembles. For example, the radial density of eigenvalues for products of quaternion-real matrices in the limit of infinite matrix dimension was obtained in \cite{ipsen2013products}, and one-matrix calculations of various statistics of interest of moduli of eigenvalues  in quaternion-real Ginibre ensemble \cite{APS2009,RiderPaper} were extended to products of independent quaternion matrices drawn from non-Gaussian matrix distributions \cite{AkemannPaper,D2021}. However, asymptotic analysis of the average eigenvalue density as function of radius and angle and that of the eigenvalue correlation functions for non-Gaussian quaternion matrices is lacking at present and this serves as the main motivation for our work.

\medskip 

{\bf Acknowledgements.}  We would to thank Gernot Akemann and Yan Fyodorov for useful discussions and helpful comments on this manuscript. Financial support provided by Queen Mary University of London is gratefully acknowledged (SL, PhD studentship).

\section{Main results}
\label{S:2}
This paper is concerned with complex eigenvalues of truncated unitary quaternion   matrices, or, equivalently, eigenvalues of truncated symplectic unitary matrices. A brief summary of facts about quaternions and matrices of quaternions can be found in Section \ref{Sec:2}. 

Let $A$ be the top-left $N\!\times \!N$ corner block of unitary quaternion   matrix drawn at random from the quaternionic unitary group $U(N\!+\!M,\mathbb{H})$. 
The matrix elements of $A$ are real quaternions and can be represented by complex $2\!\times \!2$ matrices. Then, the complex matrix representation of $A$ can be identified with the top left $2N\!\times \!2N$ corner block of the symplectic unitary matrix drawn at random from the compact symplectic group $\Sp(2(N+M))$. 
The two groups are isomorphic and we will use them interchangeably. The complex eigenvalues of $A$ appear in complex conjugated pairs $(z_j, z_j^*)$ and are identical to those of its complex matrix representation. All of the $z_j$ lie in the unit disk $\mathbb{D}$ in the complex plane. The joint density of distribution of eigenpairs $p(z_1, \ldots , z_N)$  can be obtained in closed form, 
%
\begin{align}\label{JPDF_I}
p (z_1, \ldots, z_N) =   \frac{1}{(2\pi)^N N! \prod_{j=1}^N B\big(2M, 2j\big)}  \prod_{i \geq j} |z_j - {z_i^*}|^2 \prod_{i > j} |z_j - z_i|^2  \prod_{j=1}^{N} (1 - |z_j|^2)^{2M - 1}.
\end{align}
\noindent Here, $B(p,q)$ is the Beta function and the normalisation factor ensures that the area integral of 
$p(z_1, \ldots, z_N)$ 
over $\mathbb{D}^N$ is 1. 
Apart from the normalisation factor, Eq.~\eqref{JPDF_I} was reported in \cite{F2016} along with a possible strategy for its derivation in the range $M\ge N$. Our Theorem \ref{Thm:JPDF} in Section \ref{S:3} gives a derivation of Eq.~\eqref{JPDF_I} which is valid for every integer $M\ge 1$. 

Eq.~\eqref{JPDF_I} allows one to obtain the eigenpair correlation functions 
\begin{align}\label{Rn_I}
R_n (z_1, ..., z_n) = \frac{N!}{(N-n)!} \int_{\mathbb{D}} ... \int_{\mathbb{D}} p(z_1, \ldots z_{n}, z_{n+1} \ldots, z_N) d^2\!z_{n+1} ... d^2\!z_{N}
\end{align}
in terms of a quaternion determinant, or equivalently, in terms of a Pfaffian. The corresponding derivation can be performed following one in \cite{KanzPaper} and its outcome, Theorem \ref{Thm:4.1} in Section \ref{Sec:4}, is that the eigenpair correlation functions are given by 
\begin{align*}
R_n (z_1, ..., z_n) & =\prod_{j=1}^n \Big[ (z_j- z_j^*) (1-|z_j|^2)^{2M-1}\Big] \,  \Pf \big(K_N(z_k, z_l)\big)_{k, l = 1}^n\, .
\end{align*} 
Here $K_N(z, z')$ is the $2\times 2$ matrix kernel
 \begin{align}\label{KN_I}
K_N(z, z') =  \begin{bmatrix} g_N(z, z') & g_N(z, {z^{\prime}}^{*} ) \\ g_N( z^* , z') & g_N(\bar z, {z^{\prime}}^{*})\end{bmatrix} 
 \end{align}
with its matrix elements given by 
\begin{align}\label{gN_I}
g_N(z, z') &= \frac{ B\big(1/2, M\big)}{\pi} \sum_{k=0}^{N-1} \sum_{i =0}^k \frac{z^{2i} (z')^{2k+1} - z^{2k+1}(z')^{2i}   }{B\big(i+1, M\big) B\big(k+\frac{3}{2}, M\big)} \, .
\end{align}
In particular, 
\begin{align}\label{R1_I}
R_1(z)= \frac{ B\big(1/2, M\big)}{\pi} \, (z-z^*)\,  (1-|z|^2)^{2M-1}\, \sum_{k=0}^{N-1} \sum_{ i =0}^k \frac{z^{2i} (z^*)^{2k+1} - z^{2k+1}(z^*)^{2i}   }{B\big(i+1, M\big) B\big(k+\frac{3}{2}, M\big)} \, .
\end{align}
Up to a trivial factor, this function is proportional to the spectral density 
\begin{align}\label{rho_def}   
\rho_{2N} (z)=\mathbf{E} \left\{\frac{1}{2N}\sum_{j=1}^N \big( \delta(z-z_j)+\delta(z- z_j^*)\big) \right\} \, ,
\end{align}
where $\delta (z)=\delta (\Re z)\delta (\Im z)$. It is apparent that 
\begin{align}
\label{spdensity_I}
\rho_{2N}(z) 
= \frac{1}{2N} \big(R_1(z) +R_1( z^*)\big)
= \frac{1}{N} R_1(z)=\frac{2}{N} R_1^+(z)\, .
\end{align}

\begin{figure}[t!]
\includegraphics[width=.4\linewidth]{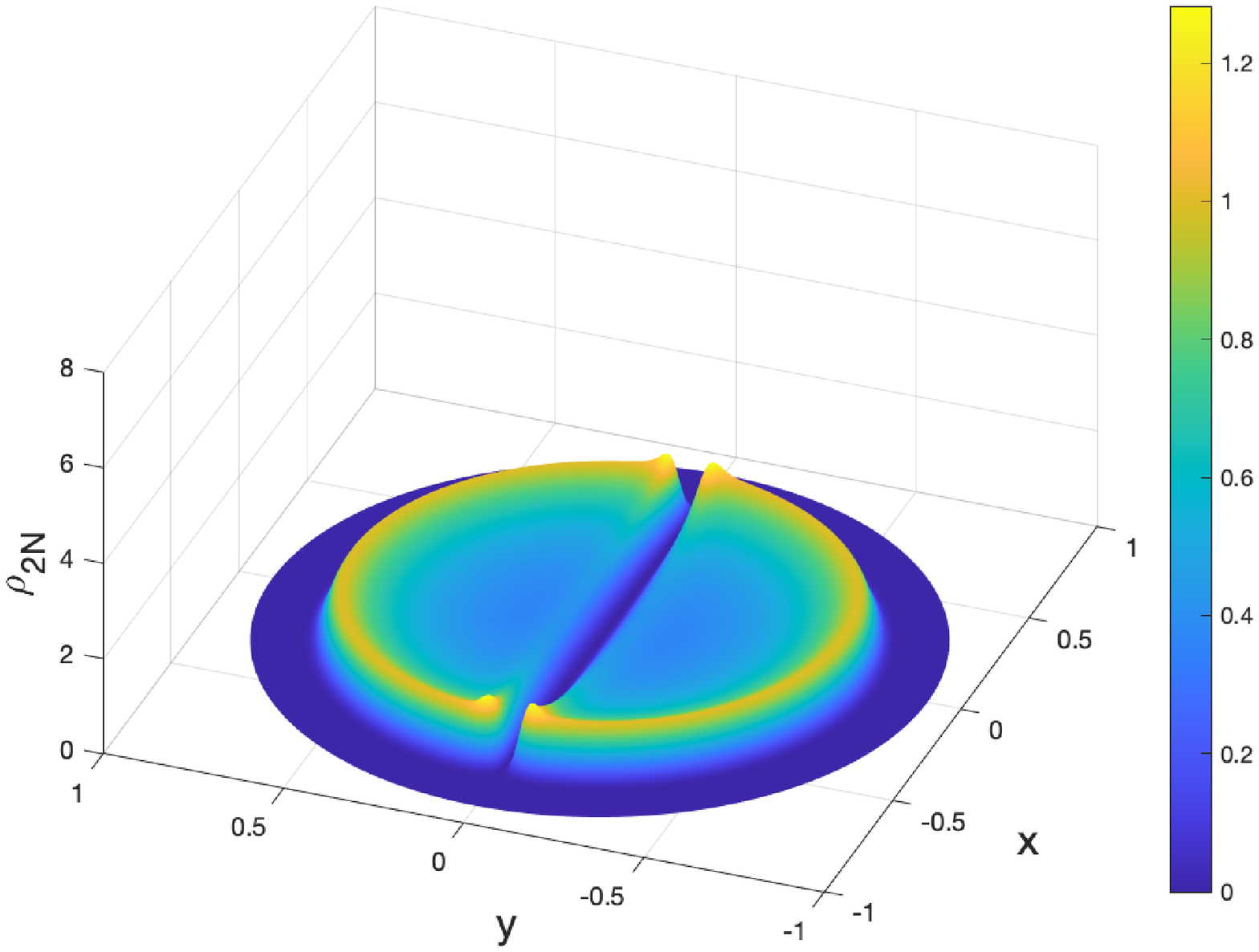}
\includegraphics[width=.4\linewidth]{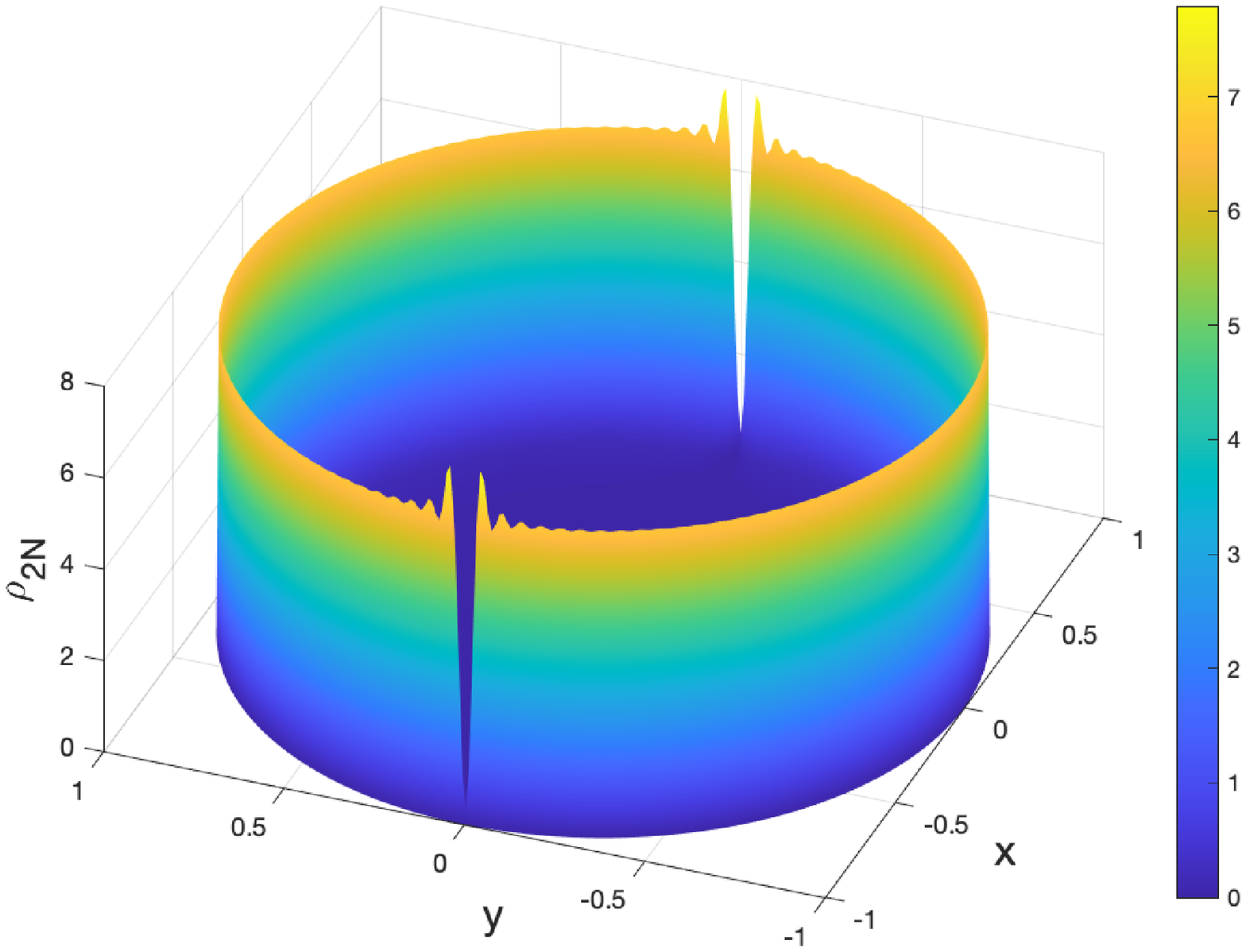}
\caption{Surface plots of the spectral density $\rho_{2N} (z)$ as function of $x=\Re z$ and $y=\Im z$. In the plot on the left-hand side $N=60$ and $M=60$ (strongly non-unitary regime). In the plot on the right-hand side $N=60$ and $M=1$ (weakly non-unitary regime).}
\label{Fig:1}
\end{figure}

Our main results are contained in Section \ref{Sec:5} where we carry out asymptotic analysis of the pre-kernel $g_N(z,z^{\prime})$ \eqref{gN_I} in the limit of infinite matrix dimension $N\to\infty$ in two regimes: the regime of \emph{strong non-unitarity} defined by the condition $M=aN, a>0$ and the regime of \emph{weak non-unitarity} defined by condition that $M$ stays finite when $N$ goes to infinity. The eigenvalue distribution on the macroscopic  (global) scale in these two regimes is visualised in Fig.~\ref{Fig:1} where we used Eqs~\eqref{R1_I} and \eqref{spdensity_I} to plot $\rho_{2N} (z)$ 
as function of $x=\Re z$ and $y=\Im z$.

\subsection{Strong non-unitarity ($M=aN, a>0, N\gg1$)}  In this regime the eigenvalues of $A$ fill the entire unit disk (except for the real line where $\rho_{2N} (z)$ vanishes at every finite $N$), and the spectral density $\rho_{2N} (z)$ is distinctively non-flat.
In the limit $N\to\infty$, $\rho_{2N} (z)$ converges to $ \frac{a}{ \pi (1-|z|^2)^2}$
 for every non-real $z$ inside the circle $|z|^2= \frac{1}{1+a}$ (Theorem \ref{DensityTheorem}) and outside this circle the spectral density vanishes exponentially fast with $N$ (Lemma \ref{lemma52}). One observes that off the real line, the limiting spectral density of truncated symplectic unitary matrices in the strongly non-unitary regime coincides with the limiting spectral density 
 \begin{align*}
\rho(z)= \frac{a}{ \pi (1-|z|^2)^2} \, \mathbb{1}_{|z|^2<\frac{1}{1+a} }
\end{align*}
 of truncated Haar complex unitary \cite{ZS2000} and real orthogonal matrices \cite{KSZ2010}. 
  
 We also obtain the eigenpair correlation functions \eqref{Rn_I} locally at any \emph{non-real} reference point $z_0$ in the bulk of the eigenvalue distribution. 
 After unfolding 
 and switching to the upper half-plane picture one recovers the Ginibre eigenvalue correlations \eqref{Gin_corr}:
\begin{align*}
\lim_{N\to\infty, M=aN}\frac{1}{\big[R_1^{+}(z_0)\big]^n}R_n^{+} \Big(z_0+\frac{s_1}{\sqrt{R_1^{+}(z_0)}}, \ldots z_0+\frac{s_n}{\sqrt{R_1^{+}(z_0)}}\Big) &=\\[1ex]
\MoveEqLeft[5] 
\det{ \left[ \exp { \left( \pi \big(s_k \overline{s_l} - \frac{1}{2} |s_k|^2 -\frac{1}{2} |s_l|^2\big) \right) } \right] }_{k, l = 1}^n.
\end{align*}
This relation, which holds for every $a>0$ and every non-real $z_0$ inside the circle $|z|^2=\frac{1}{1+a}$,  is an immediate consequence of our Theorem \ref{KernelTheorem} and it provides evidence in support of the universality of the Ginibre eigenvalue correlations. 

Our asymptotic analysis of the pre-kernel $g_N(z,z^{\prime})$ in the regime of strong non-unitarity is carried out by extending the summation over $k$ in \eqref{gN_I} to $N=\infty$ (justified in Lemma \ref{lemma51}) and 
employing an integral representation for the extended sum (Lemma \ref{lemma53}) for the  evaluation of this sum via the saddle point method.  This approach is different to the ones used previously in the literature on quaternion-real ensembles 
and is well suited for the asymptotic analysis of the correlation functions in the complex bulk\footnote{
E.g.,  in \cite{Thesis,KL_Gin} it was was used to obtain the spectral density  in the quaternion-real Ginibre ensemble and the eigenpair correlation functions in the complex bulk.}. Near the real line it leads to a saddle point analysis which is hard to perform.  The task of obtaining the spectral density and correlation functions near in this region  may require different approaches and is left for further research. 

\subsection{Weak non-unitarity ($M=O(1), N\gg 1$)}  In this regime, the truncated matrix $A$ is a finite rank perturbation of a unitary quaternion   matrix of growing dimension, and  the eigenvalues of $A$ lie close to the unit circle, typically at a distance proportional to the average separation $\Delta_{\Sp(2N)}=\frac{1}{\pi}$ between consecutive eigenvalues of symplectic unitary matrices drawn at random from $\Sp (2N)$. 
%
Correspondingly, 
we write the spectral variable as 
$z=(1-\Delta_{\Sp(2N)} \, r)e^{i\phi}$, 
where $r>0$ is the scaled radial deviation from the unit circle and $\phi$ is the angular variable.  Our Theorem  \ref{thm:610} asserts that away from the real line the spectral density $\rho_{2N}(z)$ 
is described by the asymptotic relation
\begin{align}\label{eq:000}
 \lim_{N\to\infty} \Delta_{\Sp(2N)} \rho_{2N} \big((1-\Delta_{\Sp(2N)}\, r)e^{i\phi}\big) = 2h_{2M}(r)\, , 
 \end{align}
 where
 \begin{align*}
h_{M}(r)= \frac{(4\pi r)^{M-1}}{(M-1)!} \int_0^{1} t^{M} e^{-4\pi r t} dt\, . 
\end{align*}
Relation \eqref{eq:000}  holds for every fixed $0< \phi <\pi$  and fixed $M\ge 1$ (the number of quaternion rows and columns removed). Note that the right-hand side in \eqref{eq:000} does not depend on the angular variable $\phi$. As we discuss below (also see the right-hand side plot in Fig. \ref{Fig:1}), this changes when the angle $\phi$ becomes comparable to $\Delta_{\Sp(2N)}$. 

\begin{figure}[b!]
\includegraphics[width=.6\linewidth]{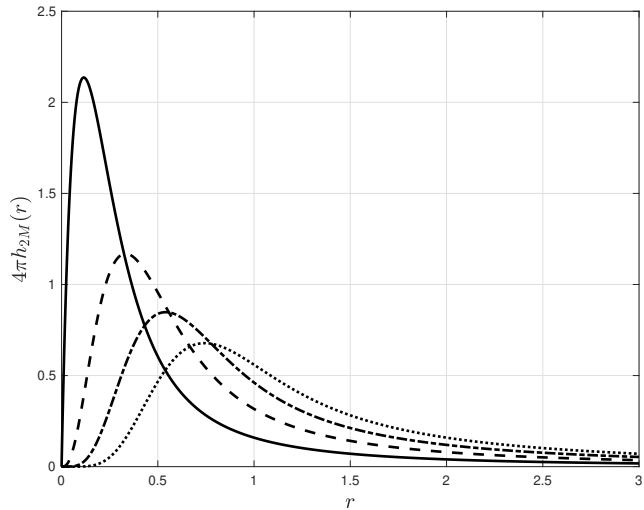} 
\caption{Plot of the rescaled radial part $4\pi h_{2M} (r)$ of the spectral density of truncations of unitary quaternion   matrices in the weakly non-unitary limit for $M=1$ (solid line), $M=2$ (dashed line), $M=3$ (dot-dashed line) and $M=4$ (dotted line). The point where each curve attains its maximum indicates the most likely radial deviation of the eigenvalues of the truncated matrices from the unit circle in units of $\Delta_{\Sp(2N)}=\pi/N$. }
\label{Fig:2}
\end{figure}

For the purpose of comparison with truncations of complex unitary and real orthogonal matrices, it is convenient to rewrite relation \eqref{eq:000} in terms of the one-point correlation function $R_1^+(z)$: 
\begin{align}\label{eq:00}
\lim_{N\to\infty}
\big(\Delta_{\Sp(2N)}\big)^2 R_1^+\big((1-\Delta_{\Sp(2N)}\, r)e^{i\phi}\big) = 4\pi  h_{2M}\!\left( r \right)\, . 
\end{align}
The scaling law \eqref{eq:00} is exactly the same as one for the one-point eigenvalue correlation function $R_1(z)$ of 
truncations of complex unitary matrices and for the smooth part of the one-point eigenvalue correlation function  $R_1^{(C)} (z)$ of truncations of real orthogonal matrices\footnote{The universality of the scaling law \eqref{eq:00} also extends to finite rank skew-Hermitian deviations from the GUE \eqref{Res}, see Eq.~102 in \cite{FSReview1997}.  }. If $A$ is an $N\times N$ block of complex unitary matrix drawn at random from $U(N+M)$ then \cite{ZS2000}
\begin{align*}
\lim_{N\to\infty}
\big(\Delta_{U(N)}\big)^2 R_1\big((1-\Delta_{U(N)}\, r)e^{i\phi}\big) = 4\pi  h_{M}\!\left( r \right)\, , 
\end{align*}
and if $A$ is an $N\times N$ block of real orthogonal matrix drawn at random from $O(N+M)$ then
\cite{KSZ2010} 
\begin{align*}
\lim_{N\to\infty}
\big( \Delta_{O(N)}\big)^2 R_1^{(C)}\big((1-\Delta_{O(N)}\, r)e^{i\phi}\big) = 4\pi  h_{M}\!\left( r \right)\, ,
\end{align*}
where $\Delta_{U(N)}=\Delta_{O(N)}=2\pi/N$ is the average separation between consecutive eigenangles of the Haar unitary and Haar orthogonal matrices of dimension $N\times N$. The doubling of index $M$ in \eqref{eq:00} can be understood on recalling that one quaternion row is equivalent to two rows of complex numbers. 

A similar picture emerges for the higher order correlation functions. Let 
\begin{align}\label{eq:0a}
z_j = \big( 1 - \Delta_{\Sp(2N)}\, r_j \big) e^{i (\phi_0 + \Delta_{\Sp(2N)}\, \phi_j)}, \quad 0 < \phi_0 < \pi\, .
\end{align}
Our Theorem \ref{thm:611} asserts that in the limit of weak non-unitarity 
\begin{align} \label{eq:0b}
	\lim_{N\to\infty} \big(\Delta_{\Sp(2N)}\big)^{2n} R^{+}_n \left(z_1, ..., z_n\right)  = & 
		 \\ \nonumber
	\MoveEqLeft[3] 
	(4\pi)^n \prod_{j=1}^n \frac{(4\pi r_j)^{2M-1}}{(2M-1)!}   \det 
	\left[
	\int_0^1 t^{2M} e^{-2\pi (r_k+r_l-i(\phi_k-\phi_l)\, t} \, dt 
	\right]_{k,l=1}^{n} \, .
\end{align}
On replacing $\Delta_{\Sp(2N)}$ by $\Delta_{U(N)}$  and $2M$ by $M$ in Eqs \eqref{eq:0a} -- \eqref{eq:0b} one arrives at the law of eigenvalue correlations of truncated Haar unitary matrices in the weakly non-unitary limit \cite{FSReview2003,KSReview2011}. Thus, away from the real line, appropriately  scaled eigenvalue correlations of truncated symplectic unitary matrices and truncated unitary (complex) matrices are given by the same probability law. In the limit of strong non-unitarity this law is universal and given by the Ginibre correlations and in the limit of weak non-unitarity this law depends on the number of rows/columns removed. 

The main analytical tool used to carry out our investigation of the weakly non-unitary limit of truncated symplectic unitary matrices employs trading the reciprocal Beta-functions in the pre-kernel \eqref{gN_I} for suitable derivatives, see, e.g.,  Eqs \eqref{psi0} -- \eqref{psi4}. 
Such an approach turns out to be very effective in the weakly non-unitary limit. For example, it gives access to the real edge of the eigenvalue distribution at $z=\pm 1$. This is an interesting region, see Fig.~\ref{Fig:1}, where the eigenvalue statistics of quaternion-real matrices differ from those of complex and real matrices. 

\begin{figure}[t!]
\includegraphics[width=.6\linewidth]{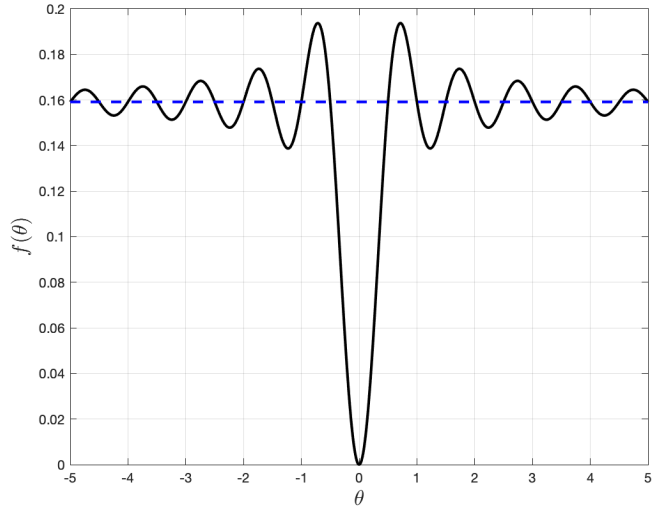} 
\caption{Plot of the microscopic density of eigenangles $f(\theta)$,  Eq.~\eqref{em3}. The dashed line is the plot of $\rho_{U(N)} ( \theta )=\frac{1}{2\pi}$. To 3 d.p.,  the local maxima of $f(\theta)$ are attained at $ \theta_1=0.715$, $\theta_2=1.735$, $\theta_3=2.741$, $\theta_4= 3.743$ and  $\theta_5=4.745$ and at $-\theta_j$ for negative $\theta$. One can observe approximately unit spacing between consecutive points of local maxima except for the pair $-\theta_1$ and $\theta_1$.}
\label{Fig:3}
\end{figure}

It can be seen in Fig.~\ref{Fig:1} that the surface plot of the spectral density $\rho_{2N} (z)$ for $N=60$ and $M=1$ (weak non-unitarity) exhibits a sawtooth profile along the ridge near $z=\pm 1$. One can develop some intuition behind this phenomenon by considering the truncated matrix as a perturbation of a matrix drawn at random from $\Sp(2N)$. Then one would expect the ridge profile of the spectral density of the truncations to bear some resemblance to the spectral density of  symplectic unitary matrices. 
The latter have their eigenvalues $e^{\pm i\phi_j}$, 
$j=1, \ldots, N$, on the unit circle. Their eigenangles form a determinantal point process on $[0,\pi]$  
and the corresponding eigenangle density 
\begin{align*}
\rho_{\Sp(2N)} (\phi) = \frac{1}{2N} \mathbf{E} \Big\{ \sum_{j=1}^N \big( \delta (\phi -\phi_j) + \delta (\phi +\phi_j)\big) \Big\}\, , \quad \phi \in [0,2\pi)\, ,
\end{align*}
can be obtained in closed form \cite{ForresterBook,Meckes_book}. One finds that 
$\rho_{\Sp(2N)} (\phi) =\frac{1}{2\pi} \big[ 1- \frac{1}{N} \sum_{j=1}^N \cos (2j\phi) \big]$. 
In the limit $N\to\infty$ the eigenangle density $\rho_{\Sp(2N)} (\phi)$  tends to $\frac{1}{2\pi}$, 
except for the two boundary points $\phi=0$ and $\phi=\pi$ where it is zero for every finite $N$. On scaling $\phi$ with 
$\Delta_{\Sp(2N)}$, one finds the eigenangle density profile in the neighbourhood of the boundary points. 
It holds that $\lim_{N\to\infty} \rho_{\Sp(2N)}\! \left( \frac{\pi \theta}{N} \right) = f(\theta)$, where 
\begin{align}\label{em3}
f(\theta)=  \frac{1}{2\pi} \left[ 1- \frac{\sin(2\pi \theta) }{2\pi \theta} \right]\, . 
\end{align}
The microscopic eigenangle density $f(\theta)$  
vanishes quadratically in $\theta$ at zero and otherwise it oscillates around value $1/(2\pi)$, see Fig.~\ref{Fig:3}. One observes the almost perfect lattice structure of the points of local maxima of $f(\theta)$. These points correspond to most probable angles of low lying eigenvalues (relative to the real line) which are spaced out due to the eigenvalue repulsion from nearby eigenvalues, including the conjugate ones. The decreasing amplitude of oscillations, as one moves further away from the boundary point at $\theta=0$, signals increasing widths of the probability distribution of the low lying eigenangles due to the waining influence of the eigenvalues in the lower half of the complex plane. 

\begin{figure}[t!]
\includegraphics[width=.4\linewidth]{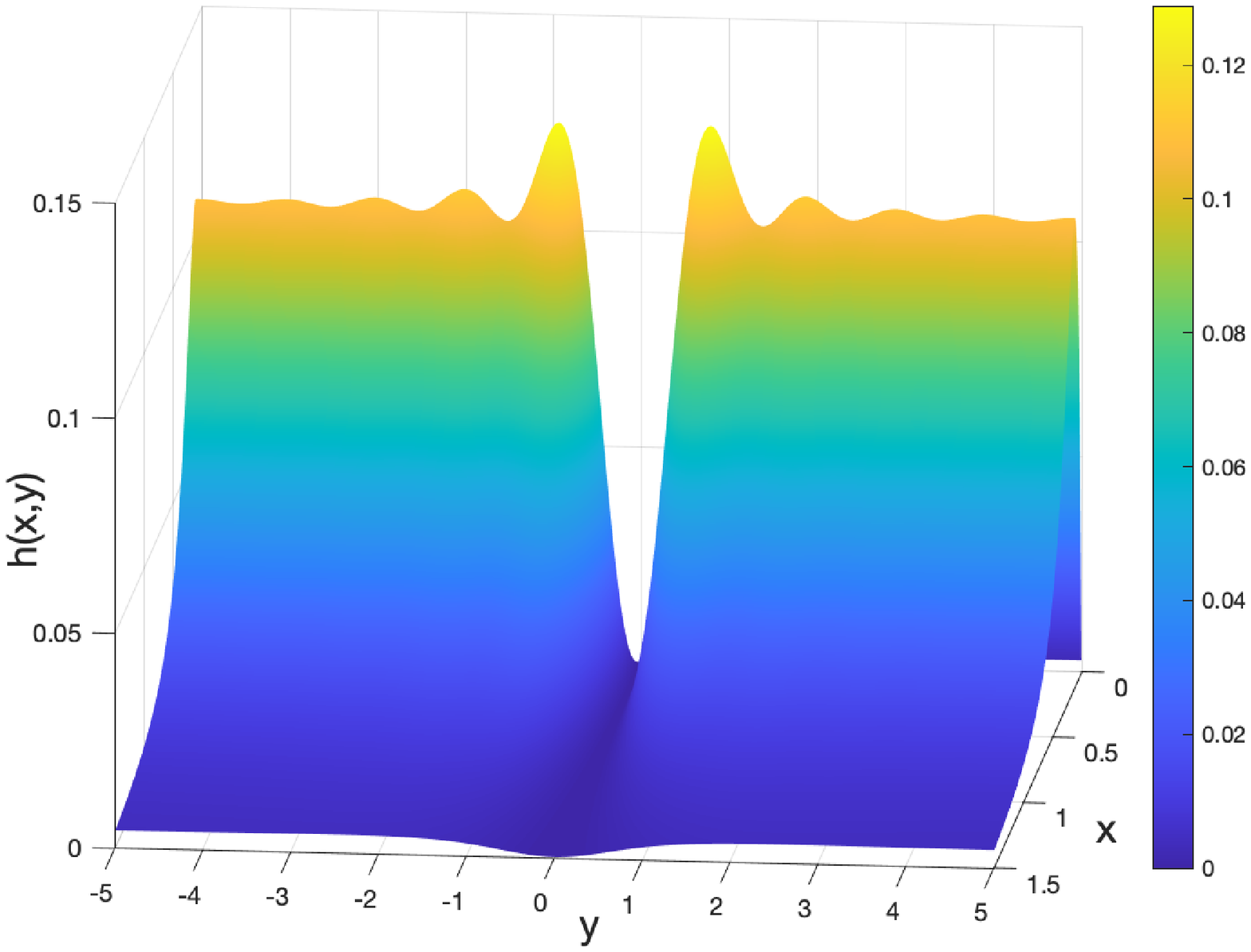}
\includegraphics[width=.4\linewidth]{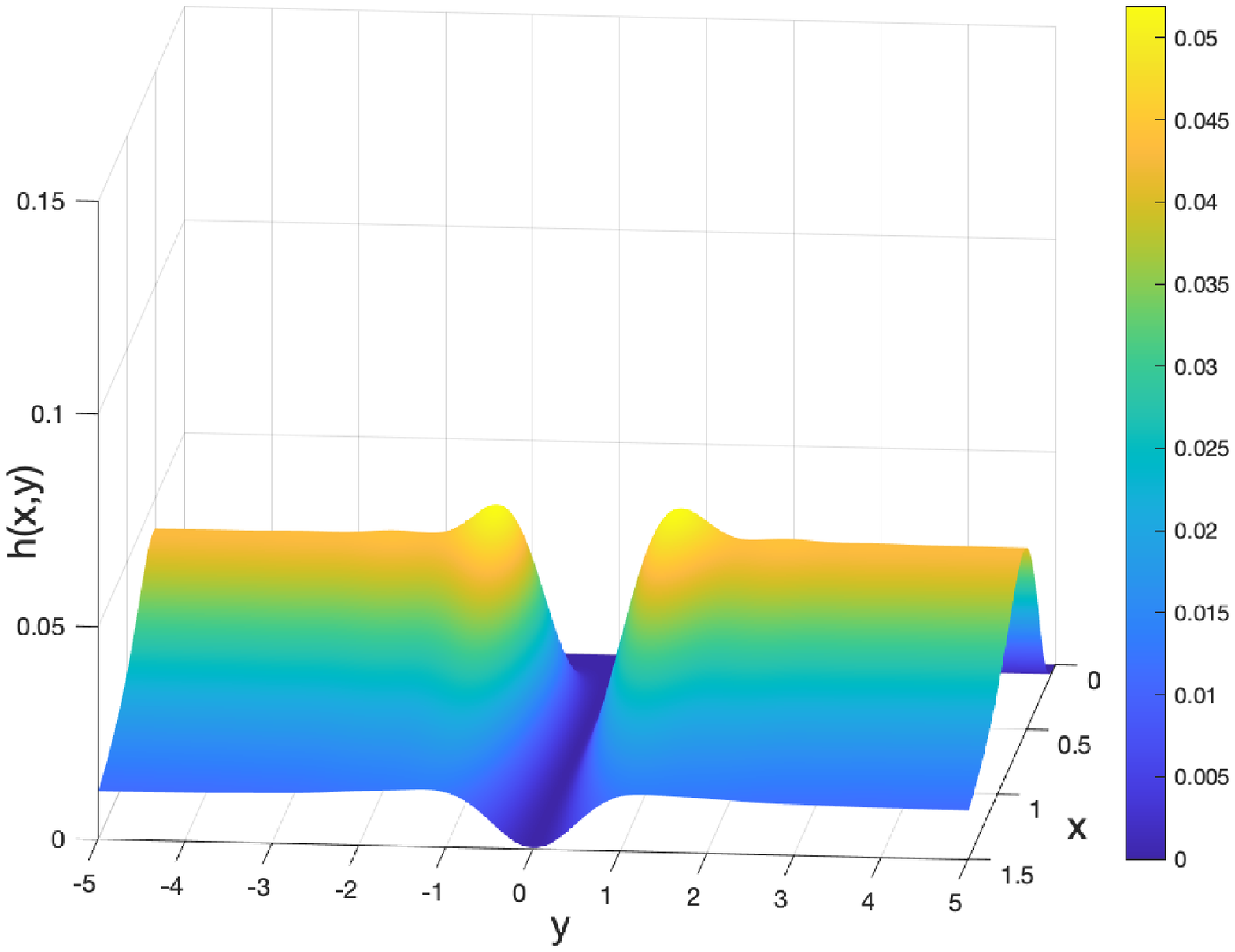}
\caption{Surface plot of the microscopic spectral density $h_{2M}(x,y)$, Eq.~\eqref{hxy}, of truncated symplectic unitary matrices  as function of $x=\Re z$ and $y=\Im z$. In the plot on the left-hand side $M=1$ and in the plot on the right-hand side $M=3$. }
\label{Fig:4}
\end{figure}

Thinking of the complex eigenvalues of the truncated matrices as perturbed eigenvalues of symplectic unitary matrices one would expect the spectral density $\rho_{2N}(z)=\frac{1}{N}R_1(z)$ of the truncated matrices near the real edge at $z=1$ to have a similar angular profile to that of $\rho_{\Sp(2N)}(\phi)$ except that the eigenvalues of the truncated matrices will have an additional degree of freedom - the radial component. One can test this intuition by computing the spectral density $\rho_{2N}(z)$ of the truncations on the microscopic scale in the vicinity of $z=1$. Our 
Theorem \ref{Thm612}
asserts that for every fixed $x>0$ and $y\in \mathbb{R}$,
\begin{align*}
\lim_{N\to\infty, M=O(1)} \frac{1}{N^2} R_1\Big(1-\frac{\pi}{N}(x+iy)\Big) = h_{2M}(x,y),
\end{align*}
where 
\begin{align}\label{hxy}
h_{2M}(x,y)=
\frac{2^{4M}\pi^{2M-1} yx^{2M-1}}{ \Gamma(2M)} \int_0^1\int_0^1 s^{2M+1}t^M e^{-2\pi s(1+t) x} \sin (2\pi s (1-t)y) dt ds\, .
\end{align}
In Fig.~\ref{Fig:4} we plot this function for two values of $M$: $M=1$ and $M=3$. One can observe that at $M=1$ the transversal profile (the profile along the $y$-direction) of $h_{2M}(x,y)$  is very similar to the shape of the microscopic density $f(\theta)$, see Fig.~\ref{Fig:3} of the eigenangles of the symplectic unitary matrices, whilst at $M=3$ the amplitude of oscillations of $h_{2M}(x,y)$ in the transversal direction is fast decaying.

The rest of this paper is organised as follows. In the next Section we provide a brief summary of facts about quaternions which are used in Sections \ref{S:3}  and  \ref{Sec:4}. In Section \ref{S:3} we derive the joint density of eigenvalues of truncated Haar unitary quaternion   matrices. The eigenvalue correlation functions at finite matrix dimension are obtained in Section \ref{Sec:4}.  Our main results are contained in Section \ref{Sec:5} where carry out an asymptotic analysis of the eigenvalue correlation functions  of truncated Haar unitary quaternion   matrices in the regimes of strong and weak non-unitarity.

\section{Quaternions and matrices of quaternions}
\label{Sec:2}

In order to obtain the joint density of eigenvalues of truncations of symplectic unitary matrices, we work with quaternion matrices in Section \ref{S:3} rather than with their complex matrix representations. Both ways are equivalent but the quaternionic derivation is more natural and straightforward. On the downside, not everyone has woking knowledge of quaternions and to make our paper self-contained, and to save reader's time on sifting through the quaternion literature, this Section provides a brief introduction to quaternion matrices and sets the notation used in Sections \ref{S:3} and \ref{Sec:4}. 

 Although the random matrices we are concerned with are composed of real quaternions their eigenvalue correlation functions are expressed in terms of a determinant of self-dual complex quaternion matrix. Hence, we start with complex quaternions.

\subsection{Quaternions} 
Complex quaternions (biquaternions) are isomorphic to the algebra of complex $2\times 2$ matrices and can be written in the form 
\begin{align}\label{standartQuat}
	q = \alpha\, \mathbf{1}  +\beta\, \mathbf{e}_1 + \gamma\, \mathbf{e}_2   + \delta \, \mathbf{e}_3 \, ,
\end{align}
where $\alpha, \beta, \gamma, \delta $ are complex numbers and 
\begin{align*}
	\mathbf{1} =  \begin{bmatrix}1 &0\\ 0 & 1 \end{bmatrix},  \, \, \mathbf{e}_1= \begin{bmatrix} i & 0 \\ 0& -i \end{bmatrix}, \, \mathbf{e}_2= \begin{bmatrix}0 & 1 \\ -1& 0 \end{bmatrix}, \, \mathbf{e}_3= \begin{bmatrix} 0 & i \\ i& 0 \end{bmatrix}\, .
\end{align*}
Here $i$ is the imaginary unit, $i=\sqrt{-1}$. Complex quaternions $q$ can be multiplied using the associative law and multiplication table 
\begin{align*}
\mathbf{1}^2 =\mathbf{1}, \, \mathbf{e}_1^2=\mathbf{e}_2^2=\mathbf{e}_3^2= \mathbf{e}_1\mathbf{e}_2\mathbf{e}_3=-\mathbf{1}, 
\end{align*}
\begin{align*}
 \mathbf{1}\mathbf{e}_j=\mathbf{e}_j\mathbf{1}=\mathbf{e}_j,\quad  j=1,2,3\, .
\end{align*}
In representation \eqref{standartQuat}, $ \alpha\, \mathbf{1} $ is called the scalar part of $q$. 
To simplify the notation we shall write $\alpha$ instead of $ \alpha\, \mathbf{1} $. The one-to-one mapping $\theta$ between 
 the complex quaternions and complex $2\times 2$ matrices is given by 
\begin{align}\label{qmat_c}
\theta (q)= \begin{bmatrix}\alpha +i \beta &\gamma +i \delta \\  -\gamma +i \delta& \alpha - i \beta \end{bmatrix} \, .
\end{align}
This mapping is a homomorphism, $\theta (q_1q_2)=\theta (q_1)\theta (q_2)$. 


Any quaternion \eqref{standartQuat} has a conjugate quaternion 
\begin{align*}
\bar q = \alpha -\beta\, \mathbf{e}_1 - \gamma\, \mathbf{e}_2   - \delta \, \mathbf{e}_3
\end{align*}
and a complex conjugate  
\begin{align*}
q^* = \alpha^* +\beta^*\, \mathbf{e}_1 + \gamma^*\, \mathbf{e}_2   + \delta^* \, \mathbf{e}_3.
\end{align*} 
The operation $\hspace*{-1ex}\phantom{z}^*$ on complex numbers  $z=x+iy$ is the usual complex conjugate, 
\begin{align*}
z^*=x-iy.
\end{align*}
 The operation of conjugation on quaternions is a function,
\begin{align}\label{conj}
	\bar q = -\frac{1}{2} (q + \mathbf{e}_1\, q\, \mathbf{e}_1 + \mathbf{e}_2\, q\,\mathbf{e}_2 + \mathbf{e}_3\, q\,\mathbf{e}_3)  \, . 
\end{align}
It can be seen from this that $\overline{q\mathbf{e}_j} = \overline{\mathbf{e}_j}\,  \bar{q}$, $j=1,2,3$. Hence, the conjugate of a product is the product of the conjugates in the reverse order.  By applying both operations of conjugation together one obtains the Hermitian conjugate 
\begin{align*}
q^{\dagger}  = \alpha^*  -\beta^*\, \mathbf{e}_1 - \gamma^*\, \mathbf{e}_2   - \delta^* \, \mathbf{e}_3.
\end{align*} 
It can be seen from \eqref{qmat_c} that operation of Hermitian conjugation of quaternions corresponds to the operation of Hermitian conjugation of matrices: $\theta (q^{\dagger})=\theta (q)^{\dagger}$ and $(q_1q_2)^{\dagger}=q_2^{\dagger}q_1^{\dagger}$.

If $\bar q = q$ then quaternion $q$ is a scalar, and if $q^*=q$ then $q$ is a real quaternion, i.e. the coefficients of the $\mathbf{e}_j$s in \eqref{standartQuat} are all real. The real quaternions are represented by complex $2\times 2$ matrices 
\begin{align}\label{qmat}
 \begin{bmatrix}a &b\\ -b^* & a^* \end{bmatrix}\, . 
\end{align}
The determinant of this matrix is $\vert a \vert^2+ \vert b \vert^2$ and it can be seen that non-zero real quaternions are invertible. Hence, the real quaternions form a noncommutative division algebra $\mathbb{H}$ over the field of real numbers,
\begin{align*}
\mathbb{H} =\{\alpha  +\beta\, \mathbf{e}_1 + \gamma\, \mathbf{e}_2   + \delta \, \mathbf{e}_3: \, \alpha,\beta,\gamma, \delta\in \mathbb{R}  \}\, .
\end{align*}
$\mathbb{H}$ is a normed space with the norm  
  \begin{align*}
 \vert q \vert  = \sqrt{q^{\dagger}  q}.
 \end{align*} 
It can be seen from \eqref{qmat} that unit real quaternions can be represented as unitary $ 2 \times2 $ matrices.  
The complex matrix representation of real quaternions of the type $\alpha +\beta  \mathbf{e}_1$ 
is 
\begin{align}\label{qmat1}
  \begin{bmatrix}\lambda &0\\ 0 & \lambda^* \end{bmatrix} 
\end{align}
with $\lambda =\alpha +i \beta$. 
 This gives a natural embedding of the complex numbers into  $\mathbb{H}$. Observe that matrix \eqref{qmat} can be reduced to the diagonal matrix  \eqref{qmat1} by a unitary similarity transformation. 
 This means that the conjugate class  $\{p^{\dagger}q\, p:\, p\in \mathbb{H},\,  \vert p\vert =1\}$ of a real quaternion $q$ \eqref{qmat} intersect $\mathbb{C}$ in two points $\lambda$ and $\lambda^*$ unless $q$ is a scalar in which case this class is just $q$.  

\subsection{Quaternion matrices} Any matrix of complex quaternions can be considered as a complex matrix of doubled dimensions by replacing each quaternion matrix element by its $2\times 2$ matrix representation. And vice versa, by cutting it into $2\times 2$ blocks, any complex matrix of even dimensions can be considered as a matrix of quaternions of half of the original dimensions. It is apparent that the homomorphism $\theta$ extends to matrices of quaternions.

Let $Q=(q_{kj})$ be a matrix of complex quaternions. The transpose of $Q$ and the Hermitian conjugate of $Q$ are defined as the matrices $Q^T$ and $Q^{\dagger}$  such that 
\begin{align}\label{Qtranspose}
(Q^T)_{kj}=-\mathbf{e}_2\bar q_{jk} \mathbf{e}_2, \quad (Q^{\dagger})_{kj} =q^{\dagger}_{jk}\, .
\end{align}
This definition corresponds to taking transpose and Hermitian conjugate transpose in the complex matrix representation of quaternion matrices: if $A$ is the complex matrix representation of quaternion matrix $Q$, 
then the complex matrix representation of $Q^T$ and $Q^\dagger$ are $A^T$ and $A^{\dagger}$, respectively. 

The dual of $Q$ is defined as the matrix  $Q^R$ such that 
\begin{align*}
(Q^R)_{kj} =\bar q_{jk}\, .
\end{align*}
It can be seen that $(Q_1Q_2)^R=Q_2^R Q_1^R$. 
A square $Q$ is called \emph{self-dual} if $Q^R=Q$. It can be seen from \eqref{Qtranspose} that $(Q^R)_{kj} =-\mathbf{e}_2 (Q^T)_{kj}\mathbf{e}_2$. Hence, introducing the notation $Z$  for the block matrix having $2\times 2$ blocks $ \mathbf{e}_2$ 
 on the main diagonal and zeros elsewhere,
\begin{align*}
Z=\mathbf{e}_2 \otimes I\, ,
\end{align*}
one observes that if $A$ is the complex matrix representation of $Q$ then $-ZA^TZ$ is the complex matrix representation of $Q^R$. Therefore, a quaternion matrix $Q$ is self-dual if and only if its complex matrix representation $A$ obeys 
\begin{align}\label{selfdual}
A=-Z\, A^T\, Z\, .
\end{align}
If $Q^R\!=\!Q^{\dagger}$ then the matrix elements of $Q$ are real quaternions. Such matrices are called \emph{quaternion-real}. Matrix $Q$ is quaternion-real if and only if its complex matrix representation $A$ obeys 
\begin{align}\label{realq}
A^{\dagger} =-Z\, A^T\, Z \, .
\end{align}
A quaternion matrix $Q$ is called \emph{Hermitian} if $Q^{\dagger}=Q$. Quaternion-real Hermitian matrices are self-dual, and Hermitian self-dual matrices are quaternion-real. 

A quaternion matrix $Q$  is called \emph{unitary} if it is real quaternion and $Q^{\dagger}Q=I$. All unitary quaternion   matrices of dimension $N$ form the group $U(N, \mathbb{H})$.
It can be seen from \eqref{realq} that if $A$ is the complex representation of a unitary quaternion   matrix  
then 
\begin{align*}
Z=A^T\, Z\, A  \, , 
\end{align*}
i.e., the matrix $A$ is symplectic and unitary. Vice versa, by cutting any $2N\times 2N$  symplectic unitary matrix  into $2\times 2$ blocks one obtains a unitary quaternion    matrix. 
The symplectic unitary matrices form the compact symplectic group $\Sp(2N)$ which is isomorphic to $U(N, \mathbb{H})$. 

\subsection{Eigenvalues of quaternion-real matrices}

If is $Q$ is a 
quaternion-real matrix then $q\in \mathbb{H} $ is called an \emph{eigenvalue} of $Q$ if there exists a non-zero real quaternion vector $ \vec{v}$ 
such that 
\begin{align}\label{qeig}
Q \vec{v} = \vec{v} q \, .
\end{align}
Note that $q$ here is the right-hand multiplier. 

If $q$ is an eigenvalue of quaternion-real $Q$ 
 then so is $p^{\dagger}q\, p$ for any unit $p\in \mathbb{H}$ as can be seen from $Q (\vec{v}p) = (\vec{v}p)\, (p^{\dagger} q\, p)$. Thus, one ought  to consider eigenclasses (conjugate classes) of quaternion-real matrices rather than eigenvalues.  As was already observed, unless $q$ is a scalar (real number), each eigenclass  $\{p^{\dagger}q\, p:\, p\in \mathbb{H},\,  \vert p\vert =1\}$ passes through two complex numbers $\lambda$ and $\lambda^*$  and if $q$ is a scalar then the eigenclass is just $q\in \mathbb{R}$. Therefore, each non-trivial eigenclass is represented 
 by a pair of complex conjugated complex numbers. This means that if the equation \eqref{qeig} for $q$ is restricted to $q\in \mathbb{C}$ then its solutions come in complex conjugated pairs, in agreement with the $2\times 2$ matrix representation \eqref{qmat1}  of complex numbers regarded as real quaternions. 
  
The precise nature of these complex conjugated eigenvalue pairs can be understood by going to the complex matrix representation of $Q$.  Introducing the notation $\Theta$ for the one-to-one mapping between quaternion-real matrices and their complex matrix representations, under this mapping the equation $Q\vec{v}=\vec{v} \lambda$ with complex $\lambda$ is transformed to 
\begin{align*}
\Theta (Q) [\vec{\psi}, Z \vec{\psi}^*] = [\vec{\psi}, Z \vec{\psi}^*]    \begin{bmatrix}\lambda &0\\ 0 & \lambda^* \end{bmatrix} 
\end{align*}
or, equivalently, $\Theta (Q) \vec{\psi} = \lambda \vec{\psi}$ and $\Theta (Q) (Z \vec{\psi}^*) = \lambda^* (Z \vec{\psi}^*)$, where 
$[\vec{\psi}, Z \vec{\psi}^*] =\Theta (\vec{v})$ is the rectangular complex matrix of two columns $\vec{\psi}$ and $Z \vec{\psi}^*$, see \eqref{qmat}. Therefore $\lambda$ and $\lambda^*$ are two eigenvalues of the complex matrix representation of $Q$. 
The constraint \eqref{realq} ensures that the non-real eigenvalues of $\Theta (Q)$ come in complex conjugate pairs and real eigenvalues have even multiplicity. The complex eigenvalues of quaternion-real matrices are the same as the eigenvalues of its complex matrix representation. 

\subsection{Schur decomposition of quaternion-real matrices \cite{L1949,B1951}}
Any square quaternion-real matrix $Q$ can be transformed to an upper-triangular matrix $T$ which has complex number on the diagonal by unitary conjugation, $Q=U T U^{\dagger}$ for some unitary quaternion   $U$. Indeed, pick a complex eigenvalue $\lambda$ of $Q$ and a unit eigenvector, $Q\vec{v}=\vec{v} \lambda$. One can construct a unitary quaternion   matrix $V$ which has $ \vec{v} $ as its first column (e.g., go to the complex matrix representation and apply Gram-Schmidt). Then all entries of the first column of $V^{\dagger} Q V$  are zeros except the first one which is $\lambda$, and one can repeat this process on the matrix obtained from $Q$ by removing its first row and the first column. Recursive repetition leads to the Schur decomposition $Q=U T U^{\dagger}$. The diagonal entries of the upper-triangular $T$ are complex eigenvalues of $Q$ which are indeterminate to the extent of changing to their complex conjugate and permutation of complex conjugated pairs. This can be verified by going to the complex matrix representation.

\subsection{Determinant of  self-dual quaternion matrix  \cite{M1922,D1970}}
The determinant of self-dual quaternion  $2\times 2$ matrices is defined in the usual way:
\begin{align*}
\det  \begin{bmatrix}~~q_{11} & q_{12} \\ \overline{q_{12}} & q_{22} \end{bmatrix}= q_{11}q_{22}-q_{12}\overline{ q_{12}}\, .
\end{align*}
No ambiguity arises due to non-commutativity of quaternions. Indeed, $q_{11}$ and $q_{22}$ scalars because the matrix is self-dual,  and $q_{12}\overline{q_{12}}$ is a scalar too. Hence $q_{12}\overline{q_{12}}=\overline{q_{12}}q_{12}$. 

The determinant of self-dual quaternion $N\times N$ matrices $Q=[q_{jk}]_{j,k=1}^N$ is defined as 
\begin{align}\label{det}
\det Q = \sum_{P} (-1)^{N-l} \prod_1^{l} (q_{ab} q_{bc} ... q_{za}),
\end{align}
where the sum is over all permutations of indices, with each permutation $P$ is assumed to consist of $ l $ exclusive cycles of the form
$(a \rightarrow b \rightarrow c \rightarrow ...  \rightarrow z)$. 
To make this definition unique, it is required that the same ordering of the cyclic factors be used for the permutation P and for the other permutations obtained from P by reversing the direction of some or all of the cycles \eqref{det}. This ensures that  $\det Q$ is a scalar and independent of the order of the $l$-cyclic factors in \eqref{det}. 

The determinant of a self-dual quaternion matrix $Q$ can be expressed in terms of the determinant of its complex matrix representation $\mathit{\Theta} (Q)$ \cite{D1970}, 
\begin{align*}
(\det Q)^2 =\det \mathit{\Theta} (Q)
\end{align*}
and, consequently, also, in terms of the Pfaffian of the complex skew-symmetric matrix $Z\mathit{\Theta} (Q)$,
 \begin{align*}
\det Q  =\Pf (Z\mathit{\Theta} (Q))\, .
\end{align*}

Finally, we observe that Hermitian quaternion-real matrices are self-dual, hence all of the above applies to this type of matrices. In particular, it holds for quaternion-real $Q$ that 
\begin{align*}
\det (Q^{\dagger} Q)  = \det \mathit{\Theta} (Q)\, .
\end{align*}


\subsection{Volume element in the space of quaternion-real matrices}

A real quaternion has four degrees of freedom (real parameters) and the calculus of real quaternions is developed in terms of the calculus of its degrees of freedom. In particular, the infinitesimal volume element $d^4q$ associated with real quaternion  \eqref{qmat} is 
$d^4q=\frac{1}{4} da \wedge da^* \wedge d b  \wedge db^*$.
Correspondingly, the infinitesimal volume element associated with a quaternion-real vector $\vec v =[q_j]$ is defined as the product $D[\vec{v}]=\wedge_{j} \ d^4\!q_{j}$ and the infinitesimal volume element $D[Q]$ associated with quaternion-real matrix $Q=[q_{jk}]$ is defined as the as the product of the infinitesimal volume elements of all independent entries. For example, if quaternion-real $Q=[q_{jk}]$ is Hermitian then 
\begin{align*}
D[Q] = \bigwedge_{j}  \ d  q_{jj}  \bigwedge_{j<k} \ d^4\!  q_{jk}\, .
\end{align*}
In this case $q_{jj}$ are just real scalars $\alpha_j$ and each $d q_{jj}$ is just the one-form $d\alpha_j$. 

We will need the Jacobian of several linear transformations of real quaternions. This information is collected in the Proposition below. Parts (a) and (b) are almost self-evident after switching to the matrix representation of quaternions, and part (c) can be derived by extending to quaternion matrices a similar calculation for real and complex matrices, see also Exercise 1.3 in \cite{ForresterBook}.  
\begin{proposition} 
\label{Prop2}
\begin{itemize}
\item[(a)] Let $\lambda$ be a fixed complex number and $q$ be a real quaternion. Then the Jacobian of the transformation from $q$ to $\lambda q$ in $\mathbb{R}^4$ is $|\lambda|^4$, i.e. $d^4(\lambda q)=|\lambda|^4 d^4q$. 
\item[(b)] Let $G$ be a fixed Hermitian matrix of real quaternions and $\vec{q}$ be a column-vector of real quaternions. Then the Jacobian of the transformation from $\vec{q}$ to $G\vec{q}$ is $|\det G|^4$. 
\item[(c)] Let $G$ and $Q$ be $N\times N$ matrices of real quaternions. Then the Jacobian of the transformation from $Q$ to $G^{\dagger}QG$ is $(\det G^{\dagger}G)^{2(N-1)+1} $, i.e., $D[G^{\dagger}QG]=(det G^{\dagger}G)^{2(N-1)+1} D[Q]$. 
\end{itemize}

\end{proposition}

Also, we will need the Jacobian of the Schur decomposition of quaternion-real matrices, for a derivation see e.g.,   \cite{AkemannPaper}:
	
\begin{proposition} \label{JacSchur}
	Let $Q$ be a square quaternion-real matrix written in the Schur form $ Q = U (S+\Lambda) U^{\dagger} $, where $S$ is strictly upper triangular, $\Lambda$ is a diagonal matrix of complex eigenvalues of $Q$ and $U$ unitary quaternion  . Then
	\begin{align} \label{EqJacFirst}
	D[Q] = \prod_{i \geq j} |\lambda_j -\lambda_i^*|^2 \prod_{i > j} |\lambda_j - \lambda_i|^2  D[\Lambda] \wedge D[S] \wedge (U^{\dagger} dU) , 
	\end{align} 
where $D[\Lambda]=  \wedge_j d^2\!\lambda_j $ and $(U^{\dagger} dU)$ is the Haar form which is the product of the 4-forms corresponding to the independent off-diagonal entries of $U^{\dagger} dU$ and the 2-forms corresponding to the diagonal elements of $U^{\dagger} dU$. 
\end{proposition}

The Haar form defines the Haar measure of the group manifold of the unitary quaternion   matrices, and is useful for Jacobian calculations. The Haar measure can also be defined in other equivalent ways. For our purposes, it is convenient to write the Haar measure as a singular measure on the space of all quaternion-real matrices. Let $Q$ be a unitary quaternion   $K\times K$ matrix. Its first $N$ columns, $N\le K$, can be regarded as a point on the Stiefel manifold $V_N(\mathbb{H}^{K})=Sp(2K)/Sp(2(K-N))$. The \emph{normalised} Haar measure on $V_N(\mathbb{H}^{K})$ can be written as a singular measure on the space $\Mat (K\times N, \mathbb{H})$ of quaternion-real  $K\times N$ matrices with density 
\begin{align}\label{Haar}
p_{\mathrm{Haar}}(V)=\frac{1}{V_{K,N}}\,  \delta (V^{\dagger}V - I_N) \, , 
\end{align}
where $V_{K,N}$ is the normalisation constant, 
\begin{align}\label{VKN1}
V_{K,N}= \int_{\Mat (K\times N, \mathbb{H})} \delta (V^{\dagger}V - I_N) D[V]\, .
\end{align}
 Here, we would like to introduce a convention about the $\delta$-functions of matrix argument. By this convention, $\delta (Q)$ is the product of $\delta$-functions of all the degrees of freedom of independent matrix elements as imposed by the symmetries of matrix $Q$. For example, if $Q=V^{\dagger}V-I_N$ which is a Hermitian matrix of real quaternions then its diagonal entries are real numbers $q^{(0)}_{jj}$  and the rest of its independent matrix entries are real quaternions  $q_{jk}=q^{(0)}_{jk} + q^{(1)}_{jk} e_1 + q^{(2)}_{jk} e_2 + q^{(3)}_{jk} e_3 $ above the main diagonal. Correspondingly, 
\begin{align*}
\delta(Q)=\prod_{j} \delta \big(q^{(0)}_{jj}\big) \prod_{j<k} \prod_{\alpha=0}^3 \delta \big(q^{(\alpha)}_{jk}\big),
\end{align*}
\begin{proposition}
\label{Prop1}
\begin{align}\label{VKN}
V_{K,N}=\prod_{k=K-N+1}^K \frac{\pi^{2k}}{\Gamma(2k)} \, .
\end{align}
\end{proposition} 
\begin{proof}
The normalisation constant $V_{K,N}$ \eqref{VKN1}
can be computed by various methods. Here is one which is probably the simplest. By making use of the multiplication by unitary quaternion   matrices, one can transform matrix $V$ in the integral in \eqref{VKN1}  to the form where each column of $V$ is orthogonal to all the preceding ones. Therefore,
\begin{align*}
V_{K,N}=\prod_{k=K-N+1}^{K} \int_{\mathbb{H}^{k}} \delta (\vec{v}^{\dagger} \vec{v}  - 1) D[\vec{v}]  = \prod_{k=K-N+1}^{K} \frac{1}{2} \big| S^{4k-1}\big|\, ,
\end{align*}
where $ \big| S^{4k-1}\big|$ is the surface area of the unit sphere in $\mathbb{R}^{4k}$. The factor $1/2$ is there because $\delta (r^2-1)=\frac{1}{2} \delta (r-1)$ for positive real $r$. On recalling $ \big| S^{n-1}\big|=2\pi^{n/2}/\Gamma (n/2)$, one arrives at \eqref{Prop1}.  
\end{proof}

\section{Joint probability density of eigenvalues }
\label{S:3}

Let $H$ be a sample from $U(N+M, \mathbb{H})$, $M\ge 1$,  the group of unitary quaternion   matrices, equipped with the normalised Haar measure, and $A$ be its top-left corner block of size $N\times N$,
\begin{align}\label{H1}
H= \begin{bmatrix} A & B \\ C & D \end{bmatrix}.
\end{align}
The $N\times N$ matrix $A$ is quaternion-real. It has $N$ pairs of complex eigenvalues which we will denote by $\lambda_1, {\lambda}_1^*, \dots, \lambda_N, {\lambda}_N^*$, where, by convention, all $\lambda_j$ lie in the upper half of the complex plane. Since the complex matrix representation of $A$ is a contraction $|\lambda_j|<1$ for all $j$. 

The joint probability density function of distribution of the complex eigenvalues of $A$ was reported in \cite{F2016}  along with an outline of its derivation in the case when $M \ge N$. In this case the probability distribution of $A$ is continuous with respect to $D[A]$, the Cartesian volume element in $\Mat (N\times N, \mathbb{H})$. 
Below we give a derivation of this density for all integer $M\ge 1$.  It follows closely a similar derivation in \cite{ZS2000} for truncations of Haar unitary matrices given. This derivation uses matrix $\delta$-function manipulations, a useful tool for Jacobian computations, see e.g. \cite{FSReview2003, F2006}.

By the way of introduction of this tool, let us first obtain the joint density of matrix entires of rectangular blocks of Haar unitary quaternion   matrices starting from the delta-function  representation of the Haar measure \eqref{Haar}.  This result is not new. The joint density of matrix elements of rectangular blocks of Haar real orthogonal matrices was first obtained in \cite{EatonBook}, and later for all three classical compact groups in \cite{F2006}. Our point here is to demonstrate that writing the unitary constraints as a matrix delta-function simplifies the matters.

\begin{proposition} [\cite{F2006}] 
Let $H$ be a unitary quaternion   matrix drawn at random from $U(N, \mathbb{H}) $  and $A$ be its top-left block of size $L\times N$. If $K\ge N+L$  then the joint density of matrix entries of $A$  is  
 \begin{align*}
p(A) =  \frac{ V_{K-L,N}}{V_{K,N}} \, \det (I_N-A^{\dagger}A)^{2(K-N-L)+1}.
\end{align*} 
\end{proposition}
\begin{proof} 
This calculation is almost verbatim version of a similar one for square sub-blocks of orthogonal matrices in \cite{KSZ2010}. With obvious modifications accounting for different number of degrees of freedom (Prop. \ref{Prop2} ) it works for all three classical compact groups.

Without loss of generality we may assume that $L\ge N$. Let $V$ be the rectangular matrix consisting of the first $N$ columns of $H$ in \eqref{H1}, $V= \begin{bmatrix} A \\ C \end{bmatrix}$. Then 
 \begin{align*}
p(A) = \frac{1}{V_{K,N}}\, \int \delta \big(A^{\dagger}A +C^{\dagger}C -I_N\big) D[C]\, .
\end{align*}
If $K\ge N+L$ then $A^{\dagger}A<I_N$. Set $X=\sqrt{I_N-A^{\dagger}A}$ and make the substitution $C=G X$ in the integral above. According to Prop. \ref{Prop2}(b),  $D[C]=\det X^{4(K-L)}D[G]$ ($G$ has $K-L$ rows), and 
 \begin{align*}
p(A) = \frac{\det X^{4(K-L)}}{V_{K,N}}\, \int \delta \big(X(G^{\dagger}G -I_N) X\big) D[G].
\end{align*}
According to Prop.\ref{Prop2}(c), $\delta \big(X(G^{\dagger}G -I_N) X\big)=\det X^{-4N +2 }\delta \big(G^{\dagger}G -I_N\big)$. Therefore, 
 \begin{align*}
p (A) &= \frac{ \det X^{4(K-L-N)+2}}{V_{K,N}}\, 
\int_{\Mat((K-L)\times N,\mathbb{H})}
\delta \big(G^{\dagger}G -I_N\big) D[G] \\
& =  \frac{ V_{K-L,N}}{V_{K,N}}\, \det (I-A^{\dagger}A)^{2(K-L-N)+1}\, .
\end{align*}
\end{proof}

If $K<N+L$ then some of the eigenvalues of $A^{\dagger}A$ are unity and the probability distribution of matrix $A$ is supported on the boundary of the matrix ball $A^{\dagger}A=I_{N}$ and is singular. In principle one can obtain the density of distribution of $A$ in this case too, see the relevant discussion of truncated Haar unitary matrices at the beginning of Section 3 in  \cite{FK2007}, but the resulting expression contains many delta-function factors and does not seem to be useful. Fortunately, the knowledge of the law of distribution of $A$ is not required for the calculation of the joint eigenvalue density of $A$ which is given below. 

\begin{theorem}
\label{Thm:JPDF}
Let $H$ be a unitary quaternion   matrix drawn at random from $U(N+M, \mathbb{H})$  and $A$ be its top-left block of size $N\times N$. If $f(\lambda_1, \ldots, \lambda_N)$ is a bounded symmetric function of the eigenvalues of matrix $A$ in the upper half of the complex plane then the average of $f$ is given by 
\begin{align}\label{JPDF1}
\mathbb{E} \{f\}= \int_{\mathbb{D}^N_{+}} f(\lambda_1, \ldots, \lambda_N) p(\lambda_1, \ldots, \lambda_N) \prod_{j=1}^N d^2\!\lambda_j,
\end{align}
where $\mathbb{D}_{+}$ is the semi-disk $\{ \lambda\in \mathbb{C}: \; |\lambda| \le 1, \; \Im \lambda \ge 0 \}$  and  
\begin{align}\label{JPDF}
p (\lambda_1, \ldots, \lambda_N) =  \frac{1}{Z_{N,M}}  \prod_{i \geq j} |\lambda_j - {\lambda_i^*}|^2 \prod_{i > j} |\lambda_j - \lambda_i|^2  \prod_{j=1}^{N} (1 - |\lambda_j|^2)^{2M - 1},
\end{align}
with the normalisation constant $Z_{N,M}= \pi^N N! \prod_{j=1}^N B\big(2M, 2j\big)$.
\end{theorem}


\begin{proof} 
One can transform $A$ to the Schur form by unitary conjugation: 
\begin{align}\label{Schur}
A=Q(\Lambda+S)Q^{\dagger}
\end{align}
where $Q$ is unitary quaternion , $Q\in U(N, \mathbb{H})$, $\Lambda$ is a diagonal matrix complex eigenvalues of $A$ in the upper half of the complex plane, $\Lambda=\diag(\lambda_1, \ldots, \lambda_N)$, and $S$ is a strictly upper triangular matrix of real quaternions. 
For a given matrix $A$ the decomposition \eqref{Schur} is unique subject to permutations of the entries of $\Lambda$ and right multiplication of $Q$ by diagonal complex unitary  matrices $U$ (note that $\Lambda U=U\Lambda$ as both diagonal matrices have complex elements),  so that in effect $Q\in U(N,\mathbb{H})/U(1, \mathbb{C})^N$. It is instructive to count the degrees of freedom on both sides in Eq.~ \eqref{Schur}. On the right-hand side, the space of strictly upper triangular matrices  $S$ of real quaternions has real dimension $2N(N-1)$, the space of complex diagonal matrices $\Lambda $ has real dimension $2N$, and the space of cosets $Sp(N)/U(1)^{N}$ has real dimension $N(2N+1)-N$. This all add up to $4N^2$ which is the real dimension of matrices $A$ on the left-hand side in \eqref{Schur}.      

The Jacobian of the transformation from $A$ to $(\Lambda, S, Q)$ 
is given by
\begin{align*}
\prod_{i \geq j} |\lambda_j - {\lambda_i^*}|^2 \prod_{i > j} |\lambda_j - \lambda_i|^2 
\end{align*} 
see e.g. \cite{AkemannPaper,Thesis}. Note that the Jacobian is independent of $S$ and $Q$. Hence,
\begin{align}\label{PNM}
p(\lambda_1, \ldots , \lambda_N)= \frac{\Vol \big[U(N, \mathbb{H})\big]}{(2\pi)^N\, N! \, V_{N+M,N}}\,  
w^2(\Lambda) \, \prod_{i \geq j} |\lambda_j - {\lambda_i^*}|^2 \prod_{i > j} |\lambda_j - \lambda_i|^2
\end{align} 
where 
\begin{align}\label{w}
w^2(\Lambda)= \int_{\mathbb{R}^{2N\!(N-1)}}  \!\!D[S]\int_{\mathbb{R}^{2N\!M}} \!\!D[C] \,\, \delta \big((\Lambda +S)^{\dagger} (\Lambda +S) +C^{\dagger}C -I_N   \big) 
\end{align} 
and
$\Vol \big[U(N, \mathbb{H})\big]$ is the volume of $U(N, \mathbb{H})$, 
\begin{align*}
\Vol \big[U(N, \mathbb{H})\big] =2^N V_{N,N}   = 2^N \prod_{k=1}^N\frac{\pi^{2k}}{\Gamma (2k)}\, .
\end{align*}
Now, we ought to integrate out $S$ and $C$ in \eqref{w}:
\begin{lemma} \label{LemID2}
	\begin{align}
	\label{w1}
	w^2(\Lambda)= \left(\frac{\pi^{2M}}{\Gamma (2M)}\right)^{\!N}  \prod_{j=1}^{N} (1 - |\lambda_j|^2)^{2M - 1}. 
	\end{align}
\end{lemma}

\begin{proof}[Proof of Lemma \ref{LemID2}]
Let $\vec{c}_1, \ldots, \vec{c}_N$ be the $N$ columns of $C$ 
and $S=(s_{jk})$. The $\vec{c}_j$ are column-vectors of real quaternions, each of length $M$, and the $s_{jk}$ are real quaternion numbers. The $N\times N$ quaternion matrix $M=(\Lambda +S)^{\dagger} (\Lambda +S) +C^{\dagger}C-I_N$ inside the $\delta$-function in \eqref{w} is Hermitian. Correspondingly, it factorises into the product of $N$ $\delta$-functions of scalar argument (diagonal entries) and $N(N-1)/2$ $\delta$-functions of (real) quaternion argument (off-diagonal entries above the diagonal). These $\delta$-functions are characterised by the constraints $M_{j,k}=0$, $1\le j\le k\le N$, which we write below row by row,

\noindent row $j=1$:
\begin{align}
\label{C1}
&  & \vert \vec{c}_1\vert^2 - (1-|\lambda_1|^2) =0\, , & &\\
\label{C2}	
&  & \vec{c}^{\dagger}_1 \vec{c}_k +  \lambda_1^* s_{1k} =0\, , & &  k=2, \ldots, N\, ;
\end{align}
rows $j=2, \ldots , N-1$:
\begin{align}
\label{C3}
& & \vert\vec{c}_j\vert^2  -(1- \vert\lambda_j\vert^2) + \sum_{l=1}^{j-1} \vert s_{lj}\vert^2 =0\, , & &\\
\label{C4}
& & \vec{c}^{\dagger}_j \vec{c}_k +  \lambda_j^* s_{jk} + \sum_{l=1}^{j-1} s_{lj}^{\dagger} s_{lk} =0\, , & & k=j+1, \ldots, N\, ;
\end{align}
row $j=N$:
\begin{align}
\label{C5}
\vert \vec{c}_N \vert^2 - (1- \vert\lambda_N\vert^2) + \sum_{l=1}^{N-1} \vert s_{lN}\vert^2 =0 \, .& &
\end{align}
The obvious hierarchical structure of these constraints which is due to the triangular shape of  $S$ can be exploited to integrate out $S$ in \eqref{w}. 

To start with this calculation, one can integrate out  the $s_{1k}$, 
see  \eqref{C2}. By Prop. \ref{Prop2}(a), 
\begin{align}
\prod_{k=2}^{N} \delta \big( \vec{c}^{\dagger}_1 \vec{c}_k +  \lambda_1^* s_{1k} \big) = 
\frac{1}{\vert \lambda_1 \vert^{4(N-1)}}\prod_{k=2}^{N} \delta \left((\lambda_1^*)^{-1}\vec{c}^{\dagger}_1 \vec{c}_k + s_{1k} \right), 
\end{align}
and the integration over the first row of $S$ yields the factor 
\begin{align} \label{jac}
1/\vert \lambda_1 \vert^{4(N-1)}
\end{align}
and amounts to substituting  $-(\lambda_1^*)^{-1}\vec{c}^{\dagger}_1 \vec{c}_k $ for   $s_{1k}$ in Eqs \eqref{C3}--\eqref{C5}. Introducing the notation $G=\big( I_M+\vec{c}_1\vert \lambda_1\vert^{-2} \vec{c}_1^{\dagger}\big)^{1/2}$, the resulting equations can be written as 

\noindent row $j=2$:
\begin{align*}
&  & \vert G\vec{c}_2 \vert^2  - (1-|\lambda_2|^2) =0 \, , & &\\	
&  & (G\vec{c}_2)^{\dagger} (G \vec{c}_k)+ \lambda_2^* s_{2k} =0\, , & &  k=3, \ldots, N \, ;
\end{align*}
rows $j=3, \ldots , N-1$:
\begin{align*}
& & \vert G \vec{c}_j\vert^2  -(1- \vert\lambda_j\vert^2) + \sum_{l=2}^{j-1} \vert s_{lj}\vert^2 =0 \, , & &\\
& &  (G\vec{c}_j)^{\dagger}  (G\vec{c}_k) + \bar \lambda_j s_{jk} + \sum_{l=2}^{j-1} s_{lj}^{\dagger} s_{lk} =0 \, ;& & k=j+1, \ldots, N
\end{align*}
row $j=N$:
\begin{align*}
\vert  G\vec{c}_N \vert^2 - (1- \vert\lambda_N\vert^2) + \sum_{l=2}^{N-1} \vert s_{lN}\vert^2 =0\, . & &
\end{align*}
Making the substitution $G^{-1}\vec{c}_j$ for $\vec{c}_j$, $j=2, \ldots , N$,  in the integral over $C$ transforms these equations to the exact form of Eqs \eqref{C1}-\eqref{C5} except that we start at $j=2$. 
The Jacobian of this transformation is 
\begin{align*}
\frac{1}{
(\det G)^{4(N-1)}
}
=
\frac{1}{
\det \big(I_M+ \vec{c}_1\vert \lambda_1\vert^{-2} \vec{c}_1^{\dagger}\big)^{2(N-1)}
}
= \frac{\vert \lambda_1\vert^{4(N-1)}}
{
\big(\vert \lambda_1\vert^2+ \vert \vec{c}_1\vert^2 \big)^{2(N-1)}
}\, .
\end{align*}
In view of the constraint \eqref{C1}, when integrating over matrix $C$ in  \eqref{w}  the denominator in the fraction on the right-hand side above will be effectively unity, $\vert \lambda_1\vert^2+ \vert \vec{c}_1\vert^2 =1$, and hence this Jacobian cancels the one in \eqref{jac}. Thus, the integration over the first row of $S$ in \eqref{w} which we just have carried out results in the first equation in \eqref{C3} corresponding to $j=2$ being replaced with 
\begin{align*}
&  & \vert \vec{c}_2\vert^2 - (1-|\lambda_2|^2) =0 & &
\end{align*}
with the rest of equations in \eqref{C3}--\eqref{C5} (and the corresponding $\delta$-functions in \eqref{w}) remaining the same. It is evident now that one can repeat this calculation and integrate out the remaining rows of $S$ one by one. This transforms the integral on the right-hand side in \eqref{w} to 
\begin{align*}
\prod_{j=1}^{N} \int_{\mathbb{R}^{4M}} \delta \big(\vert \vec{c}_j \vert^2  - (1-\vert \lambda_j\vert^2 \big) D[\vec{c}_j] \, .
\end{align*}
Switching to the spherical coordinates,
\begin{align*}
\int_{\mathbb{R}^{4M}}\!\! \delta \big(\vert \vec{c} \vert^2  - (1-\vert \lambda \vert^2) \big) D[\vec{c}] =\vert S^{4M-1}\vert \int_0^{+\infty}\!\!  \delta \big(r^2-(1-\vert \lambda \vert^2 )\big) r^{4M-1} dr = \frac{\pi^{2M}(1-\vert \lambda \vert^2 )^{2M-1} }{\Gamma(2M)}\, ,
\end{align*}
and the statement of Lemma \ref{LemID2} follows.  
\end{proof}

We can now complete our proof of Theorem \ref{Thm:JPDF}.  It follows from Eqs \eqref{PNM},  \eqref{w1} and \eqref{VKN} that the joint density of eigenvalues $P_{N,M}(\lambda_1, \ldots , \lambda_N)$ is given by Eq. \eqref{JPDF} with 
\[
Z_{N,M}=\pi^N N! \prod_{k=1}^N 
 \frac{\Gamma(2M)}{\pi^{2M}} \frac{\Gamma(2k) }{\pi^{2k}}\frac{\pi^{2(M+k)}}{\Gamma(2(M+k))} =\pi^N N! \prod_{k=1}^N B(2M,2k)\, ,
 \]
exactly as was claimed in the statement of Theorem \ref{Thm:JPDF} .
\end{proof}

\section{Eigenvalue correlation functions at finite matrix dimension}
\label{Sec:4}
It is apparent that the joint density of complex eigenvalues of truncated unitary quaternion   matrices \eqref{JPDF} is invariant with respect to reflection about the real axis in each of its variables.  Therefore the integration in \eqref{JPDF1} can be trivially extended from the unit semi-disk $\mathbb{D}_{+}$ to the entire unit disk $\mathbb{D}\{z\in \mathbb{C}: \, \vert z\vert \le 1 \}$ provided the test function $t$ has the same symmetry as the joint eigenvalue density. For such test functions, Theorem \ref{Thm:JPDF} asserts that 
\begin{align}\label{JPDF3}
\mathbb{E} \{t\}= \int_{\mathbb{D}^N} t(z_1, \ldots, z_N) p(z_1, \ldots, z_N) \prod_{j=1}^N d^2\!z_j,
\end{align}
where 
\begin{align}\label{JPDF4}
p (z_1, \ldots, z_N) =  \frac{1}{Z_{N,M}} \prod_{j=1}^{N} w^2(z_j )\prod_{i \geq j} \vert z_j -  z_i^* \vert ^2 \prod_{i > j} \vert  z_j - z_i\vert^2  ,
\end{align}
with 
weight function 
\begin{align}\label{wf}
w^2(z )=(1 - \vert z \vert^2)^{2M - 1}
\end{align}
and 
the normalisation constant
\begin{align}\label{norm}
Z_{N,M}=(2 \pi)^N N! \prod_{j=1}^N B\big(2M, 2j\big)\, .
\end{align}
This is the same expression as in \eqref{JPDF} except for the normalisation constant $Z_{N,M}$ which is adjusted to the integration over the entire unit disk in \eqref{JPDF3}. 

Eq. \eqref{JPDF3} can be thought of as the average of a test function of $N$ pairs of complex conjugated eigenvalues of the complex representation of truncated unitary quaternion   matrices. Obviously, this is an equivalent picture as was discussed in Introduction which we will use from now on for convenience of comparison with previous studies of quaternion-real random matrix ensembles and also for convenience of integration over the entire unit disk. 

In this section we obtain a closed form expression for the eigenpair correlation functions \eqref{Rn_I}.

%
%

\begin{theorem}
\label{Thm:4.1}
	The $n$-point correlation function of the point process defined by \eqref{JPDF3}--\eqref{norm} is
\begin{align}\label{Rn}
R_n (z_1, ..., z_n) & =\prod_{j=1}^n  \left( (z_j- z_j^*) w^2( z_j ) \right) \,  \det \big(\kappa_N(z_k, z_l)\big)_{k, l = 1}^n\, ,
\end{align} 
where $ \big(\kappa_N(z_k, z_l)\big)_{k, l = 1}^n$ 
is an $n\times n$ self-dual complex quaternion matrix with kernel $\kappa_N(z,z') $ which is given by its $2\times 2$ complex matrix representation  
\begin{align}
\label{kappa}
\mathit{ \Theta} (\kappa_N(z,z') ) &= \begin{bmatrix} -g_N({z^*}, z') & -g_N({z^*}, {z^{\prime}}^{*}) \\ g_N(z, z') & g_N(z, {z^{\prime}}^{*})\end{bmatrix} 
\end{align}
and 
\begin{align}\label{gN}
g_N(z, z') &= \frac{ B\big(1/2, M\big)}{\pi} \sum_{0 \le i \le k < N} \frac{z^{2i} (z')^{2k+1} - z^{2k+1}(z')^{2i}   }{B\big(i+1, M\big) B\big(k+\frac{3}{2}, M\big)}.
\end{align}
\end{theorem}

%
%
%
%
%

\begin{proof}
The $n$-point correlation function of the point process  \eqref{JPDF3}--\eqref{JPDF4} can be expressed as a quaternion determinant by the method of skew-orthogonal polynomials. This calculation was carried out  in \cite{KanzPaper} for general weight $w(z)$. Following it one arrives at \eqref{Rn}--\eqref{kappa} with the pre-kernel
\begin{align}\label{gN:SOP}
g_N(z, z') &= \sum_{k = 0}^{N - 1} \frac{q_{2k+1}(z) q_{2k}(z') - q_{2k+1}(z') q_{2k}(z)}{r_k}
\end{align}
expressed in terms of skew-orthogonal polynomials $q_{k}(z)$, $k=0,1,\ldots, 2N-1$.
These are monic polynomials of degree $k$ obeying the relations 
\begin{align}\label{RSOP}
\langle q_{2k+1}, q_{2l}\rangle_s = - \langle q_{2l}, q_{2k+1}\rangle_s = r_k \delta_{kl}\, ,
&&\langle q_{2k+1}, q_{2l+1}\rangle_s = \langle q_{2k}, q_{2l}\rangle_s = 0
\end{align} 
with the skew-product 
\begin{align*}
\langle f, g\rangle_s = \int_{\mathbb{D}} (z - {z^*})\,  w^2(z  )\,  [f(z) g({z^*}) - g(z) f({z^*})] d^2\!z\, .
\end{align*} 


In order to obtain skew-orthogonal polynomials $q_k(z)$, let us first calculate the skew-product of monomials. We have for $k,m \geq 0 $ and $w^2(z)=(1 - \vert z \vert^2)^{2M - 1}$:
\begin{align}
\nonumber
\langle z^{k+m}, z^k\rangle_s &= 
-4\int_0^1 r^{2k+2 +m} (1 - r^2)^{2M-1} dr \int_0^{2\pi}  cos(\phi)\ cos(m \phi) d\phi \\
\label{skewpr}
& = 
\begin{cases} -2\pi B(k+2, 2M)\ \  & m = 1, \\ 0 & m \ge 2. \end{cases}
\end{align} 
Now, setting $q_{2k+1} (z) = z^{2k+1}$ one can construct the desired $q_{2k}(z) =z^{2k}+ \sum_{i=0}^{k-1} c_{k,i} z^{2i} $ by determining the coefficients $c_{k,i}$ with the help of \eqref{skewpr}. The resulting expressions are  
\begin{align}\label{SOP}
q_{2k+1}(z) = z^{2k+1}, \quad
q_{2k} (z) = z^{2k} + \sum_{i=0}^{k-1} z^{2i} \prod_{j=i+1}^{k} \frac{B(2j+1, 2M)}{B(2j, 2M)}\, 
\end{align} 
and one can easily verify that these polynomials obey \eqref{RSOP}. Indeed, it is apparent that $\langle q_{2k+1}, q_{2l+1}\rangle_s  = \langle q_{2k}, q_{2l}\rangle_s  = 0$ and 
\begin{align}\label{r_k}
r_k =\langle q_{2k+1}, q_{2k}\rangle_s  = \langle z^{2k+1}, z^{2k}\rangle_s = -2\pi B(2k+2, 2M).
\end{align} 
It remains to verify that $\langle q_{2k}, q_{2l+1}\rangle_s = 0 $ for $ k \neq l $. The case of $ k < l $ is obvious. Consider $ k > l $. Let $ c_{k, l} $ be the coefficient of $ z^{2l} $ in $ q_{2k}(z) $. Note that $ c_{k, l}= c_{k, l+1}\frac{B(2l+3, 2M)}{B(2l+2,2M)}$. Then $\langle q_{2k}, q_{2l+1}\rangle_s $ is 
\begin{align*}
c_{k, l+1} \langle z^{2l+2}, z^{2l+1}\rangle_s  + c_{k, l} \langle z^{2l}, z^{2l+1}\rangle_s  
=  2\pi [-B(2l+3, 2M) c_{k, l+1} + B(2l+2, 2M) c_{k, l}] =0\, ,
\end{align*}
as was required. 

Using the obvious identity 
\begin{align*}
\prod_{j=i+1}^{k} \frac{B(2j+1, 2M)}{B(2j, 2M)}=\prod_{j=i+1}^{k} \frac{j}{M+j}=\frac{B(k+1,2M)}{B(i+1,2M)}\, ,
\end{align*}
one can write $q_{2k}(z)$ in a form more convenient for the computation of the pre-kernel $g_N(z,z')$:
\begin{align}\label{SOP1}
q_{2k+1} (z) = z^{2k+1}, \quad
q_{2k} (z) = \sum_{i=0}^{k}  \frac{B(k+1, 2M)}{B(i+1, 2M)}\, z^{2i}\, .
\end{align} 
On substituting \eqref{r_k} and \eqref{SOP1} into \eqref{gN:SOP}, one obtains 
\begin{align*}
g_N(z, z') 
&=  \frac{1}{2 \pi} \sum_{0 \le i \le k < N} \frac{B(k+1, M)}{B(i+1, M) B(2k+2, 2M)} (z^{2i} (z')^{2k+1} - (z')^{2i} z^{2k+1} )
\end{align*}
and Eq. \eqref{gN} follows from 
\begin{align*}
 \frac{B(k+1, M)}{B(2k+2, 2M)} &= \frac{2\, B(1/2, M)}{B(k + 3/2, M)}.
\end{align*}
This completes our proof of Theorem \ref{Thm:4.1}.
\end{proof}

The one- and two-point correlation functions can easily be obtained from \eqref{Rn} by expanding the quaternion determinant on the right-hand side. Indeed, the quaternion $k_N(z,z)$ is just the scalar $g_N(z, z^*)$ and the $2\times 2$ quaternion determinant is 
\begin{align*}
\det  \begin{bmatrix} \kappa_N(z_1, z_1) & \kappa_N(z_1, z_2) \\ \bar \kappa_N(z_1, z_2) & \kappa_N(z_2, z_2) \end{bmatrix} &= \kappa_N(z_1, z_1)\kappa_N(z_2, z_2) - \bar  \kappa_N(z_1, z_2)   \kappa_N(z_1, z_2) \\
& = g_N(z_1, z_1^*)g_N(z_2, z_2^*) - \vert g_N(z_1,z_2)\vert^2 +\vert g_N(z_1, z_2^* ) \vert^2 \, .
\end{align*}
Therefore, 
\begin{align} \label{EqDensity1}
R_1 (z) &= (z - {z^*}) w^2( z ) g_N(z, z^*) 
\end{align} 
and 
\begin{align} 
\label{R2}
\begin{split}
R_2 (z_1,z_2) = R_1(z_1)R(z_2) - & \\[1ex]
\MoveEqLeft[5] 
 (z_1 - {z_1^*})(z_2 - {z_2^*})\,  w^2(z_1  )\, w^2( z_2  ) \left( \vert g_N(z_1,z_2)\vert^2 -\vert g_N(z_1, z_2^* ) \vert^2 \right)\, .
\end{split}
\end{align} 
The quaternion determinant  on the right-hand side in \eqref{Rn} can also be written as a Pfaffian:  
 \begin{align}\label{Pf}
  \det \big(\kappa_N(z_k, z_l)\big)_{k, l = 1}^n =\Pf \left(Z_{2n} \Theta \big(\kappa_N(z_k, z_l) \big)_{k, l = 1}^n\right) = \Pf \left[ K_N(z_k,z_l) \right]_{k,l=1}^n \, ,
 \end{align}
where $K_N(z, z')$ is the $2\times 2$ matrix kernel
 \begin{align}\label{KN}
K_N(z, z') =  \begin{bmatrix} g_N(z^{\phantom{\prime}}, z') & g_N(z^{\phantom{\prime}}, {z^{\prime}}^{*} ) \\ g_N( z^* , z') & g_N(z^*, {z^{\prime}}^{*})\end{bmatrix} \, .
 \end{align}
The Pfaffian representation in \eqref{Pf} will come in handy for calculating the higher order eigenvalue correlation functions in the complex bulk, see the next section.

\section{Infinity approximations}
\label{Sec:5}
In this section we investigate the spectral density and eigenpair correlation functions of truncated unitary quaternion  matrices in two regimes. 
%
One is the limit of strong non-unitarity when the number $M$ of the removed rows is proportional to the dimension $N$ of the truncated matrix:
 \begin{align}\label{a}
a= \lim_{N\to\infty }\frac{M}{N}>0\, .
 \end{align}
And the other one is the limit of weak non-unitarity when $M$ stays finite as $N$ grows:
 \begin{align}\label{m}
M=O(1), \quad N\to\infty. 
 \end{align}

\subsection{Strong non-unitarity}
In order to obtain the spectral density \eqref{EqDensity1} and the $n$-point correlation functions \eqref{R2} -- \eqref{KN} one needs to evaluate the prekernel $g_{N}(u,v)$ \eqref{gN} in two cases: (i) $u- v^*$ is  zero (spectral density) or asymptotically small and (ii) $u-v$ is asymptotically small. 


The pre-kernel $g_{N}(u,v) $ is given by a double sum over a finite triangle.  First, we want to be able to extend the summation in  \eqref{gN}  to an infinite triangle. 
Define
\begin{align}\label{g}
g(u, v) &= \frac{B(1/2, M)}{\pi}  \sum_{k=0}^{\infty} \sum_{i=0}^k\,  \frac{ u^{2i} v^{2k+1} - v^{2i} u^{2k+1}  }{B(i+1, M) B(k+3/2, M)}.
\end{align}
and 
\begin{align*}
\Delta_N (u,v) = w (u) w ( u) (g_{N}(u, v) - g(u, v))\, .
\end{align*}
The series on the right-hand side in \eqref{g} is absolutely convergent if $\max (\vert u \vert,  \vert v \vert)<1$. The following Lemma asserts that for the purpose of calculating the spectral density and eigenvalue correlation functions  in the limit of strong non-unitarity  one can replace $g_N(u, v)$ with $g(u, v)$ at a cost of an exponentially small error.

\begin{lemma} (extension of summation in the limit of strong non-unitarity \eqref{a})
\label{lemma51}
Let
\begin{align} \label{D}
\mathbb{D}_{a,\epsilon} =\{ z\in \mathbb{C}: \, \vert z \vert^2 \le (1-\epsilon)/(1+a) \} \, .
\end{align}
 and consider sequences $(u_N,v_N)_N$ in $\mathbb{D}_{a,\epsilon}^2$ such that 
\begin{align}\label{uvN}
\limsup_{N\to\infty,\, M=aN}  \Big(\sqrt{N}\,  \big\vert \vert u_N \vert - \vert v_N \vert\big\vert  \Big) \le c \, . 
\end{align}
Then for any fixed $\varepsilon \in (0,1) $ and $c<+\infty$
\begin{align}\label{b:1}
\limsup\limits_{N\to\infty,\, M=aN} \frac{1}{N} \ln \left\vert \Delta_N \left(u_N, v_N \right)\right\vert \le \ln (1-\varepsilon) + a\ln \left( 1+\frac{\varepsilon}{a} \right) <0
\end{align}
uniformly in  $(u_N,v_N)_N$. 
%
%
\end{lemma}


\begin{proof}  Let $ |z_N| = \max(|u_N|, |v_N|) $.  It follows from \eqref{uvN} that 
\begin{align*}
w(u_N ) =  (1 - |z_N|^2)^M e^{\mathcal{O}(\sqrt{M})} , \, w( v_N  ) =  (1 - |z_N|^2)^M e^{\mathcal{O}(\sqrt{M})} , 
\end{align*}
and 
\begin{align*}
\left\vert \Delta_N (u_N,v_N) \right\vert \le \vert z_N\vert \big(1 - \vert z_N \vert^2\big)^{2M - 1} e^{\mathcal{O}(\sqrt{M})}  \frac{B(1/2, M)}{\pi} \sum_{n = N}^{\infty} \sum_{k = 0}^{\infty} \frac{ |z_N|^{2n} |z_N|^{2k} }{B(k+1, M) B(n+3/2, M)}.
\end{align*}
The sum over $k$ is the binomial series 
\begin{align}\label{binom}
\sum_{k=0}^{\infty} \frac{x^k}{B(k+1,M)} = M \sum_{k=0}^{\infty} \binom{k+M}{k} x^k =  \frac{M}{(1-x)^{M+1}}.
\end{align}
Therefore
\begin{align*}
\left\vert \Delta_N (u_N,v_N) \right\vert  & \le M  \vert z_N\vert  (1 - |z_N|^2)^{M - 2} e^{\mathcal{O} (\sqrt{M})}  \frac{B(1/2, M)}{\pi} \sum_{n = N}^{\infty} \frac{ \vert z_N\vert^{2n} }{ B(n+3/2, M)}.
\end{align*}
By making repeated use of $ B(p+1, M) = \frac{p}{p+M} B(p, M) $, 
\begin{align*}
&\sum_{n = N}^{\infty} \frac{ |z|^{2n} }{ B(n+3/2, M)} = \frac{|z|^{2N} }{B(N+3/2,M)} \sum_{n = 0}^{\infty} \frac{ |z|^{2n} B(N + 3/2, M) }{ B(N + n+3/2, M)} 
\\ &=  \frac{|z|^{2N} }{B(N+3/2,M)} \left(1 + |z|^2 \frac{N+3/2+M}{N+3/2} + |z|^4 \frac{N+3/2+M}{N+3/2} \frac{N+3/2+1 + M}{N+3/2+1} + ...\right)
\\ &=  \frac{|z|^{2N} }{B(N+3/2,M)} \left(1 + |z|^2 \left(1 + \frac{M}{N+3/2}\right) + |z|^4 \left(1 + \frac{M}{N+3/2}\right) \left(1 + \frac{M}{N+3/2}\right) + ...\right)
\\ & =  \frac{|z|^{2N} }{B(N+3/2,M)} \frac{1}{1 - |z|^2 \big(1 + \frac{M}{N+3/2}\big)} \le   \frac{|z|^{2N} }{B(N+3/2,M)} \frac{1}{1 - |z|^2 (1 + a)} \, .
\end{align*}
It now follows that  
\begin{align} \label{Est1}
\left\vert   \Delta_N(u_N,v_N) \right\vert  & \le  \frac{MB(1/2, M)}{\pi B(N+3/2,M) }
\frac{|z_N|^{2N+1}  (1 - |z_N|^2)^{M - 2} e^{\mathcal{O} (\sqrt{M})} }{1 - |z_N|^2 (1 +a)}.
\end{align}
Using Stirling's  approximation of the Gamma function of large argument, 
\begin{align}
\label{s1}
B(p,q) &\sim \Gamma(p) q^{-p} & &\text{$q\to\infty$}  \\
\label{s2}
B(p,q) &\sim \sqrt{2\pi} \frac{p^{p-1/2} q ^{q-1/2}}{(p+q)^{p+q-1/2}} & &\text{ $p,q\to\infty$} , 
\end{align}
one gets in the limit \eqref{a} 
\begin{align*}
&\frac{M B(1/2, M) }{B(N+3/2,M)} \sim \frac{a(1+a)}{\sqrt{2}} N (1+a)^N \left( \frac{1+a}{a}\right)^{aN}\, .
\end{align*}
It now follows that 
\begin{align}\label{aa}
\limsup\limits_{N\to\infty,\, M=aN} \frac{1}{N} \ln \left\vert \Delta_N \left(u_N, v_N \right)\right\vert \le \max_{z\in \mathbb{D}_{a,\varepsilon}} \ln \left[ |z|^2 (1 - |z|^2)^{a} (1+a) \left( \frac{1+a}{a} \right)^a\right] \, .
\end{align}
Consider function $ f(x) = x(1-x)^a $ on the interval $0\le x\le 1$. This function is monotone increasing on the interval  $0<x<1/(1+a)$, monotone decreasing on the interval  $1/(1+a)<x<1$, attaining its maximum $\frac{a^a}{(1+a)^{1+a}}$ at $x=1/(1+a)$.  Therefore 
\begin{align*}
 \max_{z\in \mathbb{D}_{a,\varepsilon}} \left[ |z|^2 (1 - |z|^2)^{a}\right] =   f\left( \frac{1-\varepsilon}{1+a} \right) =   \frac{1-\varepsilon}{1+a} \left( \frac{a+\varepsilon}{a+1} \right)^a \, .
\end{align*}
This and \eqref{aa} imply \eqref{b:1}.

\end{proof}

Our next step is to demonstrate that in the limit of strong non-unitarity \eqref{a} the spectral density $\rho_{2N}(z)$
\eqref{spdensity_I} is exponentially small if $\frac{1}{1+a}< \vert z \vert^2<1$. 

\begin{lemma} 
\label{lemma52}
For each $z$ in the annulus $\frac{1}{1+a} < \vert z \vert^2 <  1 $ it holds that 
\begin{align}
\label{bound1}
\limsup_{N\to\infty, \, M=aN} \frac{1}{N} \ln R_1(z) \le r\big(\vert z\vert^2 \big) \quad\text{with} \,\,r(x)= \ln \left[ x(1 - x)^{a} \,  \frac{(1+a)^{(1+a)}}{a^{a}} \right]\, .
\end{align}
The function $r(x)$ vanishes at $x=\frac{1}{1+a}$, is negative and monotonically decreasing on $\left(\frac{1}{1+a}, 1\right)$.
\end{lemma}
\begin{proof}
Define 
\begin{align*}
S_{N,M}(x) = \sum_{j=0}^{N-1} \frac{x^j}{B(j+3/2, M)}\, .
\end{align*}
On making the substitution $j \!=\! N \!-\! 1\! -\! i $ and then using $ B(p, M) = (1 + \frac{M}{p}) B (p+1, M) $,  
\begin{align*}
\left\vert S_{N,M}(x)\right\vert  = \frac{|x|^{N-1}}{B(N+\frac{1}{2}, M)} \sum_{i=0}^{N-1} \frac{|x|^{- i} B(N + \frac{1}{2}, M)}{B(N - i + \frac{1}{2}, M)} \le \frac{|x|^{N-1}}{B(N+\frac{1}{2}, M)} \sum_{i=0}^{N-1} \frac{1}{(|x| (1 + \frac{M}{N-1/2}))^{i} }\, .
\end{align*}
Note that $\frac{1}{1+ \frac{M}{N-1/2}} > \frac{1}{1+ a}$. Therefore, for each $ |x| > \frac{1}{1+ a}$, 
\begin{align}\label{S}
|S_{N,M}(x)| \le  \frac{|x|^{N-1}}{B(N+\frac{1}{2}, M)} \sum_{i=0}^{\infty} \frac{1}{(|x| (1 + a))^{i} } = \frac{|x|^{N}}{B(N+\frac{1}{2}, M)} \frac{1 + a}{|x| (1 + a) - 1}.
\end{align}
Recall that $R_1 (z) = (z - {z^*}) (1 - |z|^2)^{2M - 1}  g_N(z,  z^*) $ with $g_N(z, z^*) $ given by \eqref{gN}. Therefore,
\begin{align*}
R_1 (z)  &\le  4|z|^2 (1 - |z|^2)^{2M - 1}  \frac{B(1/2, M)}{\pi} \sum_{k = 0}^{N-1} \sum_{i = 0}^{\infty} \frac{|z|^{2i} }{B(i+1, M)}  \frac{ |z|^{2k}}{ B(k+3/2, M)}
\\ &\le \frac{4M \, |z|^2(1 + a)}{\pi (1 - |z|^2)^{2} [|z^2| (1 + a) - 1]} \,\,  |z^2|^{N} (1 - |z|^2)^{M } \, \frac{B(1/2, M)}{B(N+\frac{1}{2}, M)}\, , 
\end{align*}
where we have used \eqref{S} and \eqref{binom}. By Stirling's approximation \eqref{s1}--\eqref{s2}, 
\begin{align*}
&\frac{B(1/2, M)}{B(N+\frac{1}{2}, M)} \sim \frac{\Gamma(1/2) (N+M+\frac{1}{2})^{M+N}}{ M^{1/2} \sqrt{2\pi} (N+\frac{1}{2})^N M^{M-\frac{1}{2}}} 
\sim \frac{1}{\sqrt{2}} \left( \frac{(1+a)^{(1+a)}}{a^{a}} \right)^N.
\end{align*}
Therefore, 
\begin{align*}
\limsup_{N\to\infty, \, M=aN} \frac{1}{N} \ln R_1(z)   
&\le  \ln \left[ |z|^2(1 - |z|^2)^{a} \,  \frac{(1+a)^{(1+a)}}{a^{a}} \right]=r\big( \vert z \vert^2\big)\, .
\end{align*}
This proves \eqref{bound1}. By calculating the derivative of $ f(x)= x(1-x)^a$ one can verify that this function is monotonically decreasing on the interval $\frac{1}{1+a} \le x  < 1 $, and so is $r(x)$.  Since $f\big(\frac{1}{
1+a} \big)=\frac{1}{1+a)}\big(\frac{a}{(1+a}\big)^{\!a}$, we have that $r\big(\frac{1}{
1+a} \big)=0$ and, hence,  $r\big( \vert z \vert^2\big)<0$ if $\frac{1}{(1+a)}< \vert z \vert^2<1$. 

%
\end{proof}

In the limit \eqref{a},  the double sum \eqref{g} can be calculated  by employing an integral representation.

 \begin{lemma} (integral representation) 
 \label{lemma53}
\begin{align}
\label{ItegrRep}
\begin{split}
g (u, v) = & \\
\MoveEqLeft[2] 
\frac{M^2\, B(1/2, M) }{2 \pi^2 i } 
\!\!\oint_{|w|=r}
\frac{dw}{w^{M+1}}  
\frac{(1\! +\! w)^{M+\frac{1}{2}}}
{[1 \!-\! (uv)^2(1 \!+ \!w)]^{M+1}} \,
\frac{(v-u) [1 + uv(1+w)]}{[1 - v^2 (1+w)] [1 - u^2 (1+w)]}
\end{split}
\end{align} 
provided both $u$ and $v$ are in the disk $ \{z\in \mathbb{C}:\, \vert z \vert^2 < 1/(1+a)  \}$. The integral is taken counterclockwise in the complex plane along the circle $\vert w \vert =r$, $r<\min (a, 1)$.
\end{lemma}
\begin{proof}
By changing the order of summations in \eqref{g}, 
\begin{align*}
g(u, v) &= \frac{B(1/2, M)}{\pi} \sum_{i=0}^{\infty} \sum_{n=0}^{\infty}  \frac{ (uv)^{2i} (v^{2n+1} - u^{2n+1}) }{B(i+1, M) B(n+i +\frac{3}{2}, M)}\, .
\end{align*}
The reciprocal of the Beta function can be written as a contour integral 
\begin{align}\label{Stade}
\frac{1}{B(p,M)} = \frac{M}{2 \pi i } \oint_{|w|=r}  \frac{(1+w)^{M+p-1}}{w^{M+1}}dw\, ,
\end{align}
where the integral is taken counterclockwise in the complex plane over the circle $\vert w \vert =r$, $r<1$.  
Now, by making use of \eqref{Stade} and assuming that both $u$ and $v$ are in the disk $ \{z\in \mathbb{C}:\, \vert z \vert^2 < 1/(1+a)  \}$ and that $r<a$ so that the binomial and geometric series below are convergent,  
\begin{align*}
\sum_{i=0}^{\infty} \sum_{n=0}^{\infty}  \frac{ (uv)^{2i} v^{2n+1}  }{B(i+1, M) B(n+i +\frac{3}{2}, M)} & \\
\MoveEqLeft[10] 
= \frac{M}{2 \pi i } \oint_{|w|=r}  \frac{(1+w)^{M+\frac{1}{2}}}{w^{M+1}}   \sum_{i=0}^{\infty}  \frac{ (uv)^{2i}(1+w)^i }{B(i+1, M)} \sum_{n=0}^{\infty}  v^{2n+1}(1+w)^n dw\\
\MoveEqLeft[10] 
= \frac{M^2}{2 \pi i } \oint_{|w|=r}  \frac{(1+w)^{M+\frac{1}{2}}}{w^{M+1}}  \frac{ 1 }{[1-(uv)^{2}(1+w)]^{M+1}} \frac{v}{1- v^2(1+w)} dw, 
\end{align*}
and Eq.~\eqref{ItegrRep} follows. 

%
\end{proof}

 Now, we are in a position to evaluate the pre-kernel $g(u,v)$ in two scaling limits of interest: one is when $u$ is close to $v^*$ and another is when  $u$ is close to $v$.  
\begin{lemma} (prekernel in the complex bulk) 
\label{lemma54}
Let $z$ be a fixed point inside the semi-disk 
\begin{align}\label{sdisk}
 \mathbb{D}_{a,+} =  \left\{ z\in \mathbb{C}: \,  \vert z \vert^2 < 1/(1+a),  \,\, \Im z >0 \right\}
\end{align}
and 
\begin{align*}
u = z + \frac{s}{N^{1/2}}, \,\, v = z^* + \frac{t}{N^{1/2}}, \quad \tilde u = z + \frac{\tilde s }{N^{1/2}},   \,\, \tilde v = z + \frac{\tilde t}{N^{1/2}}\,. 
\end{align*} 
Then in the limit of strong non-unitarity  \eqref{a} it holds for any fixed complex $s$, $t$, $ \tilde s$ and  $ \tilde t $ that 
\begin{align} \label{EqAsymGi}
&g (u, v) \sim  \frac{M}{\pi \, (u - v)} \frac{1}{(1 - uv)^{2M + 1}}
\end{align} 
and 
\begin{align} \label{EqAsymGi1}
\lim_{N\to\infty, \, M=aN} \frac{1}{N} \ln \left\vert \frac{g (\tilde u, \tilde v)}{g (u, v)} \right\vert <0\, .
\end{align} 
\end{lemma}

\begin{proof}
Lemma \ref{lemma53} supplies an integral representation of $g(u,v)$ which is suitable for an asymptotic analysis via the saddle point method \cite{deBruijn1961}. We have 
\begin{align}\label{sp1}
g(u,v)  &
= \frac{M^2\, B(1/2, M) }{2 \pi^2 i }  \oint_{|w|=r}  e^{-M f(w)} h(w) dw,
\end{align} 
were 
\begin{align*}
f(w) = \ln \frac{w [1 - (uv)^2(1+w)]}{1+w}
\end{align*}
and 
\begin{align}\label{fh}
h(w) =  \frac{(v-u)[1 + uv(1+w)] \sqrt{1+w} }{[ 1 - v^2 (1+w)] [1 - u^2 (1+w)][1 - (uv)^2(1 + w)]w} \, . 
\end{align}
The saddle points of $f(w)$ are the roots of the equation $ f'(w)=0 $. Noting that 
\begin{align*}
f'(w) =  \frac{1 - (uv)^2(1+w)^2} {w(1+w)[1-(uv)^2(1+w)]}, \quad  f''(w) =  -\frac{2w + 1}{(w^2 + w)^2} - \frac{1}{((uv)^{-2} - (1+w))^2}\, , 
\end{align*}
the function $f(w)$ has two saddle points
\begin{align*}
w_{\pm} = -1 \pm \frac{1}{uv}\, .
\end{align*}
At these saddle points
\begin{align} \label{sp2}
e^{-Mf(w_{\pm })} = \frac{1}{(1 \mp  uv)^{2M}}, \quad   f''(w_{\pm}) =  \frac{\mp 2(uv)^3}{(1  \mp uv)^2}\, , 
\end{align}
and
\begin{align}\label{sp3}
h(w_{+}) = \frac{2(uv)^{3/2}}{(1 - uv)^2 (u - v)}, \quad  h(w)\sim (w-w_{-})\frac{v-u}{(u+v)^2}\frac{\sqrt{-uv}}{(1+uv)^2}\,\, \text{as} \,\, w\to w_{-}\, .
\end{align}
 
Now, we are ready to carry out our saddle point analysis. Let us assume first that, that $ u $ and  $ v $ are asymptotically complex conjugated and not real, 
\begin{align}\label{gc}
u = z + \frac{s}{N^{1/2}}, \quad v = z^* + \frac{t}{N^{1/2}}, \quad z\in \mathbb{D}_{a,+}\, ,
\end{align}
so that $uv=|z|^2+O\left(M^{-1/2}\right)$. 

Since 
\begin{align*}
e^{-M\Re f(w_{+})} = \left| \frac{1}{1 - uv} \right|^{2M} \gg \left| \frac{1}{1 + uv} \right|^{2M} = e^{-M \Re f(w_{-})}\, ,
\end{align*}
the saddle point at $w_{+}$ is dominant. Hence, the other one needs not be considered provided we can deform the contour of integration in \eqref{sp1} to a suitable contour $\gamma$ such that the integral does not change its value, $\gamma$ crosses $w_{+}$ in the direction (almost) perpendicular to the real axis and $\Re f(w) > \Re f(w_{+}) $ everywhere on $\gamma$  except $w=w_{+}$.  We, therefore, now turn to investigating whether we can identify such a suitable contour $\gamma$ or not.  To this end, it is more convenient to rewrite the integral in \eqref{sp1} in an equivalent form as 
\begin{align}\label{sp2}
 \oint_{|w|=r} \left[F(w)\right]^M  h(w) dw,
\end{align} 
were 
\begin{align}\label{F}
F(w) =  \frac{1+w}{w [1 - (uv)^2(1+w)]}
\end{align}
and $h(w)$ being the same as in \eqref{sp1}. We have to be mindful of the pole of $F(w)$ at $w_2=-1+\frac{1}{(uv)^2}$, the poles of $h(w)$ at 
\begin{align}\label{poles}
w_2=-1+\frac{1}{(uv)^2}, \quad w_3=-1+\frac{1}{u^2}, \quad w_4=-1+\frac{1}{v^2}, 
\end{align}
and a branch cut required to make the square root $\sqrt{1+w}$ single valued. 

For the clarity of our argument, we will assume that $u=z$, $v=z^*$ so that $uv=|z|^2$. If we can identify a suitable contour of integration in this case, then, the existence of a suitable contour for for sufficiently large $M$ in the general case \eqref{gc} will follow from the continuity of $F(w)$.

Consider the circle 
\begin{align*}
\gamma = \left\{ w: \quad w=\frac{1-r }{r}\, e^{i \theta}, \; r=|z|^2, \theta\in [0,2\pi) \right\}\, .
\end{align*}
On this circle 
\begin{align*}
|F(w)|^2= \frac{1}{(1-r)^4}\frac{\big| r+(1-r )e^{i\theta} \big|^2}{\big| 1+r - re^{i\theta}) \big|^2}
\le \frac{1}{(1-r)^4}=|F(w_{+})|^2\, ,
\end{align*}
and the equality is only attained if $w=w_{+}$. Therefore, it holds that  $|F(w)| < |F(w_{+})|$ everywhere on $\gamma$ except at $w=w_{+}$, and, obviously, $\gamma$ crosses the saddle point at $w_{+}$ in the direction perpendicular to the real axis (the direction of the steepest descent at $w_{+}$ on the surface of $|F(w)|$. 

Since $z$ is assumed to be not real, the poles $w_3=-1+\frac{1}{z^2}$ and $w_4=w_3^*$ lie outside the circle $\gamma$, and the pole $w_2=-1+\frac{1}{|z|^4}$ lies outside this circle regardless of whether $z$ is real or not. Therefore, ignoring the issue of the branch cut of $\sqrt{1+w}$, the circle $\gamma$ is an accessible suitable contour of integration for application of the saddle point method. 

Now, set the branch cut of $\sqrt{1+w}$ along the ray $(-\infty, -1]$. If $|z|^2>1/2$ then the circle $\gamma$ lies entirely to the right of this branch cut, and therefore deform the original contour of integration in \eqref{sp1} to $\gamma$ without changing the value of the integral. If  $|z|^2=1/2$ then we can slightly dent $\gamma$ to avoid the branch point $w=-1$ retaining all of the desired properties of $\gamma$. And if $|z|^2<1/2$ then the circle $\gamma$ will intersect the branch cut at $w_c= 1-\frac{1}{|z|^2}<-1$. In this case we can choose the boundary of the disk encircled by $\gamma$ with the cut along the interval $[w_c, -1]$ as our new contour of integration $\gamma'$. For every real $x\in [w_c, -1]$, 
\begin{align*}
\big| F(x) \big|  =  \frac{1+x}{x [1 - |z|^4(1+x)]} = F(x), 
\end{align*}
and the value of $ |F(x)|$ is monotonically decreasing on the interval $ [w_c, -1]$ from $F(w_c)$ at $x=w_c$ to zero at $x=-1$. Since $\big| F(w_c)\big| < \big| F(w_{+})\big| $, we conclude that $\big| F(w)\big| < \big| F(w_{+})\big| $ everywhere on $\gamma'$ except at the saddle point $w_{+}$. Obviously, by continuity of $F(w)$ this conclusion does not change if we let $\gamma'$ to run slightly above the upper bank and slightly below the lower bank of the branch cut to get inside the domain of analyticity of the square root. This $\gamma'$ will be the desired new contour of integration  suitable for the application of saddle point point method.  

Having established the existence of a suitable contour of integration, 
we can now complete the asymptotic evaluation of $g(u,v)$ for $u$ and $v$ as in \ref{gc}.
By making use of \eqref{sp2} -- \eqref{sp3}, 
\begin{align*}
g(u,v) & \sim \frac{M^2\, B(1/2, M) }{2 \pi^2 i } \sqrt{\frac{2 \pi}{M f''(w_{+})}} e^{-Mf(w_{+})} h(w_{+}) =
\frac{M^{3/2} B(1/2, M)}{\pi^{3/2} }  \frac{1}{(1 - uv)^{2M + 1} (u - v)}.
\end{align*} 
Noting that for large $M$
\begin{align*}
\frac{M^{3/2} B(1/2, M)}{\pi^{3/2} } \sim \frac{M}{\pi}, 
\end{align*} 
one then arrives at 
\begin{align*} 
g_{M} (u, v) \sim \frac{M}{\pi} \frac{1}{(1 - uv)^{2M + 1} (u - v)},
\end{align*} 
as was claimed in (\ref{EqAsymGi}).

Now, let us consider two asymptotically close points $\tilde u$ and $\tilde v$ in the semidisk \eqref{sdisk}
\begin{align}\label{acp}
\tilde u = z + \frac{\tilde s}{N^{1/2}}, \quad \tilde v = z + \frac{\tilde t}{N^{1/2}}, \quad z\in \mathbb{D}_{a,+}\, .
\end{align}
We have 
\begin{align*}
g(\tilde u,\tilde v)  &
= \frac{M^2\, B(1/2, M) }{2 \pi^2 i }  \oint_{|w|=r}  \big[\tilde F(w) \big]^M \, \tilde h(w) dw,
\end{align*} 
where $\tilde F(w)$ and $\tilde h(w) $ are given by the right-hand sides of, respectively,  \eqref{fh} and \eqref{F} with $u$ and $v$ there replaced by $\tilde u$ and $\tilde v$.  

If $u$ and $v$ are asymptotically complex conjugated, $u=z+s/\sqrt{N}$, $v=z^*+t/\sqrt{N}$ and $z\in \mathbb{D}_{a,+}$,  then it follows from \eqref{EqAsymGi} that 
\begin{align*} 
\lim_{M\to\infty} \frac{1}{M} \ln \left\vert g (u, v) \right\vert = \frac{1}{(1-|z|^2)^2}\, .
\end{align*} 
Therefore, in order to prove \eqref{EqAsymGi1} it will suffice to establish that 
\begin{align} \label{sptilde}
\limsup_{M\to\infty} \frac{1}{M} \ln \left\vert \oint_{|w|=r}  \big[\tilde F(w) \big]^M \, \tilde h(w) dw \right\vert < \ln  \frac{1}{(1-|z|^2)^2}\, .
\end{align} 
Observe that the function $\tilde F(w)$ has two saddle points $\tilde w_{\pm} = -1 \pm \frac{1}{\tilde u\tilde v}$, and  $\tilde F(\tilde w_{\pm }) = \frac{1}{(1 \mp  \tilde u \tilde v)^{2}}$.
Since $z$ is not real, it holds that $|1-|z|^2| < |1- z^2|$ and, therefore, 
\begin{align*} 
\lim_{M\to\infty} \left\vert \tilde F(\tilde w_{\pm })  \right\vert = \left\vert  \frac{1}{(1 \mp  z^2 )^{2}}  \right\vert  <  \frac{1}{(1-|z|^2)^2}\, .
\end{align*} 
This inequality suggests a  strategy of proving \eqref{sptilde} which we will follow below. Considering the contour integral on the left-hand side in \eqref{sptilde}, deform the circular contour  of integration $|w|=r$ into a new contour $\gamma$  of finite length such that  (i) the value of the integral does not change, 
\begin{align} \label{ci1}
\oint_{|w|=r}  \big[\tilde F(w) \big]^M \, \tilde h(w) dw = 
\oint_{\gamma}  \big[\tilde F(w) \big]^M \, \tilde h(w) dw \, ,
\end{align} 
(ii) 
$ \big| \tilde F(w)\big| \le C_1(z) <  \frac{1}{(1-|z|^2)^2}$ 
and $ \big| \tilde h (w) \big| \le C_2 (z) <+\infty $ on $\gamma$. The existence of such a contour  $\gamma$ will imply the asymptotic relation \eqref{sptilde}. 

In order to satisfy ourselves that such a contour $\gamma$ exists, we will examine the landscape generated by  $\big|  \tilde F (w)\big| $ in a three dimensional space. It will suffice to consider the case of $\tilde u=\tilde v=z$. The existence (for large enough $N$) of a suitable contour of integration $\gamma$ for asymptotically equal $u$ and $v$  \eqref{acp}  will follow  by the continuity argument.  

It is more convenient to work in the complex plane $q=w+1$. We put $q=x+iy$ and consider the surface in the three dimensional space $(x,y,f)$ whose equation is $f=\big| \tilde F (1+q)\big|$. This surface has several characteristic features. It touches the $(x,y)$-plane at $q =0$ and, also, asymptotically at infinity ($|q|\to\infty$) where the function $\tilde F (1+q)$ vanishes in every direction. There are two infinitely high peaks (singularities) on the surface. These correspond to the two poles of $\tilde F (1+q)$,  one at $q_1 =1$ and another at $q_2=\frac{1}{z^4}$. There are also two saddles on the surface corresponding to $q_{\pm}=\pm \frac{1}{z^2}$.  Writing $z$ in the polar coordinates as $z=\tau e^{i\theta}$, 
\begin{align*} 
q_1=1, \quad q_2 = \frac{1}{\tau^4}e^{-4 \theta i}, \quad q_{+}=\frac{1}{\tau^2}e^{-2 \theta i}, \quad q_{-}=\frac{1}{\tau^2}e^{-(\pi+2 \theta)i}, 
\end{align*} 
where 
\begin{align*} 
\tau<\sqrt{1/(1+a)}<1\quad \text{and} \quad 0< \theta < \pi\, .
\end{align*} 
In the $(x,y)$-plane, the distance between the pole at $q_1$ and the saddle point which is closest to it will always be shorter than the distance between the two poles $q_1$ and $q_2$. Hence, this is the saddle that lies in between the two peaks. The other saddle lies away from the two peaks in the region of low altitude across the straight line through the origin partitioning the $(x,y)$-plane into two halves, one with the peaks and another without, see Fig.~\ref{Fig:5}. This saddle is of no interest to us. Other than these peaks and saddles the landscape generated by the surface is gently slopping down to the sea level ($(x,y)$-plane).

\begin{figure}
\includegraphics[width=.4\linewidth]{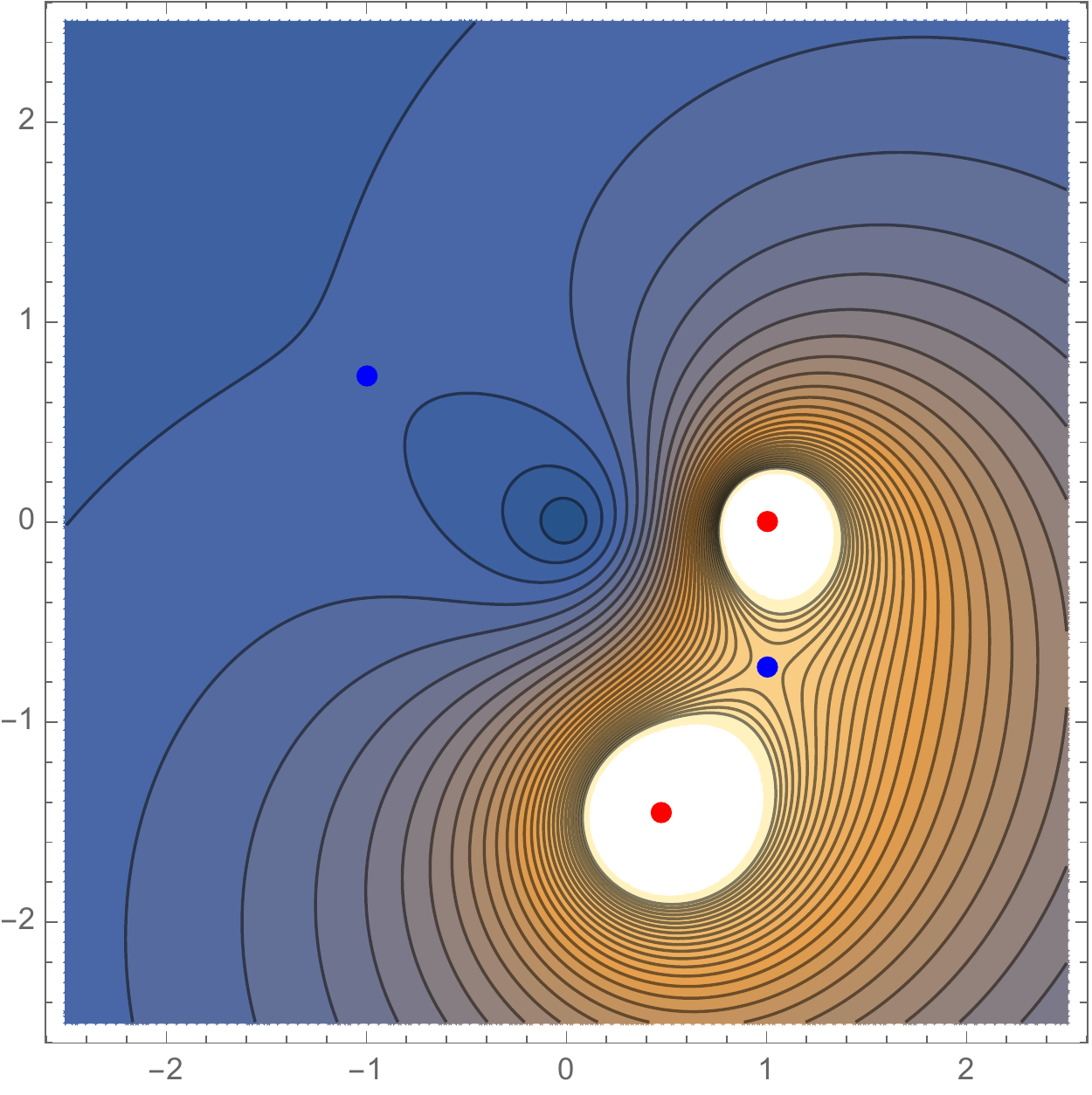}
\includegraphics[width=.4\linewidth]{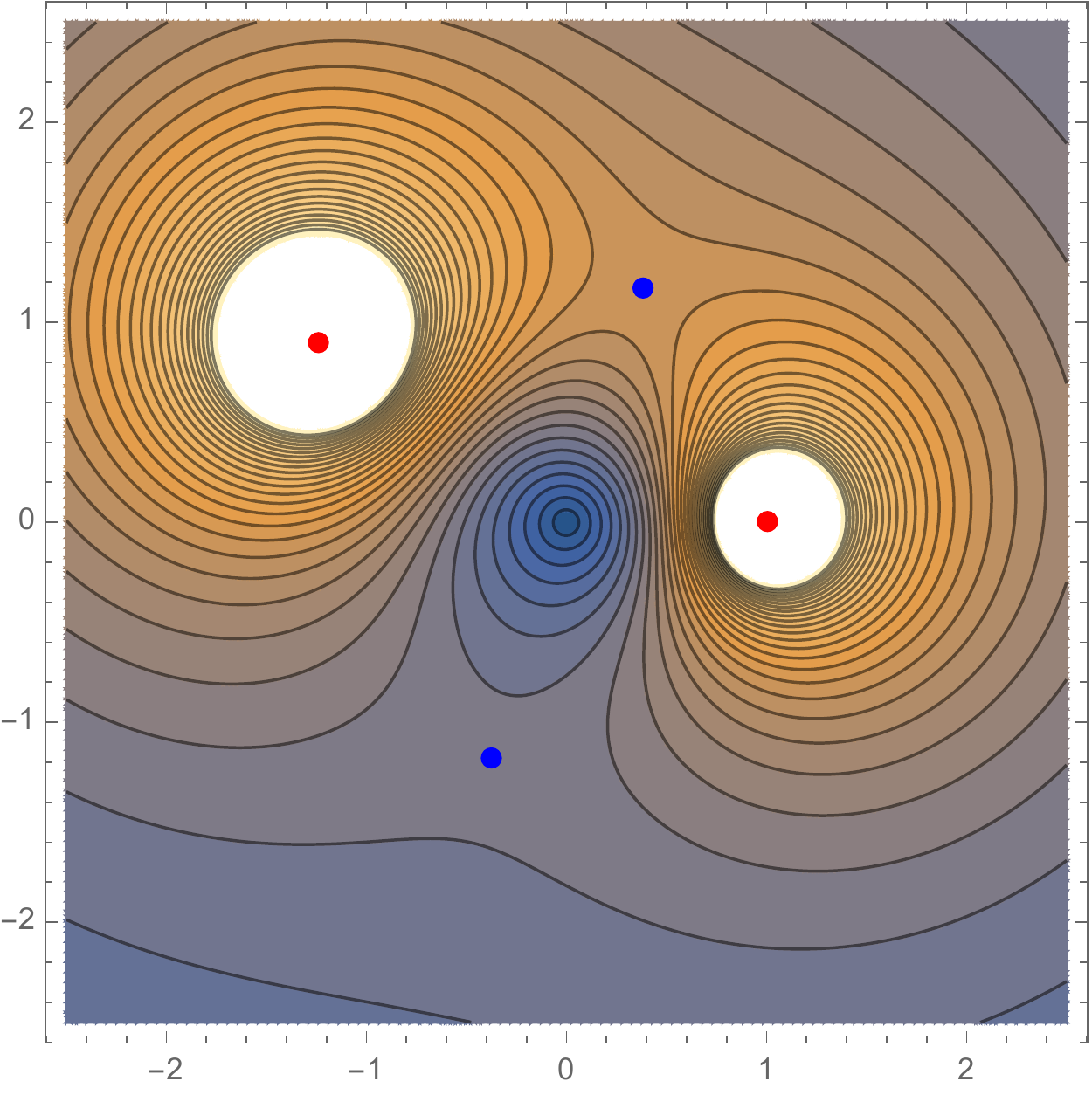}
\caption{Contour plot of function $|\tilde F (1+q)|$, $q = x+iy$. In the plot on the left $z=0.9e^{\frac{\pi}{10}i}$ and in the plot on the right $z=0.9\, e^{\frac{4\pi}{5}i}$. The red dots in white spaces are the poles and the blue dots are the saddle points of $|\tilde F (1+q )|$.}
\label{Fig:5}
\end{figure}

Now we turn to our contours of integration. In the $(x,y)$-plane, the contour in the integral on the left-hand side in \eqref{ci1} is the circle of radius $r$, $r<\min(a,1)$, centred at the pole of $F (1+q)$ at $q_1$. Since $\tau^2<1/(1+a)$, the other pole (and the two saddle points) lie outside this circle. Consider now the pass of constant altitude on our surface going through the saddle which lies in between the two peaks. The projection of this path on the $(x,y)$-plane is a level line of $|\tilde F (1+q)|$ which consists of two joined up loops, one around the pole at $q_1$, which we call $\gamma$ and the other around the pole at $q_2$. Obviously, 
$|\tilde F (1+q)|<1/(1-|z|^2)^2$ on $\gamma $  and one can deform the original circular path of integration into $\gamma$ staying in the domain of analyticity of $\tilde F (1+q)$. 

Now we need to take care of the pre-exponential factor $\tilde h (1+q)$. To make this function single valued we set the branch cut along the negative real semiaxis. Since the function $\tilde F (1+q)$ vanishes at $q=0$,  $\gamma$ loops around the pole of $\tilde F (1+q)$ at $q_1$ before it comes in the vicinity of the branch cut. Therefore the branch cut of $\tilde h (1+q)$ is not an obstacle for our purposes. The function $\tilde h (1+q)$ has also two poles, one at $q_2$ and another one at $q_3$ which in our case of $\tilde u=\tilde v=z$ coincides with one of the saddle points of $\tilde F (1+q)$. If our new contour of integration $\gamma$ goes through this saddle point then we can circumvent the singularity of $\tilde h (1+q)$ by lowering our path of constant altitude in the vicinity of the saddle. Such a new path will be exactly what we are looking for as it satisfies conditions (i) and (ii). 
This proves \eqref{EqAsymGi1} and hence \eqref{sptilde} too. 
\end{proof}

With all the preparatory work above, we can now obtain the spectral density and the eigenpair correlation functions locally in the complex bulk. We start with the spectral density $\rho_{2N}(z)$ \eqref{spdensity_I}. 

 \begin{theorem} 
 \label{DensityTheorem} 
In the limit of strong non-unitarity \eqref{a},
\begin{align}\label{d}
\lim_{N\to\infty, \, M=aN} \rho_{2N}(z) = 
\begin{cases}
\displaystyle{ 
\frac{a}{\pi (1 - |z|^2)^{2}}
}, & 
\displaystyle{ 
\text{if \,  $\vert z\vert^2 < \frac{1}{1+a}$ and $\Im z \not= 0$}
} \\
 0 & 
 \displaystyle{ 
 \text{if \, $ \frac{1}{1+a}< \vert z\vert^2 <1$}
 }
\end{cases}
\end{align}
\end{theorem}

\begin{proof}  
Lemma \ref{lemma52} asserts that  $R_1(z)$ is exponentially small in the annulus $ \frac{1}{1+a}< \vert z\vert^2 <1$ for $N$ large. The derived bound \eqref{bound1} on the rate of decay is uniform on compact sets in this annulus. Since  $\rho_{2N}(z) =\frac{1}{N} R_1(z)$,  the same is true of   $\rho_{2N}(z)$.

On the other hand, $R_1(z)$ is given by Eq.~\eqref{EqDensity1} and if $\vert z\vert^2 < \frac{1}{1+a}$ then one can use  Lemma \ref{lemma51} to approximate $R_1(z)$ 
in the limit of strong non-unitarity by 
\begin{align*}
(z- z^*) (1 - \vert z \vert^2)^{2M - 1} g(z, z^*) \, .
\end{align*}
This approximation is uniform on compact sets in the disk $\vert z\vert^2 < \frac{1}{1+a}$. Lemma \ref{lemma54} asserts that 
\begin{align*}
g(z, z^*)  \sim \frac{M}{\pi \, (z -  z^*)}
\frac{1}
{ (1 - \vert z \vert^2)^{2M + 1} } \, .
\end{align*}
for each $z$ in the semi-disk $\mathbb{D}_{a,+}$ \eqref{sdisk}. Therefore
\begin{align}\label{R1_a}
R_1(z) \sim \frac{M}{\pi (1 - |z|^2)^{2}}, \quad \text{if }\vert z \vert^2 < \frac{1}{1+a}\, \text{ and }\, \Im z \not= 0\, ,
\end{align}
and $\rho_{2N}(z)\to \frac{a}{\pi (1 - |z|^2)^{2}}$ everywhere in the disk $\vert z\vert < \frac{1}{1+a}$ except the real line where $\rho_{2N}(z)$ is zero.  
\end{proof}

{\bf Remark.} Theorem \ref{DensityTheorem} implies the weak convergence of the expectation of normalised empirical eigenvalue counting measure $d\mu_{2N}(z) = \frac{1}{2N}\sum_j \big( \delta(z-z_j)+\delta(z- z_j^*)\big)d^2\!z$ to the limiting measure $d\mu$  with density given by the right-hand side of  \eqref{d}. The convergence of  $d\mu_{2N}$ to $d\mu$ in probability and also almost-sure convergence on a suitable probability space can be established by adapting the potential theory analysis undertaken in \cite{B-GCh2012} for the quaternion-real Ginibre ensemble to truncations, and the probabilities of large deviations of  $d\mu_{2N}$ from $d\mu$ were obtained in \cite{Thesis}. 

\medskip

Now, we compute the two-point correlation function $R_2(z_1,z_2)$ locally in the complex bulk. Let the reference point $z$ be in the semi-disk $\mathbb{D}_{a,+}$ \eqref{sdisk} and consider 
\begin{align*} 
z_1=z+\frac{s_1}{\sqrt{M} }, \quad z_2=z+\frac{s_2}{\sqrt{M}}
\end{align*} 
with complex $s_1$ and $s_2$. Lemma \ref{lemma51} asserts  that one can replace $g_N$ in Eq.~\eqref{R2} with $g$ for the purpose of calculating $R_2(z_1, z_2)$, so that 
\begin{align*} 
R_2 (z_1,z_2) \sim R_1(z_1)R(z_2) - & \\[1ex]
\MoveEqLeft[5] 
 (z_1 - {z_1^*})(z_2 - {z_2^*})\,  w^2( z_1)\, w^2( z_2 ) \left( \vert g(z_1,z_2)\vert^2 -\vert g(z_1,  z_2^* ) \vert^2 \right)\, ,
\end{align*} 
with the weight function $w(z)$ and the pre-kernel $g(z,z')$ given by Eqs. \eqref{wf} and \eqref{g}. 

Now, it follows from Lemma \ref{lemma54} that $g(z_1,z_2) \ll g(z_1, z_2^*) \sim   \frac{M}{\pi \, (z_1 -  z_2^*)} \frac{1}{(1 - z_1 z_2^*)^{2M + 1}}$. Therefore
\begin{align}\label{R2_u} 
R_2 (z_1,z_2) 
 & \sim R_1(z_1)R(z_2)\!\left(1  - e^{ -2M \frac{\vert z_1-z_2\vert^2}{\vert 1- z_1 z_2^* \vert^2}}\right)
\end{align} 
where we have used \eqref{R1_a} and 
\begin{align*} 
\left[\frac{(1-\vert z_1 \vert^2)(1-\vert z_2 \vert^2)}{ (1- z_1 z_2^*)(1-  z_1^* z_2)}\right]^{2M}= \left[1- \frac{\vert z_1-z_2\vert^2}{\vert 1-z_1z_2^*\vert^2}\right]^{2M}\sim e^{ -2M \frac{\vert z_1-z_2\vert^2}{\vert 1- z_1  z_2^* \vert^2}}\, .
\end{align*} 
Eq.~\eqref{R2_u} can be rewritten in a form which exhibits the universal form of the local two-point correlation function.
Recalling that $R_1(z)\sim \frac{M}{\pi (1 - |z|^2)^{2}}$ in $\mathbb{D}_{a,+}$, one observes that Eq.~\eqref{R2_u} is equivalent to  Eq.~\eqref{eq:5}. 
The same universality holds true for the $n$-point correlation function:

\begin{theorem} \label{KernelTheorem}
Let $z$ be in the semi-disk $\mathbb{D}_{a,+}$ \eqref{sdisk}. Set 
$z_j= z+ {s_j}/{\sqrt{R_1(z)}}$
where $s_1, \ldots , s_n$ are complex numbers. Then in the limit of strong non-unitarity \eqref{a},
\begin{align} \label{fe}
\lim_{N\to \infty, \, M=aN} \frac{1}{(R_1 (z))^n}R_n (z_1, ..., z_n) = \det{ \left[e^{ 2\pi \big(s_k \overline{s_l} - \frac{1}{2}|s_k|^2 - \frac{1}{2} |s_l|^2\big)  } \right] }_{k, l = 1}^n.
\end{align}
\end{theorem}
\begin{proof}
We shall use the Pfaffian representation of the $n$-point correlation function, see Eqs  \eqref{Rn} and \eqref{Pf}. 
Since $z\in \mathbb{D}_{a,+}$ it follows from Lemma \ref{lemma51} that the pre-kernel $g_N$ in \eqref{KN} can be replaced by $g$. Lemma \ref{lemma54} then asserts that the diagonal entires of the resulting matrix kernel are effectively zeros, so that in the strongly non-unitary limit \eqref{a} we have 
\begin{align}\label{Rn2}
R_n(z_1, \ldots , z_n) \sim \prod_{j=1}^n \big( (z_j-z_j^*) w^2(z_j) \big) \Pf \left( K(z_k,z_l) \right)_{k,l=1}^n
\end{align}
where 
 \begin{align*}
K(z_k, z_l) =  \begin{bmatrix} 0 & g(z_k, z_l^*) \\ g(z_k^* , z_l) & 0\end{bmatrix} \, .
 \end{align*}
The $2n\times 2n$ skew-symmetric matrix $\left( K(z_k,z_l) \right)_{k,l=1}^n$ has a checkerboard structure which makes it possible to reduce the Pfaffian in  \eqref{Rn2} to an $n\times n$ determinant, $ \Pf \left( K(z_k,z_l) \right)= \det \left( g(z_k, z_l^*) \right)$, see for example Section 4.6 of \cite{SinclArt}. It follows from this and Lemma \ref{lemma54}, that 
\begin{align*}
R_n(z_1, \ldots , z_n) \sim \prod_{j=1}^n \big( (z_j- z_j^*) w^2(z_j) \big) \det \left( g(z_k, z_l^*) \right)_{k,l=1} =
\det \big[T(z_k, z_l)  \big]_{k,l=1}\, , 
\end{align*}
where 
\begin{align*}
T(z_k, z_l) = \frac{M}{\pi} \frac{(1-|z_k|^2)^{M-\frac{1}{2}} (1-|z_l|^2)^{M-\frac{1}{2}}}{(1-z_k {z_l^*})^{2M + 1}}= \frac{M}{\pi (1-z_k {z_l^*})^2} \exp{ \left( (M-\frac{1}{2}) \ln{A} (z_k,z_l) \right) }\, ,
\end{align*}
and 
\begin{align*}
A (z_k,z_l) = \frac{(1-|z_k|^2) (1-|z_l|^2)}{(1-z_k {z_l^*})^2}.
\end{align*}
Recalling that  $ z_i = z + \frac{s_i}{\sqrt{R_1(z)}} $, 
\begin{align*}
\ln A(z_1, z_2) = &\sqrt{ \frac{\pi}{M} } [({z^*} s_1 - z {s_1^*}) - ({z^*} s_2 - z {s_2^*})] + \frac{\pi}{M} (2 s_1 {s_2^*} - |s_1|^2 - |s_2|^2) + \\
&+ \frac{\pi}{2M} [({z^*}^2 s_1^2 - z^2 {s_1^*}^2) - ({z^*}^2 s_2^2 - z^2 {s_2^*}^2)] + O(M^{-\frac{3}{2}}).
\end{align*}
For every diagonal matrix $D$, $ R_n = \det{D T D^{-1}}$. Choosing $ D = \diag[d(s_1), d(s_2), ..., d(s_n)] $ with $ d(s) = \exp \left( -\sqrt{ \frac{\pi}{M} } ({z^*} s - z {s^*}) - \frac{\pi}{2M} ({z^*}^2 s^2 - z^2 {s^*}^2) \right) $,  
\begin{align*}
T(z_k, z_l)d(s_k)\left( d(s_l) \right)^{-1}= \frac{M}{\pi (1-z_k {z_l^*})^2} \exp{ \left( M \frac{\pi}{M} (2 s_k {s_l^*} - |s_k|^2 - |s_l|^2) +O(M^{-\frac{1}{2}}) \right) }.
\end{align*}
This implies \eqref{fe}. \end{proof}

\subsection{Weakly non-unitarity limit}

Let us now turn to the second regime, $N\to\infty$, $M$ is finite. 
%
For the purpose of calculating the pre-kernel $g_{N} (u, v)$ \eqref{gN} in this regime, it is convenient to change to new variables $ x = u^2 $ and $ y = v^2 $, so that 
\begin{align}\label{psi0}
g_{N} (u, v)=\psi_N(u^2,v^2)\, ,
\end{align}
where 
\begin{align}\label{psi1}
\psi_N (x, y) &= \frac{B(\frac{1}{2},M)}{\pi} \sum_{k = 0}^{N - 1} \sum_{n = 0}^{k} \frac{(n+1)_M }{ \Gamma (M) } \, \frac{(k+3/2)_M}{\Gamma (M) }\,  \big(x^{n} y^{k+\frac{1}{2}} - y^{n} x^{k+\frac{1}{2}}\big)
\\ \nonumber 
 &= \frac{\Gamma (\frac{1}{2})}{\pi\, \Gamma (M+\frac{1}{2})\Gamma (M)}\,\, \frac{\partial^{2M}}{\partial x^M \partial y^M} \sum_{k = 0}^{N - 1} 
 \sum_{n = 0}^{k} \big(x^{M+n}y^{M+k+\frac{1}{2}} -    x^{M+k+\frac{1}{2}}y^{M+n}\big)\, .
\end{align} 
Adding up the geometric sequences and using the duplication formula for the Gamma function, $ \Gamma (z)\Gamma(z+\frac{1}{2})  = 2^{1-2z} \Gamma(\frac{1}{2}) \Gamma (2z)$, one arrives at 
\begin{align}\label{psi3}
\psi_N (x, y) &= \frac{2^{2M-1}}{\pi\, \Gamma (2M)}\,  \frac{\partial^{2M}}{\partial x^M \partial y^M} \big[A_N(x,y) - A_N(x,y) \big]
\end{align}
where 
\begin{align}\label{psi4}
A_N(x,y) = \frac{x^M y^{M+\frac{1}{2}}}{1 - x} \left[ \frac{1 - y^{N}}{1 - y} - x \frac{1 - (xy)^{N}}{1 - xy} \right].
\end{align}
These  are useful equations in the regime of weak non-unitarity. It is instructive to consider first the case of $M = 1$, i.e. of Haar unitary quaternion   matrices of size $(N+1)\times(N+1)$ with one quaternion row and one quaternion column removed.  

\begin{lemma} \label{lemma1RowTrSE}
	Let $M=1$ and $z_0=e^{i\phi_0}$ be a fixed point on the unit circle in the upper half of the complex plane, $ 0 < \phi_0 < \pi $. Set 
	\begin{align*}
	u = \left( 1 - \frac{q_1}{N} \right) e^{i \phi_0 + i \frac{\phi_1}{N}}, \ \ \ \  v =  \left( 1 - \frac{q_2}{N} \right) e^{ - i \phi_0 + i \frac{\phi_2}{N}},
	\end{align*}
	where  $ q_j >0 $. Then, in the limit $ N \to \infty $, $g_{N} (u, v^*)=O(N^2)$ and   
	\begin{align}\label{gNM=1}
	g_{N} (u, v) =  \frac{N^3}{\pi t^3 (e^{i \phi_0 } - e^{-i \phi_0 })}  \left(1 - e^{-2t} \left(1 + 2t + 2 t^2 \right) \right)+ O(N^2),
	\end{align}
	where $ t = q_1 + q_2 - i (\phi_1 + \phi_2) $. 
\end{lemma}
\begin{proof}  By setting $M=1$ in \eqref{psi3} -- \eqref{psi4}, 
\begin{align*}
g_{N} (u, v)=\psi_N(x,y),
\end{align*}
 where 
	\begin{align}\label{uv}
		x = \left( 1 - \frac{q_1}{N} \right)^2 e^{2i \phi_0 + 2i \frac{\phi_1}{N}}, \ \ \ \  y =  \left( 1 - \frac{q_2}{N} \right)^2 e^{ - 2i \phi_0 +2 i \frac{\phi_2}{N}},
	\end{align}
and  
\begin{align*}
\psi_N (x,y) = \frac{2}{\pi} \frac{\partial^2}{\partial x \partial y} \big[A_N(x, y) - A_N(y, x)\big], \quad A_N(x, y) &= \frac{x y^{\frac{3}{2}} }{1 - x} \left[ \frac{1 - y^{N}}{1 - y} - x \frac{1 - (xy)^{N}}{1 - xy} \right].
\end{align*}
By our assumptions $0<2\phi_0<2\pi$. Hence, $|1-x| |1-y| \ge \epsilon$ for some $\epsilon >0$. On the other hand, 
\begin{align}\label{xy}
 1 - xy \sim \frac{2(q_1 + q_2) - 2i(\phi_1 + \phi_2)}{N}.
  \end{align}
 Therefore,
\begin{align*}
 \frac{\partial^2}{\partial x \partial y} \left( \frac{x y^{\frac{3}{2}} }{1 - x} \frac{1 - y^{N}}{1 - y} \right) = O(N^2), \quad \frac{\partial^2}{\partial x \partial y} \left( \frac{x^2 y^{\frac{3}{2}} }{1 - x} \frac{1 - (xy)^{N}}{1 - xy} \right) = O(N^3).
 \end{align*}
Focusing on the computation of $\psi_N (x,y) $ to leading order, we have 
%
\begin{align*}
\frac{\partial^2}{\partial x \partial y} A_N(x, y) &= \frac{-x y^{\frac{1}{2}}}{1 - x} \frac{3 - \frac{3}{2} xy - 3 (xy)^{N} - \frac{3}{2} (xy)^{N} - (N^2 + \frac{7}{2}N + \frac{3}{2}) (xy)^N (1 - xy)}{(1 - xy)^2} \\
&- \frac{x^2 y^{\frac{1}{2}}}{(1 - x)^2} [1 + 2 y - 3 xy] \frac{\frac{3}{2} - \frac{1}{2} xy - (N + \frac{3}{2}) (xy)^{N} + (N + \frac{1}{2}) (xy)^{N+1} }{(1 - xy)^3} \\
&+ \frac{\partial^2}{\partial x \partial y} \left( \frac{x y^{\frac{3}{2}} }{1 - x} \frac{1 - y^{N}}{1 - y} \right),
\end{align*}
and, by making use of $ (1 - xy) = O(N^{-1}) $,
\begin{align*}
\frac{\partial^2}{\partial x \partial y} A_N(x, y) = &O(N^2) + \frac{x y^{\frac{1}{2}}}{1 - x} \cdot \frac{N^2 (xy)^N}{1 - xy} \\
&- \frac{x^2 y^{\frac{1}{2}}}{(1 - x)^2} [1 + 2 y - 3 xy] \frac{\frac{3}{2} - \frac{1}{2} xy - (xy)^{N} - N (xy)^{N} (1 - xy) }{(1 - xy)^3}.
\end{align*}
Recalling \eqref{uv} and \eqref{xy} and collecting all terms of order $N^3$, 
\begin{align*}
\frac{\partial^2}{\partial x \partial y} A_N(x, y) = O(N^2) + \frac{e^{i \phi_0 }}{1 - e^{2i \phi_0}} \frac{2 N^3}{(2t)^3} \left(-1 + e^{-2t} \left(1 + 2t + 2 t^2 \right) \right).
\end{align*}
where $ t = (q_1 + q_2) - i (\phi_1 + \phi_2) $. Similarly, 
\begin{align*}
\frac{\partial^2}{\partial x \partial y} A_N(y, x) = O(N^2) + \frac{e^{-i \phi_0 }}{1 - e^{-2i \phi_0}} \frac{2 N^3}{(2t)^3} \left(-1 + e^{-2t} \left(1 + 2t + 2 t^2 \right) \right),
\end{align*}
and \eqref{gNM=1} follows. Now, $g_{N} (u, v^*)=\psi_N(x,y^*)$. Since $1-xy^*=O(1)$, $g_{N} (u, v^*)=O(N^2)$. 
%
%
\end{proof}

As an immediate corollary of Lemma \ref{lemma1RowTrSE}, one gets a closed form expression for the eigenpair correlation functions of truncated Haar unitary quaternion  matrices with one quaternion row and one quaternion column removed.

\begin{theorem} \label{1rowThRnTrSE}
	Set $ z_j = \left( 1 - \frac{q_j}{N} \right) e^{i \phi_0 + i \frac{\phi_j}{N}} $, where $ 0 < \phi_0 < \pi $ and  $ q_j > 0 $. Then 		in the limit $ N \to \infty $, $ M = 1$:
	\begin{equation} \label{1rowResRnTrSE}
	R_n (z_1, ..., z_n) \sim \det \left[\frac{N^2}{\pi t_{ij}^2}  
	\left(1 - e^{-2t_{ij}} \left(1 + 2t_{ij} + 2 t_{ij}^2 \right) \right) \right]_{i, j = 1}^n,
	\end{equation}
	where $  t_{ij} = q_i + q_j - i (\phi_i - \phi_j) $. In particular, 
	\begin{equation} \label{1rowResTrSE}
	R_1 \left( \left( 1 - \frac{q}{N} \right) e^{i \phi_0} \right) \sim \frac{N^2}{4 \pi q^2} (1 - e^{-4q} (1 + 4q + 8q^2)). 
	\end{equation}
\end{theorem}

\begin{proof} In terms of the pre-kernel, the one-point correlation function $R_1(z)$ is given by (see \eqref{EqDensity1}) 
\begin{align} \label{EqDensity2}
R_1 (z) = (z - z^*) (1 - |z|^2)^{2M - 1} g_N (z, z^*).
\end{align}
Setting here $z=(1-q/N)e^{i \phi_0}$ and applying Lemma \ref{lemma1RowTrSE} one arrives at \eqref{1rowResTrSE}. 

For higher order correlation functions, Lemma \ref{lemma1RowTrSE} asserts that the diagonal entries of the $2\times 2$ matrix kernel $K_N(z_j,z_k)$ in the Pfaffian representation  \eqref{Pf} -- \eqref{KN} of $R_n(z_1, \ldots, z_n)$ are of a sub-leading order compared to the off-diagonal entries. Therefore, in the limit $N\to\infty$, 
\begin{align}\label{Rn2a}
R_n(z_1, \ldots , z_n) \sim \prod_{j=1}^n \big( (z_j-z_j^*) w^2(z_j) \big) \Pf \left( K_N(z_k,z_l) \right)_{k,l=1}^n
\end{align}
where 
 \begin{align*}
K_N(z_k, z_l) =  \begin{bmatrix} 0 & g_N(z_k, z_l^*) \\ g_N(z_k^* , z_l) & 0\end{bmatrix} \, .
 \end{align*}
The $2n\times 2n$ skew-symmetric matrix $\left( K_N(z_k,z_l) \right)_{k,l=1}^n$ has a checkerboard structure. Hence,  $ \Pf \left( K_N(z_k,z_l) \right)= \det \left( g_N(z_k, z_l^*) \right)$, see for example Section 4.6 of \cite{SinclArt}, and  
\begin{align}\label{wnu}
R_n(z_1, \ldots , z_n) \sim \prod_{j=1}^n \big( (z_j- z_j^*) w^2(z_j) \big) \det \left( g_N(z_k, z_l^*) \right)_{k,l=1} \, .
\end{align}
Lemma \ref{lemma1RowTrSE} provides also an asymptotic expression for $g_N(z_k, z_l^*)$ and passing on to the limit $N\to\infty$ on the right-hand side of \eqref{wnu} one arrives at \eqref{1rowResRnTrSE}.
%
\end{proof}

The method which was used to prove Lemma \ref{lemma1RowTrSE} can be extended to cover the entire range of $M$ in the weakly non-unitary regime, see for example \cite{Thesis} where the one-point correlation function $R_1(z)$ was computed in this way for arbitrary fixed $M$. Here we use a variation of this method, which is less formal but simpler, to compute the pre-kernel, and hence all the correlation functions in the weakly non-unitary regime. 

\begin{lemma} \label{lemmaMRowTrSE}
	Let $z_0=e^{i\phi_0}$ be a fixed point on the unit circle in the upper half of the complex plane, $ 0 < \phi_0 < \pi $. Set 
	\begin{align}\label{uv1}
	u = \left( 1 - \frac{q_1}{N} \right) e^{i \phi_0 + i \frac{\phi_1}{N}}, \ \ \ \  v =  \left( 1 - \frac{q_2}{N} \right) e^{ - i \phi_0 + i \frac{\phi_2}{N}},
	\end{align}
	where  $ q_j >0 $. Then, for any fixed $M$ in the limit $ N \to \infty $, $g_{N} (u, v^*)=O(N^{2M})$ and   
	\begin{align}\label{gNM=1}
	g_{N} (u, v) \sim 
	\frac{2^{2M}}
	{\pi \Gamma (2M)} 
	\frac{N^{2M+1}}{e^{i \phi_0 } - e^{-i \phi_0 }}  \int_0^1 t^{2M}  e^{-2[q_1+q_2-i(\phi_1+\phi_2)]\, t} dt\, .
	\end{align}
\end{lemma}

\begin{proof} The pre-kernel $g_N(u,v)$ can be calculated in the weakly non-unitary regime with the help of representations \eqref{psi0} and \eqref{psi3} -- \eqref{psi4}. Effectively, we need to calculate $\psi_N(x,y)$  and $\psi_N(x,y^*)$ with 
\begin{align*}
x=\left(1-\frac{s_1}{N} \right) e^{i\varphi_1}, \quad y=\left(1-\frac{s_2}{N} \right)e^{i\varphi_2}\, ,
\end{align*}
where $s_1=2q_1$, $s_2=2q_2$ and 
\begin{align}\label{varphi}
\varphi_1=2\phi_0+ \frac{2\phi_1}{N},\,\,\, \varphi_2=-2\phi_0+ \frac{2\phi_2}{N},\quad  0 < \varphi_0 < \pi. 
\end{align} 
To this end, note that 
\begin{align*}
\frac{\partial^{2M}}{\partial x^M \partial y^M} = N^{2M} e^{-iM(\varphi_1+\varphi_2)} \frac{\partial^{2M}}{\partial s_1^M \partial s_2^M}\, .
\end{align*} 
Therefore,
\begin{align*}
\psi_N\left( \left(1-\frac{s_1}{N} \right) e^{i\varphi_1}, \left(1-\frac{s_2}{N} \right) e^{i\varphi_2}\right)&= \\[2ex]
\MoveEqLeft[5] 
 \frac{2^{2M-1} N^{2M}}{\pi \Gamma (2M)}\,   \frac{\partial^{2M}}{\partial s_1^M \partial s_2^M}  \left[ a_N(s_1,\varphi_1; s_2, \varphi_2) - a_N(s_2,\varphi_2; s_1, \varphi_1) \right]\, ,
\end{align*}
where 
\begin{align*}
a_N(s_1,\varphi_1; s_2, \varphi_2) = &  
\frac{\left( 1-\frac{s_1}{N} \right)^M\!\!  \left( 1-\frac{s_2}{N} \right)^{M+\frac{1}{2}} e^{i\frac{\varphi_2}{2}} }{1- \left( 1-\frac{s_1}{N} \right)e^{i\varphi_1}} \times  \\[2ex]
\MoveEqLeft[3] 
\left[ 
\frac{
1- \left( 1-\frac{s_2}{N} \right)^N \!\! 
e^{i N \varphi_2}
}
{
1- \left( 1-\frac{s_2}{N} \right)e^{i\varphi_2}
} 
- \left( 1-\frac{s_1}{N} \right)
e^{i \varphi_1} \, 
\frac{
1- \left( 1-\frac{s_1}{N} \right)^N\!\! \left( 1-\frac{s_2}{N} \right)^N \!\! e^{i N (\varphi_1+\varphi_2)} 
}
{1-  \left( 1-\frac{s_1}{N} \right) \left( 1-\frac{s_2}{N} \right) e^{i (\varphi_1+\varphi_2)} }
\right].
\end{align*}
It follows from \eqref{varphi} that $|1-e^{i\varphi_2}|>0$ and $ e^{i(\varphi_1+\varphi_2)}=1+{(\phi_1+\phi_2)}/{N}$ in the limit $N\to\infty$ and therefore the first term in the square brackets above  and its derivatives are of order 1 while the second term and its derivatives are of order $N$. Hence, 
%
\begin{align}\label{lt}
a_N(s_1,\varphi_1; s_2, \varphi_2) &\sim - N
\frac{e^{i \phi_1}}{1- e^{2 i\phi_1}} 
\frac{
1- e^{-s_1-s_2 +2 i (\phi_1+\phi_2)}
}
{s_1+s_2 - 2 i (\phi_1+\phi_2)} \\
& = \frac{N}{e^{i \phi_1}- e^{-i\phi_1}} \int_0^1 e^{-(s_1+s_2-2i(\phi_1+\phi_2)t} dt
\end{align}
and 
\begin{align*}
\psi_N\left( \Big(1-\frac{2q_1}{N} \Big) e^{2i\big(\phi_0 +\frac{\phi_1}{N}\big)}, \big(1-\frac{2q_2}{N} \big) e^{2i\big(\phi_0 +\frac{\phi_1}{N}\big)}\right)
&\sim  \\[2ex]
\MoveEqLeft[10] 
 \frac{2^{2M-1} N^{2M}}{\pi \Gamma (2M)}\,  \frac{2N}{e^{i \phi_1}- e^{-i\phi_1}}  \int_0^1 t^{2M} e^{-2[q_1+q_2-i(\phi_1+\phi_2)]t} dt\, .
\end{align*}
Recalling \eqref{psi0}, this implies \eqref{gNM=1}. 

When calculating 
$\psi_N\left( \left(1-\frac{s_1}{N} \right) e^{i\varphi_1}, \left(1-\frac{s_2}{N} \right) e^{-i\varphi_2}\right)$ one encounters the same 
$a_N(s_1,\varphi_1; s_2, \varphi_2) $ only with $\varphi_2$ replaced by $-\varphi_2$. In this case the denominator of the second term in the square brackets is bounded away from zero and $a_N(s_1,\varphi_1; s_2, \varphi_2) $ and its derivatives are of order of unity. Therefore $\psi_N\left( \left(1-\frac{s_1}{N} \right) e^{i\varphi_1}, \left(1-\frac{s_2}{N} \right) e^{-i\varphi_2}\right)$ and, consequently, $g_N(u,v^*)$ are $O(N^{2M})$.
\end{proof}

We can now calculate the spectral density $\rho_{2N}(z)$, Eq.~\eqref{spdensity_I}, of the truncated unitary quaternion  matrices in the regime of weak non-unitarity and also all the eigenvalue correlation functions $R_n (z_1, ..., z_n)$, 
locally away from the real line. The universality of the obtained expressions is discussed in Section \ref{S:2}. 
\begin{theorem}\label{thm:610}
Away from the real line, the spectral density of the truncated Haar unitary quaternion  matrices in the weakly non-unitary limit is given by 
\begin{align} \label{rho_wnu}
\lim_{N\to\infty} \, \frac{1}{N}\, \rho_{2N} \Big(\!\big(1-\frac{q}{N}\big)e^{i\phi}  \Big) = \frac{2}{\pi} \frac{(4 q)^{2M-1}}{(2M-1)!} \int_0^{1} t^{2M} e^{-4 q t} dt\, , \quad 0<\phi <\pi, 
\end{align}
\end{theorem}
\begin{proof}
It follows from \eqref{EqDensity2} and \eqref{uv1} -- \eqref{gNM=1} that  to leading order in $N$
\begin{align}
R_1\left( \Big(1-\frac{q}{N} \Big)e^{i\phi}  \right) = \frac{2N^2}{\pi} \frac{(4q)^{2M-1}}{(2M-1)!} \int_0^{1} t^{2M} e^{-4qt} dt
\end{align}
for every fixed $\phi \in (0,\pi)$. It is straightforward to verify that this relation reduces to Eq.~\eqref{1rowResTrSE} if $M=1$. Recalling now that $\rho_{2N}(z)=R_1(z)/N$, one arrives at  \eqref{rho_wnu}.

\end{proof}

\begin{theorem} \label{thm:611}
Let $ z_j = \left( 1 - \frac{q_j}{N}\, \right) e^{i \phi_0 + i \frac{\phi_j}{N}} $, where $ 0 < \phi_0 < \pi $, $ q_j > 0 $.  
Then in the weakly non-unitary limit 		
\begin{equation*}
\lim_{N\to\infty}  \frac{1}{N^{2n}} R_n (z_1, ..., z_n) = \left(\frac{2}{\pi}\right)^{\! \! n}
\!\! \left( \prod_{j=1}^n \frac{(4 q_j)^{2M-1}}{(2M-1)!} \right) \, 
	\det \!
	\left[
	\int_0^1 t^{2M} e^{-2 (q_k+q_l-i(\phi_k-\phi_l)\, t} \, dt
	\right]_{k,l=1, \ldots , n}
\end{equation*}
\end{theorem}

\begin{proof} Proof of this theorem is similar to the proof of Theorem \ref{1rowThRnTrSE} and we omit it.
\end{proof}

Using Eqs \eqref{psi0}--\eqref{psi4} one can also work out the pre-kernel $g_N(u,v)$ near the real line in the limit of weak non-unitarity. At $z_0=1$, instead of using the spectral variable $z=(1-\frac{r}{N})e^{i\frac{1}{N} \phi}$, it is more convenient to use the spectral variable $z=1-\frac{u}{N}$ with complex $u$ restricted by the condition $\Re u>0$.  To leading order in $N$, $\Im u=\phi$. 

 \begin{lemma} \label{z=1} For every complex $u$, $v$ with positive real parts it holds in the limit $N\to\infty$ that 
	\begin{align*}
	g_{N} \left( 1- \frac{u}{N} , 1- \frac{v}{N} \right)  \sim 
	\frac{N^{2M+2} \, 2^{2M-1} }
	{\pi \Gamma (2M)} 
	\int_0^1\int_0^1 s^{2M+1} t^M e^{-2s (u+v)} \left( e^{2s (1-t)u}-e^{2s  (1-t) v}\right)\, d s  dt \, .
	\end{align*}
\end{lemma}

\begin{proof}
We have 
\begin{align*}
g_{N} \left( 1- \frac{u}{N} , 1- \frac{v}{N} \right) \sim \psi_N \left( 1- \frac{2u}{N} , 1- \frac{2v}{N} \right),
\end{align*}
where $\psi_N(x,y)$ is given by Eqs \eqref{psi3}--\eqref{psi4}. Introducing $a_N(x,y)=\frac{1}{N^2}A_N\left( 1- \frac{x}{N} , 1- \frac{y}{N} \right)$, 
\begin{align*}
\psi_N \left( 1- \frac{x}{N} , 1- \frac{y}{N} \right) =\frac{N^{2M+2} \, 2^{2M-1}}{\pi \Gamma (2M)} \frac{\partial^{2M}}{\partial x^M \partial y^M} \big(a_N(x,y)-a_N( y,x)\big)\, .
\end{align*}
In the limit $N\to\infty$, $a_N(x,y) \to a(x,y)$, where 
\begin{align*}
a(x,y)= \frac{1-e^{-y}}{xy} - \frac{1-e^{-(x+y)}}{x(x+y)} =\frac{1}{x} \int_0^1 e^{-s y} ds - \frac{1}{x} \int_0^1 e^{-s(x+y)}ds \, ,
\end{align*}
so that 
\begin{align*}
\psi_N \left( 1- \frac{x}{N} , 1- \frac{y}{N} \right) \sim \frac{N^{2M+2} \, 2^{2M-1}}{\pi \Gamma (2M)} \frac{\partial^{2M}}{\partial x^M \partial y^M} \big(a(x,y)-a( y,x)\big)\, .
\end{align*}
On differentiating, 
 \begin{align}\nonumber 
 \frac{\partial^{2M}a(x,y)  }{\partial x^M \partial y^M}  & = (-1)^M \frac{\partial^{M}}{\partial x^M } \left( \frac{1}{x} \int_0^1 \sigma^M e^{-s  y} d\sigma - \frac{1}{x} \int_0^1 s ^M e^{-s (x+y)}d\sigma  \right)\\ \nonumber 
 & = \frac{M!}{x^{M+1}} \int_0^1 s^M e^{-s y} ds - \frac{M!}{x^{M+1}} \int_0^1 s^M  e^{-s (x+y)} \sum_{k=0}^M  \frac{(sx)^k}{k!} \,  ds\\
 \label{eq6a:1}
 &= \frac{1}{x^{M+1}} \int_0^1 s^M e^{-s y} \gamma (M+1, sx)\,   ds \, ,
 \end{align}
 where $\gamma (n, s)$ is the lower incomplete Gamma function,
 \begin{align*}
 \Gamma (M+1) -\Gamma (M+1) e^{-sx} \sum_{k=0}^{M} \frac{(sx)^k}{k!}=\gamma (M+1, sx) = \int_0^{sx} \tau^{M} e^{-\tau} d\tau.
\end{align*}
On substituting the integral on the right hand side back into \eqref{eq6a:1} one gets 
 \begin{align*}
 \frac{\partial^{2M}a(x,y)  }{\partial x^M \partial y^M} =  \int_0^1  \int_0^1 s^{2M+1} t^M e^{-s (y+tx)} ds dt \, .
 \end{align*}
 Hence,
 \begin{align*}
\psi_N \left( 1- \frac{x}{N} , 1- \frac{y}{N} \right) \sim \frac{N^{2M+2} \, 2^{2M-1}}{\pi \Gamma (2M)}  \int_0^1  \int_0^1 s^{2M+1} t^M e^{-s (x+y)} \left( e^{s (1-t)x} - e^{s (1-t)y} \right) ds dt \, ,
\end{align*}
and Lemma follows. \end{proof}

With a closed form expression for the pre-kernel $g_N$ near $z_0=1$ in hand, one also gets the eigenpair correlation functions of truncated Haar unitary symplectic matrices in the limit of weak non-unitarity n terms of a Pfaffian. In the general case of $n$-point correlation function the resulting expression is cumbersome, but the one-point correlation function can be put in a relatively compact form:   

\begin{theorem} 
 \label{Thm612} 
In the limit of weak non-unitarity \eqref{a}, for every $x>0$ and $y\in \mathbb{R}$
\begin{align*}
\lim_{N\to\infty} \frac{1}{N^2} R_1\left(1-\frac{x+iy}{N} \right)= \frac{2^{4M} yx^{2M+1}}{\pi \Gamma (2M)}  \int_0^1  \int_0^1 s^{2M+1} t^M e^{-2s(1+t) x} \sin (2s(1-t)y) ds dt \, .
\end{align*}
\end{theorem}

\begin{proof}
This is an immediate consequence of Lemma \ref{z=1}  and Eq.~\eqref{EqDensity1}. 
\end{proof}

Until now, when investigating eigenvalue correlations in the regime of weak non-unitary we considered the $1/N$-neighbourhood of the unit circle where the bulk of the eigenvalues of the truncated unitary matrices are. Consider now the closed disk $\mathbb{D}_r$  of radius $r$ centred at the origin. 
We will see below that, on average, the number of the eigenvalues in $\mathbb{D}_{r}$ in the 
weakly non-unitary limit stays finite for every $r <1$. The eigenpair correlation functions in $\mathbb{D}_{0,\varepsilon}$ can be easily obtained by using the same technique as before.

\begin{theorem} 
 \label{Thm613} 
Let $z_j\in \mathbb{D}_{r}$, $r<1$. Then in the limit of weak non-unitarity, 
\begin{align*}
\lim_{N\to\infty} R_n(z_1, \ldots, z_n) = \prod_{j=1}^n \left( (z_j-z_j^*) (1-|z_j|^2)^{2M-1} \right) \Pf \left[ K(z_j, z_k) \right]_{j,k=1}^n\, ,
\end{align*}
where
\begin{align*}
K(u, v) =  \begin{bmatrix} \psi(u^{2\phantom{*}}, v^2) & \psi(u^{2\phantom{*}}, v^{*2} ) \\ \psi( u^{*2} , v^2) & \psi(u^{*2}, v^{*2})\end{bmatrix} \, .
\end{align*}
and 
\begin{align}\label{psi7}
\psi(x,y)=\frac{2^{2M-1}}{\pi \Gamma (2M)}  \frac{\partial^{2M}}{\partial x^{M} \partial y^M} \left[\frac{x^My^M}{(1-x)(1-y)} \frac{\sqrt{y}-\sqrt{x}}{1-\sqrt{xy}}\, \right] \, .
\end{align}
\end{theorem}

\begin{proof}
We shall use the Pfaffian representation of the $n$-point correlation function, see Eqs  \eqref{Rn} and \eqref{Pf}. 
Since all $z_j\in \mathbb{D}_{r}$ it follows from Lemma \ref{lemma51} that the pre-kernel $g_N$ in \eqref{KN} can be replaced by the pre-kernel $g$ \eqref{g} in the limit $N\to\infty$. By letting $N$ to to infinity in \eqref{psi1} one obtains $g(u,v)=\psi(u^2,v^2)$, where $\psi(x,y)$ is given by \eqref{psi7}, and the Theorem follows.  
\end{proof}
  
Eq.~\eqref{psi7} provides a closed form expression for the pre-kernel $g$ although it is not immediately obvious how useful this representation is. For example, using it one can write a compact expression for the one-point correlation function  $R_1(z)$ in the case when $M=1$:
\begin{align*}
R_1(z)=-\frac{1}{\pi} \frac{(z-z^*)^2}{|1-z^2|^2}\left[ 3-\frac{(z-z^*)^2}{(1-|z|^2)^2} \right]\, .
\end{align*}
But then already for $M=2$ the expression for $R_1(z)$ is not so compact. Using Eq.~\eqref{psi7} one can also obtain the correlation functions to leading order for small values of $|z_j|$. But whether Eq.~\eqref{psi7} is useful beyond this remains to be seen.

\bibliography{Truncations_Refs} 

\begin{thebibliography}{10}

\bibitem{AB2012}
G.~Akemann and Z.~Burda.
\newblock Universal microscopic correlation functions for products of
  independent {G}inibre matrices.
\newblock {\em J. Phys. A}, 45(46):465201, 18, 2012.

\bibitem{ABKN2014}
G.~Akemann, Z.~Burda, M.~Kieburg, and T.~Nagao.
\newblock Universal microscopic correlation functions for products of truncated
  unitary matrices.
\newblock {\em J. Phys. A}, 47(25):255202, 26, 2014.

\bibitem{akemann2021scaling}
G.~Akemann, S.-S. Byun, and N.-G. Kang.
\newblock Scaling limits of planar symplectic ensembles, 2021.

\bibitem{AEP2021}
G.~{Akemann}, M.~{Ebke}, and I.~{Parra}.
\newblock {Skew-orthogonal polynomials in the complex plane and their
  {B}ergman-like kernels}.
\newblock {\em arXiv:2103.1214}, 2021.

\bibitem{AkemannPaper}
G.~Akemann, J.~R. Ipsen, and E.~Strahov.
\newblock Permanental processes from products of complex and quaternionic
  induced {G}inibre ensembles.
\newblock {\em Random Matrices Theory Appl.}, 3(4):1450014, 54, 2014.

\bibitem{akemann2019universal}
G.~Akemann, M.~Kieburg, A.~Mielke, and T.~Prosen.
\newblock Universal signature from integrability to chaos in dissipative open
  quantum systems.
\newblock {\em Phys. Rev. Lett.}, 123(25):254101, 6, 2019.

\bibitem{APS2009}
G.~Akemann, M.~J. Phillips, and L.~Shifrin.
\newblock Gap probabilities in non-{H}ermitian random matrix theory.
\newblock {\em J. Math. Phys.}, 50(6):063504, 32, 2009.

\bibitem{AHM2011}
Y.~Ameur, H.~Hedenmalm, and N.~Makarov.
\newblock Fluctuations of eigenvalues of random normal matrices.
\newblock {\em Duke Math. J.}, 159(1):31--81, 2011.

\bibitem{BFK2021}
G.~Ben~Arous, Y.~V. Fyodorov, and B.~A. Khoruzhenko.
\newblock Counting equilibria of large complex systems by instability index.
\newblock {\em Proceedings of the National Academy of Sciences}, 118(34), 2021.

\bibitem{B-GCh2012}
F.~Benaych-Georges and F.~Chapon.
\newblock Random right eigenvalues of {G}aussian quaternionic matrices.
\newblock {\em Random Matrices Theory Appl.}, 1(2):1150009, 18, 2012.

\bibitem{BorodinPaper}
A.~Borodin and C.~D. Sinclair.
\newblock The {G}inibre ensemble of real random matrices and its scaling
  limits.
\newblock {\em Comm. Math. Phys.}, 291(1):177--224, 2009.

\bibitem{B1951}
J.~L. Brenner.
\newblock Matrices of quaternions.
\newblock {\em Pacific J. Math.}, 1:329--335, 1951.

\bibitem{SBE2021}
S.-S. Byun and M.~Ebke.
\newblock Universal scaling limits of the symplectic elliptic {Ginibre}
  ensemble.
\newblock {\em arXiv:2108.05541}, 2021.

\bibitem{CZ1998}
L.-L. Chau and O.~Zaboronsky.
\newblock On the structure of correlation functions in the normal matrix model.
\newblock {\em Comm. Math. Phys.}, 196(1):203--247, 1998.

\bibitem{deBruijn1961}
N.~G. de~Bruijn.
\newblock {\em Asymptotic methods in analysis}.
\newblock Bibliotheca Mathematica, Vol. IV. North-Holland Publishing Co.,
  Amsterdam; P. Noordhoff Ltd., Groningen, second edition, 1961.

\bibitem{D2021}
G.~Dubach.
\newblock Symmetries of the quaternionic {G}inibre ensemble.
\newblock {\em Random Matrices Theory Appl.}, 10(1):2150013, 19, 2021.

\bibitem{D1970}
F.~J. Dyson.
\newblock Correlations between eigenvalues of a random matrix.
\newblock {\em Comm. Math. Phys.}, 19:235--250, 1970.

\bibitem{EatonBook}
M.~L. Eaton.
\newblock {\em Group invariance applications in statistics}, volume~1 of {\em
  NSF-CBMS Regional Conference Series in Probability and Statistics}.
\newblock Institute of Mathematical Statistics, Hayward, CA; American
  Statistical Association, Alexandria, VA, 1989.

\bibitem{Edelman1997}
A.~Edelman.
\newblock The probability that a random real {G}aussian matrix has {$k$} real
  eigenvalues, related distributions, and the circular law.
\newblock {\em J. Multivariate Anal.}, 60(2):203--232, 1997.

\bibitem{JonitThesis}
J.~Fischmann.
\newblock {\em Eigenvalue distributions on a single ring}.
\newblock PhD thesis, Queen Mary University of London, 2012.

\bibitem{FBKSZ2012}
J.~Fischmann, W.~Bruzda, B.~A. Khoruzhenko, H.-J. Sommers, and
  K.~\.{Z}yczkowski.
\newblock Induced {G}inibre ensemble of random matrices and quantum operations.
\newblock {\em J. Phys. A}, 45(7):075203, 31, 2012.

\bibitem{F2006}
P.~J. Forrester.
\newblock Quantum conductance problems and the {J}acobi ensemble.
\newblock {\em J. Phys. A}, 39(22):6861--6870, 2006.

\bibitem{ForresterBook}
P.~J. Forrester.
\newblock {\em Log-gases and random matrices}, volume~34 of {\em London
  Mathematical Society Monographs Series}.
\newblock Princeton University Press, Princeton, NJ, 2010.

\bibitem{F2016}
P.~J. Forrester.
\newblock Analogies between random matrix ensembles and the one-component
  plasma in two-dimensions.
\newblock {\em Nuclear Phys. B}, 904:253--281, 2016.

\bibitem{FI2016}
P.~J. Forrester and J.~R. Ipsen.
\newblock Real eigenvalue statistics for products of asymmetric real {G}aussian
  matrices.
\newblock {\em Linear Algebra Appl.}, 510:259--290, 2016.

\bibitem{FIK2020}
P.~J. Forrester, J.~R. Ipsen, and S.~Kumar.
\newblock How many eigenvalues of a product of truncated orthogonal matrices
  are real?
\newblock {\em Exp. Math.}, 29(3):276--290, 2020.

\bibitem{FM2012}
P.~J. Forrester and A.~Mays.
\newblock Pfaffian point process for the {G}aussian real generalised eigenvalue
  problem.
\newblock {\em Probab. Theory Related Fields}, 154(1-2):1--47, 2012.

\bibitem{FK99}
Y.~V. Fyodorov and B.~A. Khoruzhenko.
\newblock Systematic analytical approach to correlation functions of resonances
  in quantum chaotic scattering.
\newblock {\em Phys. Rev. Lett.}, 83:65--68, Jul 1999.

\bibitem{FK2007}
Y.~V. Fyodorov and B.~A. Khoruzhenko.
\newblock A few remarks on colour-flavour transformations, truncations of
  random unitary matrices, {B}erezin reproducing kernels and {S}elberg-type
  integrals.
\newblock {\em J. Phys. A}, 40(4):669--699, 2007.

\bibitem{FK2016}
Y.~V. Fyodorov and B.~A. Khoruzhenko.
\newblock Nonlinear analogue of the {May-Wigner} instability transition.
\newblock {\em Proceedings of the National Academy of Sciences},
  113(25):6827--6832, 2016.

\bibitem{FSReview1997}
Y.~V. Fyodorov and H.-J. Sommers.
\newblock Statistics of resonance poles, phase shifts and time delays in
  quantum chaotic scattering: random matrix approach for systems with broken
  time-reversal invariance.
\newblock {\em J. Math. Phys.}, 38(4):1918--1981, 1997.

\bibitem{FSReview2003}
Y.~V. Fyodorov and H.-J. Sommers.
\newblock Random matrices close to {H}ermitian or unitary: overview of methods
  and results.
\newblock {\em J. Phys. A}, 36(12):3303--3347, 2003.

\bibitem{Gin1965}
J.~{Ginibre}.
\newblock Statistical ensembles of complex, quaternion, and real matrices.
\newblock {\em Journal of Mathematical Physics}, 6(3):440--449, Mar. 1965.

\bibitem{HaakeBook}
F.~Haake, S.~Gnutzmann, and M.~Ku\'{s}.
\newblock {\em Quantum signatures of chaos}.
\newblock Springer Series in Synergetics. Springer, Cham, 2018.
\newblock Fourth edition of [ MR1123481], With a foreword to the first edition
  by Hermann Haken.

\bibitem{H2000}
M.~B. Hastings.
\newblock Fermionic mapping for eigenvalue correlation functions of weakly
  non-{H}ermitian symplectic ensemble.
\newblock {\em Nuclear Phys. B}, 572(3):535--546, 2000.

\bibitem{HN1996}
N.~Hatano and D.~R. Nelson.
\newblock Localization transitions in non-{Hermitian} quantum mechanics.
\newblock {\em Phys. Rev. Lett.}, 77:570--573, Jul 1996.

\bibitem{ipsen2013products}
J.~R. Ipsen.
\newblock Products of independent quaternion {Ginibre} matrices and their
  correlation functions.
\newblock {\em Journal of Physics A: Mathematical and Theoretical},
  46(26):265201, 2013.

\bibitem{IK2014}
J.~R. Ipsen and M.~Kieburg.
\newblock Weak commutation relations and eigenvalue statistics for products of
  rectangular random matrices.
\newblock {\em Phys. Rev. E}, 89:032106, Mar 2014.

\bibitem{KanzPaper}
E.~Kanzieper.
\newblock Eigenvalue correlations in non-{H}ermitean symplectic random
  matrices.
\newblock {\em J. Phys. A}, 35(31):6631--6644, 2002.

\bibitem{KL_Gin}
B.~A. Khoruzhenko and S.~Lysychkin.
\newblock Scaling limits of the quaternion-real {Ginibre} ensemble.
\newblock {\em in preparation}, 2021.

\bibitem{KSReview2011}
B.~A. Khoruzhenko and H.-J. Sommers.
\newblock Non-{H}ermitian ensembles.
\newblock In {\em The {O}xford handbook of random matrix theory}, pages
  376--397. Oxford Univ. Press, Oxford, 2011.

\bibitem{KSZ2010}
B.~A. Khoruzhenko, H.-J. Sommers, and K.~\.{Z}yczkowski.
\newblock Truncations of random orthogonal matrices.
\newblock {\em Phys. Rev. E (3)}, 82(4):040106, 4, 2010.

\bibitem{K2009}
M.~Krishnapur.
\newblock From random matrices to random analytic functions.
\newblock {\em Ann. Probab.}, 37(1):314--346, 2009.

\bibitem{L1949}
H.~C. Lee.
\newblock Eigenvalues and canonical forms of matrices with quaternion
  coefficients.
\newblock {\em Proc. Roy. Irish Acad. Sect. A}, 52:253--260, 1949.

\bibitem{LW2016}
D.-Z. Liu and Y.~Wang.
\newblock Universality for products of random matrices {I}: {G}inibre and
  truncated unitary cases.
\newblock {\em Int. Math. Res. Not. IMRN}, (11):3473--3524, 2016.

\bibitem{Thesis}
S.~Lysychkin.
\newblock {\em Complex Eigenvalues of High Dimensional Quaternion Random
  Matrices}.
\newblock PhD thesis, Queen Mary University of London, 2021.

\bibitem{May1972}
R.~M. May.
\newblock Will a large complex system be stable?
\newblock {\em Nature}, 238(5364):413--414, 1972.

\bibitem{Meckes_book}
E.~S. Meckes.
\newblock {\em The random matrix theory of the classical compact groups},
  volume 218 of {\em Cambridge Tracts in Mathematics}.
\newblock Cambridge University Press, Cambridge, 2019.

\bibitem{Mehta2ed}
M.~L. Mehta.
\newblock {\em Random matrices}.
\newblock Academic Press, Inc., Boston, MA, second edition, 1991.

\bibitem{MehtaShri1965}
M.~L. {Mehta} and P.~K. {Srivastava}.
\newblock {Correlation functions for eigenvalues of real quaternian matrices}.
\newblock {\em Journal of Mathematical Physics}, 7(2):341--344, Feb. 1966.

\bibitem{M1922}
E.~Moore.
\newblock On the determinant of a {H}ermitian matrix of quaternionic elements.
\newblock {\em Bull. Amer. Math. Soc.}, 28:161--162, 1922.

\bibitem{PS2018}
M.~Poplavskyi and G.~Schehr.
\newblock Exact persistence exponent for the $2d$-diffusion equation and
  related {Kac} polynomials.
\newblock {\em Phys. Rev. Lett.}, 121:150601, Oct 2018.

\bibitem{RiderPaper}
B.~Rider.
\newblock A limit theorem at the edge of a non-{H}ermitian random matrix
  ensemble.
\newblock {\em J. Phys. A}, 36(12):3401--3409, 2003.
\newblock Random matrix theory.

\bibitem{SinclArt}
C.~D. Sinclair.
\newblock Averages over {G}inibre's ensemble of random real matrices.
\newblock {\em Int. Math. Res. Not. IMRN}, (5):Art. ID rnm015, 15, 2007.

\bibitem{Som2007}
H.-J. Sommers.
\newblock Symplectic structure of the real {G}inibre ensemble.
\newblock {\em J. Phys. A}, 40(29):F671--F676, 2007.

\bibitem{S96}
M.~A. Stephanov.
\newblock Random matrix model of {QCD} at finite density and the nature of the
  quenched limit.
\newblock {\em Phys. Rev. Lett.}, 76:4472--4475, Jun 1996.

\bibitem{TaoVu2015}
T.~Tao and V.~Vu.
\newblock Random matrices: universality of local spectral statistics of
  non-{H}ermitian matrices.
\newblock {\em Ann. Probab.}, 43(2):782--874, 2015.

\bibitem{TWT2004}
M.~{Timme}, F.~{Wolf}, and T.~{Geisel}.
\newblock Topological speed limits to network synchronization.
\newblock {\em Physical Review Letters}, 92(7):074101, Feb. 2004.

\bibitem{ZS2000}
K.~\.{Z}yczkowski and H.-J. Sommers.
\newblock Truncations of random unitary matrices.
\newblock {\em J. Phys. A}, 33(10):2045--2057, 2000.

\end{thebibliography}
\bibliographystyle{abbrv}

\end{document}